\documentclass[preprint]{aastex}

\usepackage{subfigure}
\usepackage{url}
\usepackage{comment}
\usepackage{color}
\usepackage{dcolumn}
\usepackage{enumitem}

\setlist[enumerate]{noitemsep}
\setlist[enumerate,1]{leftmargin=*}
\setlist[itemize]{noitemsep}
\setlist[itemize,1]{leftmargin=*}
\setlist[description]{noitemsep}
\setlist[description,1]{leftmargin=*}
\setenumerate[0]{label=(\arabic*)}

\shorttitle{$H$-band Integrated Light Abundances of M31 Clusters}
\shortauthors{Sakari et al.}

\begin{document}

\title{Infrared High-Resolution Integrated Light Spectral Analyses of
  M31 Globular Clusters from APOGEE}

\author{Charli M. Sakari\email{E-mail:sakaricm@u.washington.edu}}
\affil{Department of Astronomy, University of Washington, Seattle WA
98195-1580, USA}

\author{Matthew D. Shetrone}
\affil{McDonald Observatory, University of Texas at Austin, HC75 Box
1337-MCD, Fort Davis, TX 79734, USA}

\author{Ricardo P. Schiavon}
\affil{Gemini Observatory, 670 N. A'Ohoku Place, Hilo, HI 96720, USA}
\affil{Astrophysics Research Institute, Liverpool John Moores University, 146 Brownlow Hill, Liverpool, L3 5RF, UK}

\author{Dmitry Bizyaev}
\affil{Apache Point Observatory and New Mexico State University, P.O. Box 59, Sunspot, NM, 88349-0059, USA }
\affil{Special Astrophysical Observatory of the Russian AS, Nizhnij Arkhyz, Russia}
\affil{Sternberg Astronomical Institute, Moscow State University, Universitetsky prosp.  13, Moscow, Russia}

\author{Carlos Allende Prieto}
\affil{Instituto de Astrof\'{i}sica de Canarias (IAC), Vía Lactea s/n, E-38205 La Laguna, Tenerife, Spain}
\affil{Departamento de Astrofísica, Universidad de La Laguna (ULL), E-38206 La Laguna, Tenerife, Spain}

\author{Timothy C. Beers}
\affil{Department of Physics and JINA Center for the Evolution of
the Elements, University of Notre Dame, Notre Dame, IN 46556,  USA}

\author{Nelson Caldwell}
\affil{Harvard-Smithsonian Center for Astrophysics, 60 Garden Street, Cambridge, MA 02138, USA}

\author{Domingo An\'{i}bal Garc\'{i}a-Hern\'{a}ndez}
\affil{Instituto de Astrofísica de Canarias (IAC), Vía Lactea s/n, E-38205 La Laguna, Tenerife, Spain}
\affil{Departamento de Astrofísica, Universidad de La Laguna (ULL), E-38206 La Laguna, Tenerife, Spain}

\author{Sara Lucatello}
\affil{INAF – Osservatorio Astronomico di Padova, Vicolo dell’Osservatorio 5, 35122 Padova, Italy.}

\author{Steven Majewski, Robert W. O'Connell}
\affil{Dept. of Astronomy, University of Virginia, Charlottesville, VA 22904-4325, USA}

\author{Kaike Pan}
\affil{Apache Point Observatory and New Mexico State University, P.O. Box 59, Sunspot, NM, 88349-0059, USA}

\author{Jay Strader}
\affil{Department of Physics and Astronomy, Michigan State University, East Lansing, MI 48824, USA}


\begin{abstract}
Chemical abundances are presented for 25 M31 globular clusters (GCs),
based on moderately high resolution ($R=22,500$) $H$-band integrated
light spectra from the Apache Point Observatory Galactic Evolution
Experiment (APOGEE).  Infrared spectra offer lines from new elements,
of different strengths, and at higher excitation potentials compared
to the optical.  Integrated abundances of C, N, and O are derived from
CO, CN, and OH molecular features, while Fe, Na, Mg, Al, Si, K, Ca,
and Ti abundances are derived from atomic features.  These abundances
are compared to previous results from the optical, demonstrating the
validity and value of infrared integrated light analyses.  The CNO
abundances are consistent with typical tip of the red giant branch
stellar abundances, but are systematically offset from optical, Lick
index abundances.  With a few exceptions, the other abundances agree
between the optical and the infrared within the $1\sigma$
uncertainties.  The first integrated K abundances are also presented,
and demonstrate that K tracks the $\alpha$-elements.  The combination
of infrared and optical abundances allows better determinations of GC
properties, and enables probes of the multiple populations in
extragalactic GCs.  In particular, the integrated effects of the Na/O
anticorrelation can be directly examined for the first time.
\end{abstract}


\section{Introduction}\label{sec:Intro}
Integrated Light (IL) spectroscopy of globular clusters (GCs) provides
valuable clues about the assembly histories of distant galaxies and
their GC systems.  An IL spectrum comes from an entire stellar
population---integrated chemical abundances therefore represent
flux-weighted averages from the individual stars observed in the IL
spectrum.  Despite the potential difficulties in modeling the
underlying stellar populations, there are certain elements, spectral
features, and/or wavelength regions that provide robust IL abundances
(see, e.g., \citealt{SchiavonHB,Sakari2013}). Low and medium
resolution ($R \lesssim 5000$) IL spectroscopy provides ages,
metallicities, and abundances of the elements with the strongest
spectral features (e.g., C, N, Mg;
\citealt{Caldwell2011,Schiavon2013}), while higher resolution
spectroscopy provides higher precision abundances of a wider variety
of elements (including neutron capture elements such as Ba and Eu in
the optical; \citealt{McWB,Colucci2009,Colucci2012,Sakari2015}).  IL
spectral observations have identified, among other things, possible
metallicity bimodalities (\citealt{Perrett2002}, though
\citealt{Caldwell2011} find no bimodality) and metallicity gradients
(e.g., \citealt{Caldwell2011}) in M31's GC population,
chemically-peculiar GCs in M31's outer halo that may have been
accreted \citep{Sakari2015}, $\alpha$-deficiencies in distant GCs that
are associated with dwarf galaxies \citep{Puzia2008}, and enhanced
[$\alpha$/Fe] ratios in metal-rich GCs associated with the early type
galaxy NGC~5128  \citep{Colucci2013}.  IL spectroscopy has also
provided insight into the nature of GCs themselves, through
comparisons with Milky Way GCs \citep{Schiavon2012}, detections
of anomalous abundances indicative of multiple populations
(e.g. \citealt{Colucci2009,Colucci2014}, \citealt{Sakari2015}) and
abundance correlations with cluster mass \citep{Schiavon2013}.

Though previous IL observations have typically been at optical
wavelengths ($\sim 3000-9000$ \AA), high resolution, infrared (IR) IL
spectroscopy is now possible thanks to recent advances in infrared
(IR) spectroscopy.  In particular, the Apache Point Observatory
Galactic Evolution Experiment (APOGEE) provides multi-object,
high-resolution ($R=22,500$) spectroscopy with coverage in the
$H$-band (from $1.51$ to $1.69\;\mu$m); this wavelength coverage has
some significant advantages for IL spectroscopy.
\begin{enumerate}
\item {\it Insensitivity to hot stars.}  IL spectra are composed of
  light from all the stars in a stellar population, including dwarfs 
  and giants, hot and cool stars, etc.  At blue wavelengths,
  contributions from hot horizontal branch (HB) stars complicate
  analyses (e.g. \citealt{SchiavonHB,Sakari2014}).  Similarly, optical
  spectra are more sensitive to turnoff stars, and therefore require
  estimates of GC age.  IR spectra are likely to be sensitive only to
  the brightest red giant branch (RGB) and asymptotic giant branch
  (AGB) stars, simplifying IL analyses.
\item {\it Additional spectral lines.}  The $H$-band offers different
  spectral lines than the optical.  In particular, strong molecular
  lines of CN, CO, and OH enable determinations of C, N, and O
  abundances \citep{Smith2013}.  There are CN indices in the optical,
  though they are in the blue and may be too weak in metal-poor
  clusters \citep{Schiavon2013}.  The $H$-band also offers
  complementary lines to the optical, including additional Mg, Al, Si,
  Ca, and Ti lines.  Stronger Al and Si lines can also be utilized in
  the IR than in the optical.
\item {\it Opportunities to probe multiple populations in GCs.}  The
  well-established chemical variations in Milky Way (MW) GCs (e.g., in
  Na/O and Mg/Al; \citealt{Carretta2009}) have been inferred to exist
  in extragalactic GCs because of their IL abundance ratios, notably
  high [Na/Fe] \citep{Colucci2014,Sakari2015}.  The $H$-band offers
  detectable lines from elements that should vary within (at least
  some) GCs, including C, N, O, Mg, and Al.  The ability to detect
  [O/Fe] and directly probe the Na/O anticorrelation makes the IR
  particularly valuable for extragalactic GC studies.
\end{enumerate}

\noindent However, IR IL spectroscopy also suffers from some
disadvantages, compared to the optical.
\begin{enumerate}
\item {\it Line blending.}  In metal-rich clusters molecular features
  dominate the $H$-band; as a result, abundances derived from lines in
  the $H$-band are sensitive to the abundances of C, N, and O.  This
  blending is especially significant in spectra whose lines are
  already blended as a result of the cluster velocity dispersion.
\item {\it Weak lines at low metallicity.}  Many of the strong features
  in IL spectra become weaker in the more metal poor GCs, and may
  disappear entirely.  For the most metal-poor clusters, $H$-band IL
  spectroscopy  may therefore not provide abundances for as many
  elements as the more metal-rich GCs.
\item {\it Sensitivity to evolved stars.}  As stated above, the
  $H$-band is most sensitive to the brightest cluster RGB and AGB
  stars.  The abundances are therefore sensitive to how the evolved
  AGB stars are modeled (in terms of the isochrones, the model
  atmospheres, the relative numbers of AGB stars, and stochastic
  sampling).
\item {\it Lack of iron lines.}  The high-resolution IL analyses of
  unresolved GCs that were developed by \citet{McWB},
  \citet{Colucci2009,Colucci2011a,Colucci2014}, and
  \citet{Sakari2013,Sakari2015} rely on \ion{Fe}{1} lines to determine
  the parameters of the underlying stellar population (specifically
  the age and metallicity of an appropriate isochrone).  However,
  there are few sufficiently strong Fe lines in the $H$-band
  (especially at low metallicities), and any detectable Fe lines may be
  blended with other features.
\end{enumerate}
\noindent Though IR IL spectroscopy may not be as informative as
optical IL spectroscopy on its own, it offers valuable complementary
information to the optical, {\it provided that IL analysis techniques
  are viable in the $H$-band.}  For highly reddened objects the IR might
also provide the only viable spectra for abundance analyses.  This
paper presents the first IR, IL detailed abundance analyses of GCs,
utilizing targets in M31.  The IR abundances are compared to those
derived from optical lines (similar to the individual stellar analysis
of \citealt{Smith2013}).  The targets cover a wide range in
metallicity and have all had previous high-resolution, detailed IL
abundances analyses in the optical.  These $H$-band IL spectra were
observed during an ancillary program of the APOGEE survey, as
described in Section \ref{sec:Observations}. The analysis techniques
are introduced in Section \ref{sec:Analysis}; $H$-band abundances for
25 clusters are then presented (Section \ref{sec:Abunds}), along with
new optical abundances for five GCs (Appendix
\ref{appendix:OpticalAbunds}).  The comparisons are discussed in
detail in Section \ref{sec:Discussion}---particular emphasis is
placed on the feasibility of future high-resolution IR IL analyses.

\section{Observations and Data Reduction}\label{sec:Observations}
The data presented in this paper are a subset of a larger sample of
M31 GCs.  The subset of GCs in this paper were targets in previous
high-resolution, optical spectroscopic analyses and therefore have
reference abundances for comparisons with the $H$-band.  The
high-resolution optical abundances from
\citet{Colucci2009,Colucci2014} are supplemented with new results for
five GCs, as described in Appendix \ref{appendix:OpticalAbunds}.

\begin{table}
\centering
\begin{center}
\caption{Target list.\label{table:Targets}\label{table:Velocities}}
  \newcolumntype{d}[1]{D{,}{\pm}{#1}}
  \newcolumntype{e}[1]{D{.}{.}{#1}}
  \begin{tabular}{@{}lccccccccd{3}@{}}
  \hline
Cluster & RA & Dec & $R_{\rm{proj}}$ & $V$ & $H$ & Observing & \# of & S/N$^{c}$ & \multicolumn{1}{c}{$v_{\rm{helio}}$} \\ 
   & \multicolumn{2}{c}{(J2000)} & (kpc) & & & Epoch$^{a}$ & Visits$^{b}$ &  & \multicolumn{1}{c}{km s$^{-1}$} \\
\hline
B006-G058  & 00:40:26.5 & $+$41:27:26.7 & 6.43 & 15.46 & 12.74 & 11, 13 & 10 & 24.94 & -251.2,0.5 \\
B012-G064  & 00:40:32.5 & $+$41:21:44.2 & 5.78 & 15.04 & 12.79 & 11, 13 & 10 & 22.70 & -361.6,0.5 \\
B034-G096$^{d}$& 00:41:28.1 & $+$40:53:49.6 & 6.05 & 15.47 & 12.61 & 11  & 7  & 23.65 & -554.4,0.5 \\
B045-G108  & 00:41:43.1 & $+$41:34:20.0 & 4.90 & 15.83 & 13.00 & 11, 13 & 10 & 17.02 & -423.3,0.8 \\
B048-G110$^{e}$  & 00:41:45.5 & $+$41:13:30.6 & 2.59 & 16.51 & 13.38 & 11, 13& 10 & 7.99 & -228.9,1.0 \\
B063-G124  & 00:42:00.9 & $+$41:29:09.5 & 3.49 & 15.73 & 12.22 & 11     & 3  & 34.34 & -306.2,0.5 \\
B088-G150  & 00:42:21.1 & $+$41:32:14.3 & 3.79 & 15.00 & 12.34 & 11     & 7  & 29.28 & -489.6,1.0 \\
B110-G172$^{e}$  & 00:42:33.1 & $+$41:03:28.4 & 2.93 & 15.28 & 12.35 & 11 & 7  & 29.26 & -236.7,0.4 \\
B163-G217  & 00:43:17.0 & $+$41:27:44.9 & 3.00 & 15.04 & 11.84 & 11         & 7  & 46.43 & -166.4,0.3 \\
B171-G222$^{d,f}$ & 00:43:25.0 & $+$41:15:37.1 & 1.77 & 15.28 & 12.24 & 11   & 10 & 24.03 & -276.3,0.6 \\
B182-G233$^{d}$ & 00:43:36.7 & $+$41:08:12.2 & 2.88 & 15.43 & 12.52 & 11, 13 & 6  & 22.82 & -358.3,0.5 \\
B193-G244  & 00:43:45.5 & $+$41:36:57.5 & 5.41 & 15.33 & 12.23 & 11 & 11 & 32.22 & -66.1,0.3 \\
B225-G280$^{d}$ & 00:44:29.8 & $+$41:21:36.6 & 4.67 & 14.15 & 11.23 & 11     & 11 & 65.72 & -162.5,0.3 \\
B232-G286$^{d}$ & 00:44:40.5 & $+$41:15:01.4 & 4.97 & 15.65 & 13.35 & 11, 13 & 10 & 10.98 & -196.1,1.0 \\
B235-G297$^{d}$ & 00:44:57.9 & $+$41:29:23.7 & 6.46 & 16.27 & 13.39 & 11, 13 & 10 & 10.69 & -94.4,1.0 \\
B240-G302$^{d}$ & 00:45:25.2 & $+$41:06:23.8 & 7.24 & 15.18 & 12.89 & 11, 13 & 10 & 17.64 & -54.4,0.9 \\
B311-G033$^{d}$ & 00:39:33.8 & $+$49:31:14.4 & 13.11 & 15.45 & 12.85 & 11, 13 & 10 & 17.66 & -502.8,1.0 \\
B312-G035$^{d}$ & 00:39:40.1 & $+$40:57:02.3 & 9.02 & 15.52 & 12.85 & 11, 13 & 10 & 13.58 & -177.7,0.9 \\
B381-G315$^{d}$ & 00:46:06.6 & $+$41:20:58.9 & 8.72 & 15.76 & 13.14 & 11, 13 & 10 & 10.13 & -82.9,1.0 \\
B383-G318$^{d}$ & 00:46:12.0 & $+$41:19:43.2 & 8.92 & 15.30 & 12.84 & 11, 13 & 10 & 16.09 & -233.6,0.8 \\
B384-G319$^{d}$ & 00:46:21.9 & $+$40:17:00.0 & 16.42 & 15.75 & 13.00 & 11, 13 & 10 & 19.98 & -364.5,0.8 \\
B386-G322      & 00:46:27.0 & $+$42:01:52.8 & 14.08 & 15.64 & 12.95 & 11     & 6  & 23.07 & -399.8,0.8 \\
B403-G348      & 00:49:17.0 & $+$41:35:08.2 & 17.34 & 16.22 & 13.47 & 11     & 6  & 16.09 & -269.3,0.8 \\
B405-G351      & 00:49:39.8 & $+$41:35:29.7 & 18.28 & 15.20 & 12.76 & 11     & 6  & 21.26 & -165.5,0.6 \\
B472-D064      & 00:43:48.4 & $+$41:26:53.0 & 3.67 & 15.19 & 12.69 & 11, 13 & 6  & 24.52 & -120.6,0.7 \\
\hline
\end{tabular}\\
\end{center}
\medskip
\raggedright {\bf References:} Positions, projected distances from the
center of M31, and magnitudes are from the Revised Bologna Catalog \citep{RBCref}.\\
{\bf Notes: } 
$^{a}$ Clusters were observed in 2011 and/or 2013.\\
$^{b}$ Each visit is 66.6 minutes of integration.\\
$^{c}$ S/N ratios are per pixel, and represent the median value across
the entire spectral range.  There are approximately 2.06 pixels per
resolution element in the blue, 2.27 in the green, and 2.66 in the
red.\\
$^{d}$ Some visits were affected by superpersistence (see text). The
superpersistence regions are masked out, leading to lower S/N in the
$1.51-1.62\;\mu$m region; some abundances are derived with unmasked
spectra, and have been flagged in subsequent tables. \\
$^{e}$ All visits were affected by superpersistence.\\
$^{f}$ A background galactic component was subtracted during data
reduction.\\
\end{table}

\subsection{Target Selection and Observations}\label{subsec:Observations}
M31 GCs were observed as part of an ancillary project in the APOGEE
survey (see \citealt{Zasowski2013}).  Confirmed GCs were selected from
the optical sample of \citet{Caldwell2009}; altogether, $\sim250$ GCs
were observed by APOGEE, all with $H<15$ mag.  Of the larger sample,
only twenty five are utilized for this paper's comparison with optical
abundances---these twenty five are some of the brightest targets.
Table \ref{table:Targets} shows that all targets in this paper have
$H<13.5$ mag.

$H$-band spectra ($1.51$ to $1.69\;\mu$m) of the target clusters were
obtained with the moderately-high resolution ($R=22,500$;
\citealt{Wilson2010,Wilson2012}) APOGEE spectrograph, on the
2.5 m Telescope at Apache Point Observatory \citep{Gunn2006}.  APOGEE
is a multi-object spectrograph, with 300 fibers placed on a variety of
science and calibration targets (including the sky).  The dispersed
spectra are fed to three detectors, providing ``blue,'' ``green,'' and
``red'' coverage within the $H$-band.  The details of the observations
can be found in \citet{Majewski2015} and \citet{Zasowski2013},
including descriptions of the plates and fibers that were utilized for
the observations.

These GCs were observed in the same manner as individual APOGEE
stellar targets, i.e., a single fiber was allocated for each cluster.
The fiber diameter of 2$\arcsec$ is large enough to cover most of the
clusters past their half-light radii \citep{RBCref}.   Several of the
targets in the larger sample are close to the center of M31 and are
affected by light from M31's unresolved field stars.  Based on
examination of the flux and S/N of background fibers, the background
subtraction is estimated to only be important for targets within
$9\arcmin$ of the center of M31 (i.e., outside $9\arcmin$ there is no
detectable background signal).  Only one cluster in this sample, B171,
had a background component subtracted.  For B171 and the other GCs
within 9$\arcmin$ that are not included in this analysis, separate
fibers were placed on background regions adjacent to the clusters.
Because the fiber collision distance of 72$\arcsec$ prohibited
simultaneous GC and background observations, two 3$^{\circ}$ SDSS
plates were used to observe these clusters \citep{Zasowski2013}: one
plate contained the GC fibers, while another had the background fibers
displaced from the target centers by $\sim$ 10$\arcsec$.  These
background fibers were placed on smooth galactic components, based on
the M31 near-IR maps from the 2 Micron All-Sky Survey
\citep{Skrutskie2006}.    Clusters without dedicated background fibers
were observed with both plates.

The first pair of SDSS GC plates was observed during 2011; one
received three visits (where one visit = 66.6 minutes) while another
received four.  In 2013 one to two more visits were obtained (see
Table \ref{table:Velocities}).   A significant number of GCs were
observed in both the 2011 and 2013 campaigns.  Table
\ref{table:Velocities} also shows the signal-to-noise (S/N) ratios of
the final spectra: all range from 10-65 per pixel.

\subsection{Data Reduction}\label{subsec:DataReduction}
The APOGEE data reduction pipeline \citep{Nidever2015} provides
reduced and flux-calibrated single-epoch spectra (so-called
``visits''), with estimates of the noise and bad-pixel masks.   Unlike
most of the optical range, $H$-band spectra are affected by the 
presence of strong sky lines in both emission and absorption.  The
APOGEE data reduction pipeline flags these regions via bits set in the
bad-pixel mask. The sensitivity to the previously accumulated signal
in some regions of the APOGEE detectors (superpersistence;
\citealt{Nidever2015}) makes some spectral ranges in certain fibers
almost useless for analysis, especially for targets with low S/N
(though note that even if a fiber was located in the superpersistence
area, the red part of its spectrum is unaffected by this effect and
is viable for further analysis). Fortunately, the main APOGEE pipeline
flags out such regions as well.

The combined spectrum of an individual target comes from a weighted
average of all available visits in 2011 and 2013.  A visit's weight is
calculated individually in each pixel as a product of the inverse
variance multiplied by the inverted mask. The latter is a binary mask,
where any non-zero pixels in the original bad-pixel mask (from the
data reduction pipeline) have been set to zero, and any original
pixels with no error bits have been set to one.  Including this weight
during the combination rejects all of the bad or problematic pixels
from being included in the final spectrum.  In particular, this will
mask out regions affected by superpersistence.  With regular APOGEE
targets, the individual, weighted visits for a given GC are then
combined into a single, higher S/N spectrum, based on the radial
velocity estimated for each individual visit.  While this works well
for typical, relatively bright APOGEE targets, most of the M31 GCs
have mediocre to poor S/N (as low as S/N$\sim 10$; see Table
\ref{table:Velocities}) and the radial velocity estimation is
bypassed.  Instead the radial velocities are assumed to be constant
between visits, and multi-epoch spectra are aligned using barycentric
correction information only. The pipeline therefore delivers the
combined spectra, the associated errors, and the combined inverted
mask.

As described in Section \ref{subsec:Observations}, special
``background fibers'' were reserved for clusters within $9\arcmin$ of
the center of M31 (only B171 in this paper's sample). The background
fibers were treated as regular objects, and were reduced and combined
using the method described above.  The galactic background light comes
from an unresolved stellar population, and it is therefore assumed
that any sharp spectral features are caused by noise or sky features;
in order to avoid adding noise during the subtraction the background
spectra were smoothed by a Gaussian kernel with FWHM~=~220~km~s$^{-1}$ 
to reproduce a mostly featureless background.  The smoothed background
was then subtracted from B171's spectrum.\footnote{Note that for the
  optical B171 spectrum (Appendix \ref{appendix:OpticalAbunds}) a
  separate background subtraction was not performed. Sky fibers were
  located 10$\arcsec$ from the central fiber, providing simultaneous
  sky observations.  These sky fibers likely contained a background
  component as well.}

\begin{figure}[h!]
\begin{center}
\centering
\includegraphics[scale=0.75,trim=0 0.75in 0in 0.0in]{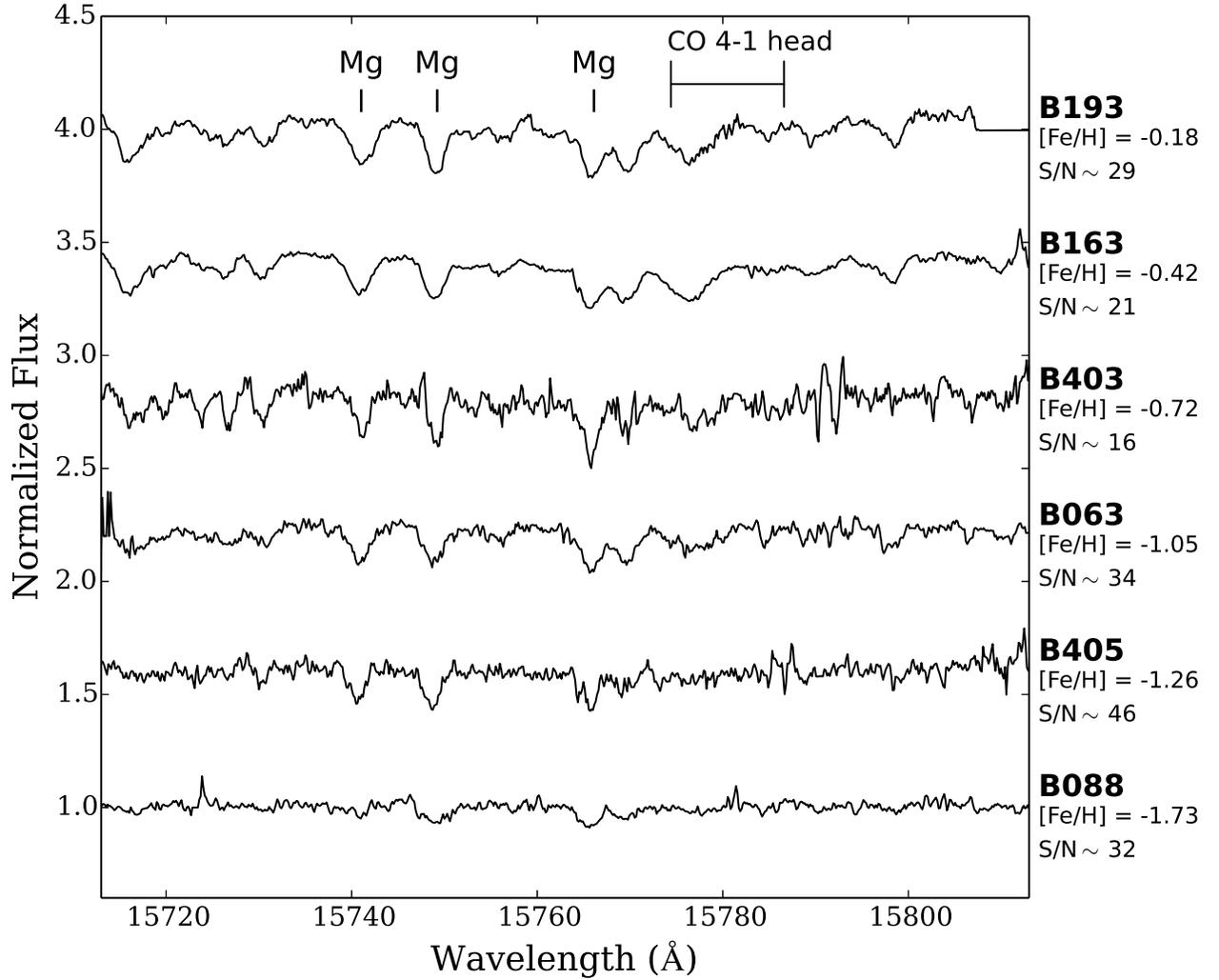}
\caption{Sample $H$-band spectra of M31 GCs, spanning a range in
  metallicity ($H$-band [Fe/H] ratios are listed; see Section
  \ref{subsec:Fe}).  Median S/N ratios (per pixel) are also given.
  The region around the strong \ion{Mg}{1} lines and a CO bandhead is
  shown.  Note that the line depths also depend on the velocity
  dispersion of the cluster.\label{fig:SampleSpectra}} 
\end{center}
\end{figure}

Samples of the final spectra are shown in Figure
\ref{fig:SampleSpectra}, for GCs with a range of metallicities.  The
region around the strong \ion{Mg}{1} lines and a CO bandhead is
shown.  Note that the features are barely detectable at B088's
metallicity and velocity dispersion.

\subsection{Radial Velocities and Line Broadening}\label{subsec:Velocities}
The spectra for the metal-rich ($[\rm{Fe/H}] \gtrsim -1.2$) GCs were
shifted to the rest frame through cross-correlations with a spectrum
of Arcturus \citep{Hinkle2003} using the Image Reduction and Analysis
Facility program (IRAF)\footnote{IRAF is distributed by the National
Optical Astronomy Observatory, which is operated by the Association of
Universities for Research in Astronomy, Inc., under cooperative
agreement with the National Science Foundation.} task {\it fxcor}.
For metal-poor GCs with low S/N and very few detectable lines this
technique did not work---in this latter case, lines were identified
and offsets were calculated manually.  The final, heliocentric
velocities are shown in Table \ref{table:Velocities}, and are in
excellent agreement with \citet{Colucci2014} and
\citet{Caldwell2011}.

\begin{figure}[h!]
\begin{center}
\centering
\includegraphics[scale=0.75]{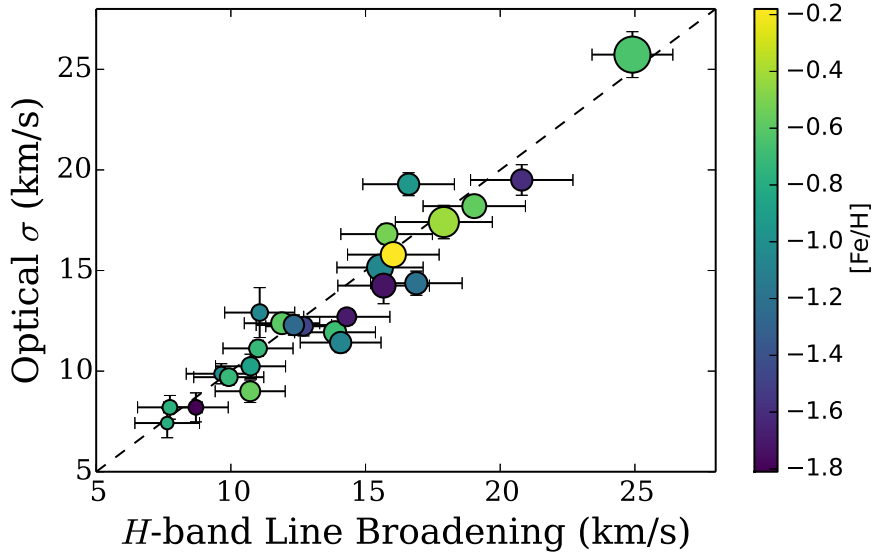}
\caption{A comparison of $H$-band line broadening (with instrumental
  broadening removed) and optical velocity dispersion.  The dashed
  line shows perfect agreement.  The points are sized according to
  S/N ratios, where the larger symbols have higher S/N, and are
  colored according to [Fe/H].\label{fig:Velocities}}
\end{center}
\end{figure}

Using spectrum syntheses to measure line strengths (see Section
\ref{subsec:Synpop}) requires modeling the line broadening due to
(primarily) the GC velocity dispersion and the instrumental
resolution.  Although there are well-constrained velocity dispersions
available from optical spectroscopy, the line broadening values in the
IR were determined by fitting the strong \ion{Si}{1} lines at 15960
and 16095~\AA.  These lines were utilized instead of full spectrum
fitting in an attempt to minimize degeneracies between assumed initial
abundances and line broadening.  The precise broadening is more
uncertain for the metal-poor GCs and/or GCs with lower S/N: high S/N,
metal-rich targets can have line broadening uncertainties around
$\sim1$ km s$^{-1}$, while lower S/N, metal-poor GCs can have
uncertainties of $\sim 2-3$ km s$^{-1}$.  The abundance errors reflect
uncertainties in fitting the full line profiles, and therefore account
for small uncertainties in the derived broadening values.  Figure
\ref{fig:Velocities} shows a comparison between the line broadening
derived from the $H$-band (with the instrumental broadening removed)
and the velocity dispersion from the optical, demonstrating that the
two are in agreement.  Note that the precise value for the velocity
dispersion will depend upon the area of the cluster covered by the
spectrograph slit or fiber.  The optical values in Figure
\ref{fig:Velocities} were obtained from spectra with a slit size of
$1.\arcsec7 \times 7\arcsec$ \citep{Colucci2014} or a $3\arcsec$ fiber
(Appendix \ref{appendix:OpticalAbunds}), while the APOGEE fibers have
$2\arcsec$ diameters.

\section{Analysis Methods}\label{sec:Analysis}
$H$-band spectra exhibit strong molecular features as well as numerous
atomic lines, particularly in metal-rich targets---this means that
nearly every spectral feature is a blend of lines from multiple
elements.  In addition, IL spectral features are broadened by the GC
velocity dispersion, exacerbating the blending problems.  As a result,
standard equivalent width analyses of individual features may not be
suitable for abundance analyses of $H$-band IL spectra (see the
discussions in \citealt{Sakari2013} and \citealt{Colucci2014}).  This
paper therefore presents abundances that were derived from spectrum
syntheses of the regions around each line of interest, utilizing a
recent modification to the 2014 version of the line analysis code
\texttt{MOOG} \citep{Sneden}.

\subsection{{\it synpop}: Spectrum Syntheses for Stellar Populations}\label{subsec:Synpop}
\texttt{MOOG} \citep{Sneden} is a Local Thermodynamic Equilibrium
(LTE) line analysis code that has been widely used for individual
stellar analyses, both in the optical and the IR. \texttt{MOOG}'s
spectrum synthesis routine {\it synth} requires a model atmosphere and
a line list; with this input, it then synthesizes a spectrum over the
desired wavelength range, and abundances can be altered until the
synthetic spectrum matches the observed one.  \texttt{MOOG}'s {\it
  synpop} routine works the same way, but for an entire stellar
population: given a list of model atmospheres (see Section
\ref{subsec:ModelAtms}) and a line list (Section
\ref{subsec:LineList}), it produces a synthetic IL spectrum, weighted
by the flux and number of stars assigned to each atmosphere.  The {\it
  synpop} routine is therefore nearly identical to the synthesis
version of the code {\tt ILABUNDS} \citep{McWB,Sakari2013}; however,
{\it synpop} is now a supported part of {\tt MOOG}, runs off current
releases, and is publicly
available.\footnote{\url{http://www.as.utexas.edu/~chris/moog.html}}
This analysis utilizes the 2014 release of {\tt MOOG}.  As with the
{\it synth} routine, {\it synpop} provides residual fits to identify
the syntheses that best match the observed spectra.

\subsection{Isochrones and Model Atmospheres}\label{subsec:ModelAtms}
Since IL spectra contain contributions from all the stars in a
stellar population, isochrones are used to model the underlying
stellar populations in each GC.  The BaSTI isochrones
\citep{BaSTIref,ODFNEWref,Cordier2007} are utilized because they
extend through the evolved horizontal branch (HB) and AGB phases;
these are the same isochrones utilized for the comparison optical
values.  The ages and metallicities determined from the optical
spectra (\citealt{Colucci2014}, Appendix \ref{appendix:OpticalAbunds})
are adopted---these values were found by minimizing trends in
\ion{Fe}{1} abundance with line wavelength, reduced equivalent width,
and excitation potential (see \citealt{McWB}), and agree reasonably
well with the values derived from Lick indices \citep{Caldwell2011}.
This procedure may not be suitable for $H$-band spectra; however, the
$H$-band spectral lines are not as sensitive to cluster age (see
Section \ref{subsec:FeDiscussion}).  The standard
horizontal branch morphologies from the BaSTI isochrones are adopted,
along with extended AGBs---the effects of these choices are discussed
in \citet{Colucci2009,Colucci2012,Colucci2014} and \citet{Sakari2014}.  

Once an appropriate isochrone has been identified, the isochrones are
populated with stars to match the observed total magnitude for each
GC, assuming a \citet{Kroupa2002} initial mass function (IMF).  The
stars are binned along the isochrones, with each box containing 3.5\%
of the total flux. (Note that in the optical the box size has only a
negligible effect on the abundances; \citealt{Sakari2014}.)
Microturbulent velocities are assigned to each box based on an
empirically-motivated  calibration with surface gravity \citep{McWB};
this relation produces similar microturbulent velocities as the
standard APOGEE relation \citep{ASPCAPref}, except at very high and
low surface gravities.  Each box is then assigned a Kurucz model
atmosphere\footnote{\url{http://kurucz.harvard.edu/grids.html}} 
\citep{KuruczModelAtmRef} with the average effective temperature,
surface gravity, and microturbulence of the stars in that box.  All
atmospheres are chosen  to be $\alpha$-enhanced, since optical
integrated [$\alpha$/Fe] ratios are enhanced.  This collection of
model atmospheres is then fed as input to {\tt MOOG}, along with a
line list.

\subsection{Line List}\label{subsec:LineList}
The line list adopted for this study includes both atomic and
molecular species. The line list version adopted, linelist.20150714,
is an updated version of what was used for DR12 results
\citep{Shetrone2015} and is the version adopted in DR13.
\citep{Shetrone2015} noted a number of problems with the DR12 line
list that have been corrected in the ``linelist.20150714'' version
adopted here and documented in Holtzman et al. (in prep).  A short
summary of how the line list was generated including the differences
with the DR12 version are detailed below.

The molecular line list is a compilation of literature sources
including transitions of CO, OH, CN, C2, H2, and SiH.  The CN line
list was updated using a compilation from C. Sneden (private
communication).   All molecular data are adopted without change, with
the exception of a few obvious typographical corrections. The atomic
line list was compiled from a number of literature sources and
includes theoretical, astrophysical, and laboratory oscillator
strength values.  A few new lines were added from
NIST\footnote{\url{http://physics.nist.gov/PhysRefData/ASD/lines_form.html}}
and other literature publication since the DR12 line list was created,
including hyperfine splitting components for Al and Co.   To calculate
the astrophysical $gf$ values {\tt Turbospectrum} was utilized
\citep{Turbo98,Plez2012} to generate synthetic spectra with varying
oscillator strengths and damping values in order to fit the solar and
Arcturus spectra.  For lines with laboratory oscillator strengths, the
astrophysical $\log(gf)$ values were not allowed to vary beyond twice
the error quoted by the source.

One change from the methodology described in \citet{Shetrone2015} is
that a  different weighting scheme was used between the solar and
Arcturus solutions. The astrophysical solutions were weighted
according to line depth in Arcturus and in the Sun, which usually
gives more weight to the Arcturus solution since the lines are
generally stronger in the cooler, low surface gravity star, Arcturus,
despite it being more metal-poor. Another difference from the
methodology adopted in DR13 was the proper use of the center-of-disk
spectral synthesis of the solar center-of-disk atlas using a
microturbulence of 0.7 km s$^{-1}$.

\subsection{Abundance Determinations}\label{subsec:AbundDeterminations}
As described in Section \ref{subsec:Synpop}, the IR abundances are
determined via spectrum syntheses; the best-fitting abundances were
derived iteratively, following \citet{Smith2013}.  The procedure
adopted for this analysis is as follows.
\begin{enumerate}
\item Adopt the best-fitting isochrones and model atmospheres from the
  optical (see Section \ref{subsec:ModelAtms}).
\item Begin with standard abundance ratios for typical MW field stars
  at the same [Fe/H], e.g., $[\alpha/{\rm Fe}]=+0.4$
  \citep{Venn2004}.  Determine line broadening parameters (see Section
  \ref{subsec:Velocities}).
\item Determine initial values for the infrared C, N, O, and Fe
  abundances based on syntheses of the lines in \citet{Smith2013}.
  The fits are done by eye over a $\sim 40$\AA \hspace{0.025in}
  region, guided by the output residual fits in {\tt MOOG}.
\item Iterate on the C, N, O, and Fe abundances until the values do
  not change.  Adjust the isochrone [Fe/H], if needed, and reiterate on
  C, N, O, and Fe.
\item Find the abundances of the other elements.
\end{enumerate}

Final [Fe/H] and [X/Fe] abundance ratios are calculated relative to
the \citet{Asplund2009} solar abundances.  All [X/Fe] ratios are
calculated with the $H$-band [\ion{Fe}{1}/H] abundance (since there
are no detectable \ion{Fe}{2} lines in the $H$-band).

\subsubsection{Strong Lines}\label{subsubsec:Strong}
\citet{McWilliam1995b} showed that abundances derived from the
strongest spectral lines are very sensitive to the treatment of the
stellar atmosphere, particularly the outer layers---they estimated
that lines with reduced equivalent widths
(REW)\footnote{$\rm{REW}~=~\log(\rm{EW}/\lambda)$, where EW is the
  equivalent width of a spectral line and $\lambda$ is its wavelength;
  both are typically expressed in \AA, and REW is dimensionless.} in
the range $\rm{REW}~>~-4.7$ should be removed from abundance analyses,
particularly when the \ion{Fe}{1} lines are utilized to determine
atmospheric parameters (or, in the case of IL analyses, the proper
single stellar population parameters: age and metallicity).  In the
optical, lines with EWs larger than $\sim100$ to
150~m\AA \hspace{0.025in} must be removed. In the $H$-band, the REWs
are smaller for lines of the same EW, and $H$-band analyses can safely
utilize lines with strengths up to 300-320 m\AA. In some cases, when
only strong lines are available for a given element, the REW limit was
pushed up to $-4.5$; these lines may introduce offsets of $\sim0.1$
dex \citep{McWilliam1995b}, and these abundances have been flagged in
all tables and figures. Similarly, some of the optical abundances were
derived with strong lines, and those clusters are also flagged in all
plots. 

\subsubsection{Abundance Errors}\label{subsubsec:Errors}
The random error in the abundance from each spectral line was
estimated from the range of abundances that can fit the line profile,
based on by-eye estimates and {\tt MOOG}'s output residuals.  Line
profiles may not be perfectly fit due to, e.g.,  S/N, uncertain
continuum placement, telluric features, emission lines, the adopted
line broadening, etc.  Individual line abundance uncertainties range
from 0.05 to 0.25 dex, are are primarily driven by S/N.  The
uncertainty in a GC's mean abundance for a given element is determined
by dividing the individual error by $\sqrt{N}$, where $N$ is the
number of lines for that element.

Potential systematic errors are more difficult to ascertain.
Generally, the largest systematic uncertainties in IL analyses are due
to uncertainties in modeling the underlying stellar populations,
particularly the age of the population, the HB morphology, and the
distribution of the brightest RGB and AGB stars (including the AGB/RGB
star ratio).  \citet{Sakari2014} present a detailed systematic error
analysis for optical spectral lines, but those quantitative estimates
may not apply to these IR lines.  For instance, age and HB morphology
can have strong effects in the optical (also see
\citealt{Colucci2014}) but the effect in the $H$-band seems to be
negligible (see Section \ref{subsec:FeDiscussion}).  The treatment of
the brightest RGB and AGB stars is likely to have the largest effect
in the $H$-band.  Section \ref{subsec:AGBratio} demonstrates that the
relative number of AGB stars, relative to RGB stars, is not likely to
have a significant effect on the IL ratios; however, the choice of AGB
models may affect some abundance ratios by as much as 0.2 dex
\citep{Sakari2014}.  Stochastic sampling of the upper RGB and AGB may
also be significant---this will be examined in a separate paper.
Systematic errors can also arise between analyses as a result of
different models, assumptions, and techniques.  Ultimately, systematic
errors in the individual errors could be as high as 0.2 dex, though
the errors in [X/Fe] ratios are likely to be smaller.  However,
the methodology, isochrones, model atmospheres, and solar abundances
are the same between the high resolution optical and infrared
analyses, which should reduce model-dependent systematic errors.

\section{Chemical Abundances}\label{sec:Abunds}
In this section the $H$-band abundances are compared to a) the
high-resolution optical abundances from
\citet{Colucci2009,Colucci2014} and Appendix
\ref{appendix:OpticalAbunds} and b) the Lick index optical abundances
from \citet{Schiavon2012,Schiavon2013}.

\subsection{Iron}\label{subsec:Fe}
Iron is a crucial element for chemical abundance analyses, since it
is typically used to represent the GC metallicity.  Table
\ref{table:Fe} shows the $H$-band \ion{Fe}{1} abundances, along with
the random errors and the number of measured lines.  The clusters have
been ordered from low to high [Fe/H].  Note that there are fewer
measurable \ion{Fe}{1} lines (2-13) in the $H$-band compared to the
optical (which has 20-60, depending on metallicity and wavelength
range). The $H$-band \ion{Fe}{1} lines are those quoted in
\citet{Smith2013}, along with several additional high excitation
potential (EP) lines which were utilized for metal-poor stars by
\citet{Lamb2015}. Note that there are no useful \ion{Fe}{2} lines in
the $H$-band.  As mentioned in Section
\ref{subsec:AbundDeterminations}, the $H$-band \ion{Fe}{1} abundances
are mildly dependent on the adopted CNO abundances as a result of
blending with CN and CO lines, particularly for the metal-rich
targets.

\begin{table}
\centering
\begin{center}
\caption{Mean $H$-band abundances and random errors: Fe, C, N, and O.\vspace{0.1in} \label{table:Fe}\label{table:CNO}}
  \newcolumntype{d}[1]{D{,}{\pm}{#1}}
  \newcolumntype{e}[1]{D{.}{.}{#1}}
  \begin{tabular}{@{}ld{3}cd{3}cd{3}cd{3}c@{}}
  \hline
 & \multicolumn{1}{c}{[Fe I/H]} & $N$ & \multicolumn{1}{c}{[C/Fe]$^{a}$} & $N$ & \multicolumn{1}{c}{[N/Fe]$^{a}$} & $N$ & \multicolumn{1}{c}{[O/Fe]$^{a}$} & $N$\\
\hline
B232     & -1.81,0.11 & 3 & -0.20,0.20 & 1 & 1.20,0.20 & 1 & 0.22,0.25 & 1 \\
B088     & -1.73,0.07 & 6 & -0.38,0.07 & 5 & 1.31,0.18 & 2 & 0.15,0.08 & 4 \\
B311     & -1.70,0.13 & 2 & -0.03,0.09 & 4 & 1.13,0.14 & 2 & 0.35,0.14 & 2 \\
B012     & -1.60,0.10 & 5 & -0.54,0.15 & 5 & 0.93,0.10 & 2 & 0.07,0.10 & 4 \\
B240     & -1.48,0.10 & 9 & -0.36,0.10 & 4 & 1.03,0.11 & 3 & 0.31,0.14 & 2 \\
B405     & -1.26,0.07 & 9 & -0.46,0.08 & 5 & 0.93,0.11 & 4 & 0.35,0.09 & 2 \\
B472     & -1.20,0.07 & 8 & -0.41,0.08 & 4 & 1.34,0.09 & 4 & 0.26,0.12 & 3 \\
B386     & -1.07,0.14 & 4 & -0.37,0.04 & 4 & 1.16,0.10 & 4 & 0.46,0.03 & 3 \\
B312     & -1.06,0.06 & 4 & -0.48,0.13 & 3 & 1.02,0.19 & 2 & 0.35,0.14 & 2 \\
B063     & -1.05,0.09 & 12& -0.27,0.09 & 5 & 1.30,0.14 & 8 & 0.42,0.07 & 5 \\
B381     & -1.03,0.13 & 5 & -0.28,0.11 & 2 & 1.32,0.14 & 2 & 0.32,0.10 & 1 \\
B182     & -0.95,0.06 & 6 & -0.54,0.08 & 5 & 1.09,0.05 & 2 & 0.11,0.05 & 3 \\
B045     & -0.88,0.07 & 11& -0.41,0.07 & 5 & 0.90,0.10 & 4 & 0.33,0.12 & 5 \\
B048$^{b}$& -0.78,0.13 & 2 & -0.17,0.10 & 1 & 1.08,0.12 & 3 & 0.53,0.10 & 1 \\
B235     & -0.77,0.06 & 8 & -0.13,0.10 & 4 & 1.21,0.10 & 4 & 0.57,0.08 & 3 \\
B383     & -0.72,0.04 & 6 & -0.18,0.11 & 3 & 1.25,0.10 & 3 & 0.47,0.77 & 4 \\
B403     & -0.72,0.09 & 9 & -0.31,0.08 & 4 & 1.06,0.07 & 8 & 0.44,0.11 & 4 \\
B006     & -0.69,0.05 & 7 & -0.32,0.05 & 5 & 1.35,0.04 & 6 & 0.32,0.04 & 5 \\
B225     & -0.64,0.05 & 10& -0.21,0.05 & 4 & 1.09,0.07 & 5 & 0.39,0.09 & 4 \\
B034     & -0.60,0.10 & 5 & -0.24,0.09 & 5 & 1.11,0.07 & 7 & 0.45,0.08 & 4 \\
B110$^{b}$& -0.57,0.09 & 3 & -0.32,0.07 & 4 & 1.09,0.12 & 6 & 0.33,0.09 & 2 \\
B384     & -0.56,0.04 & 11& -0.26,0.07 & 5 & 1.11,0.07 & 6 & 0.42,0.05 & 4 \\
B171     & -0.52,0.03 & 8 & -0.28,0.05 & 4 & 1.24,0.04 & 8 & 0.36,0.04 & 5 \\
B163     & -0.42,0.09 & 12& -0.24,0.05 & 5 & 0.94,0.12 & 5 & 0.35,0.08 & 4 \\
B193     & -0.18,0.09 & 13& -0.10,0.05 & 4 & 1.10,0.09 & 7 & 0.45,0.06 & 4 \\
 & & & & & & & & \\
\hline
\end{tabular}\\
\end{center}
\medskip
\raggedright $^{a}$ CNO abundances are determined from the CO, CN, and
OH molecular lines in \citet{Smith2013}, assuming
$^{12}$C/$^{13}\rm{C} = 6$ (see the text).\\
 $^{b}$ Cluster's spectral range is limited due to
superpersistence; see Section \ref{subsec:DataReduction}.
\end{table}

\begin{figure}[h!]
\begin{center}
\centering
\subfigure[]{\includegraphics[scale=0.55,trim=1.25in 0in 0.25in 0.0in]{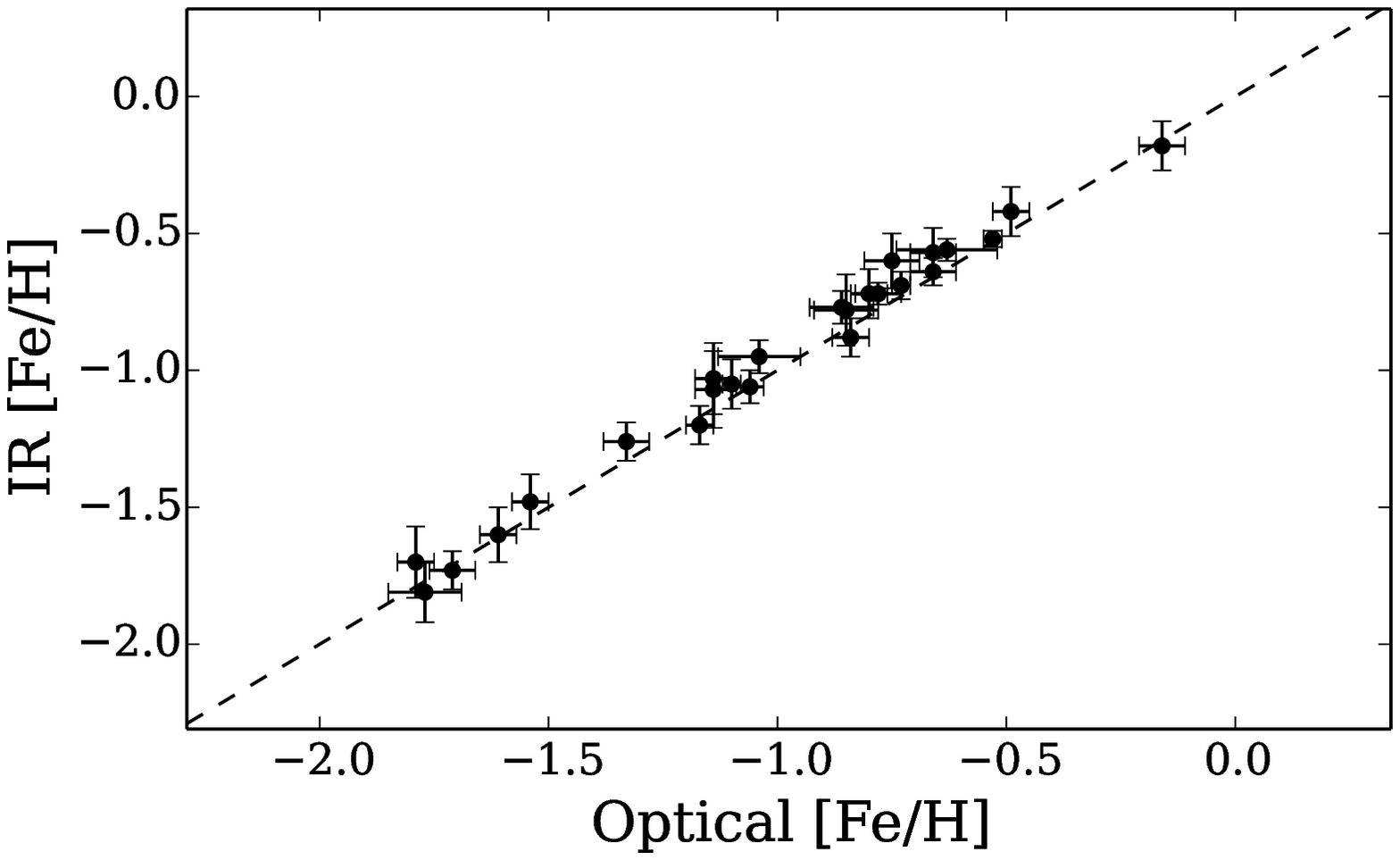}\label{fig:FeCompA}}
\subfigure[]{\includegraphics[scale=0.55,trim=0.5in 0in 1.25in 0.0in]{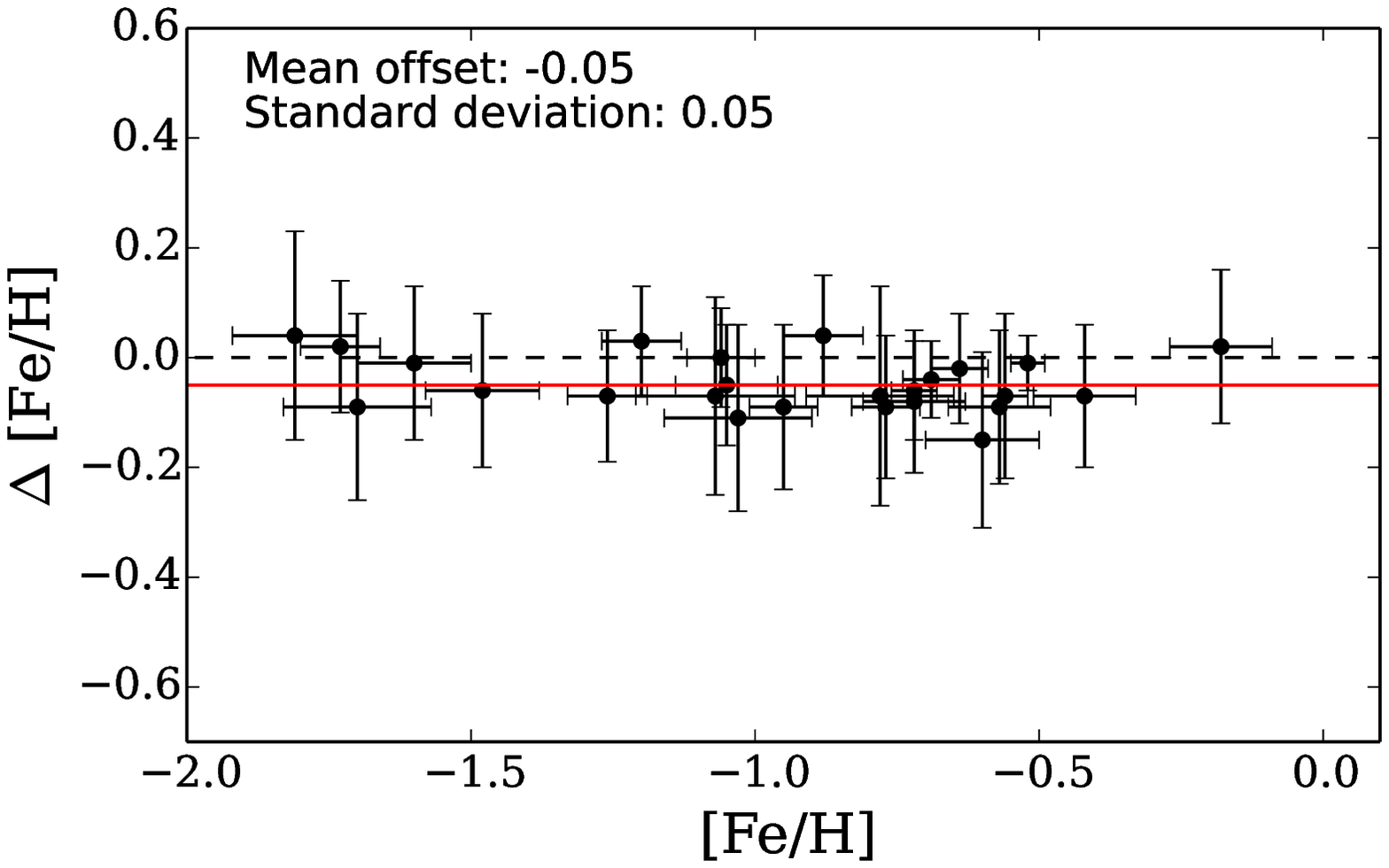}\label{fig:FeCompB}}
\caption{Comparisons of the $H$-band [\ion{Fe}{1}/H] abundances with
  the optical, high-resolution abundances from Appendix
  \ref{appendix:OpticalAbunds} and \citet{Colucci2009,Colucci2014}.
  The error bars represent $1\sigma$ random errors, while the dashed
  lines show perfect agreement.  {\it Left: } The two values plotted
  against each other.  {\it Right: } The difference in
  [\ion{Fe}{1}/H] (optical $-$ IR) as a function of $H$-band
  [Fe/H]. The solid red line shows the average offset between the
optical and $H$-band, which is also quoted in the upper left corner.\label{fig:FeComp}}
\end{center}
\end{figure}

Figure \ref{fig:FeComp} compares the $H$-band [\ion{Fe}{1}/H]
abundances with the high resolution, optical [\ion{Fe}{1}/H] ratios
from \citet{Colucci2014} and Appendix \ref{appendix:OpticalAbunds}.
Figure \ref{fig:FeCompA} compares the values to each other, while Figure
\ref{fig:FeCompB} shows the differences versus $H$-band [Fe/H].
Though the $H$-band values are generally slightly higher than the
optical (by 0.05 dex on average), all agree within the $1\sigma$
errors and there is no trend with [Fe/H].  A small offset between IR
and optical [\ion{Fe}{1}/H] may be expected due to NLTE effects (e.g.,
\citealt{KraftIvans2003}).  Strong NLTE effects may plague the optical
\ion{Fe}{1} lines, especially in the brightest AGB stars (e.g.,
\citealt{Lapenna2015}), which would then affect the combined,
integrated \ion{Fe}{1} lines.  NLTE corrections to the \ion{Fe}{1}
lines in the $H$-band are expected to be smaller than for those in the
optical \citep{GarciaHernandez2015}, which could naturally lead to
slight offsets between the optical and IR.  \ion{Fe}{2} is expected to
be a more reliable Fe indicator; however,  in the optical there are
fewer \ion{Fe}{2} lines (1-16 for these GCs; \citealt{Colucci2014}),
they are often weaker, most are located further in the blue, and they
are more sensitive to systematic uncertainties in the underlying
stellar population \citep{Sakari2014}.  For these reasons, the
$H$-band \ion{Fe}{1} abundances are not always in agreement with the
optical \ion{Fe}{2}, although the {\it average} offset is very small.

The small offset between optical and $H$-band [\ion{Fe}{1}/H] suggests
that the NLTE offsets are likely to be small.  The $H$-band therefore
provides robust, integrated [\ion{Fe}{1}/H] ratios down to at least
$[\rm{Fe/H}] \sim -1.8$, depending on S/N and GC velocity dispersion.

\clearpage
\subsection{Light Elements: Carbon, Nitrogen, and Oxygen}\label{subsec:CNO}
\begin{figure}[h!]
\begin{center}
\centering
\includegraphics[scale=0.6,trim=1.5in 0.in 0in 0.0in]{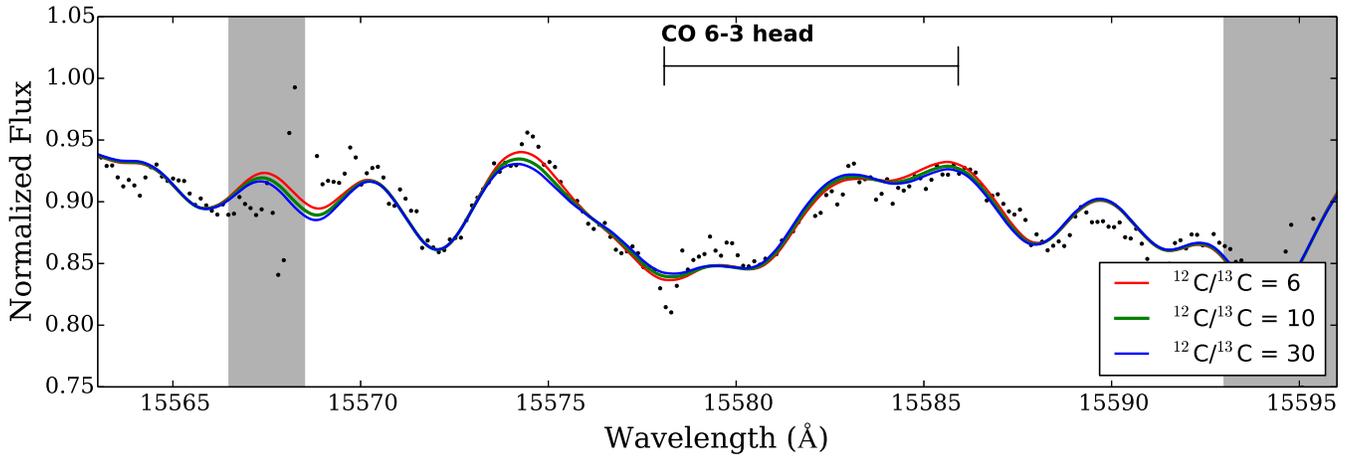}
\caption{Sample syntheses of a C-sensitive CO feature in the most
  metal-rich GC, B193, demonstrating the small effects of the assumed
  $^{12}$C$/^{13}$C ratio (the effects of varying $^{12}$C$/^{13}$C
  are likely to be even smaller in more metal-poor GCs).  The grey
  areas show regions that were masked out in the data reduction
  pipeline (see Section \ref{subsec:DataReduction}).  Three
  $^{12}$C$/^{13}$C ratios are shown: 6, 10, and 30.  Though the lower
  $^{12}$C$/^{13}$C is slightly preferred, the differences are quite
  small.\label{fig:C12C13}}
\end{center}
\end{figure}

\begin{figure}[h!]
\begin{center}
\centering
\includegraphics[scale=0.75,trim=0.0 0.5in 0in 0.0in]{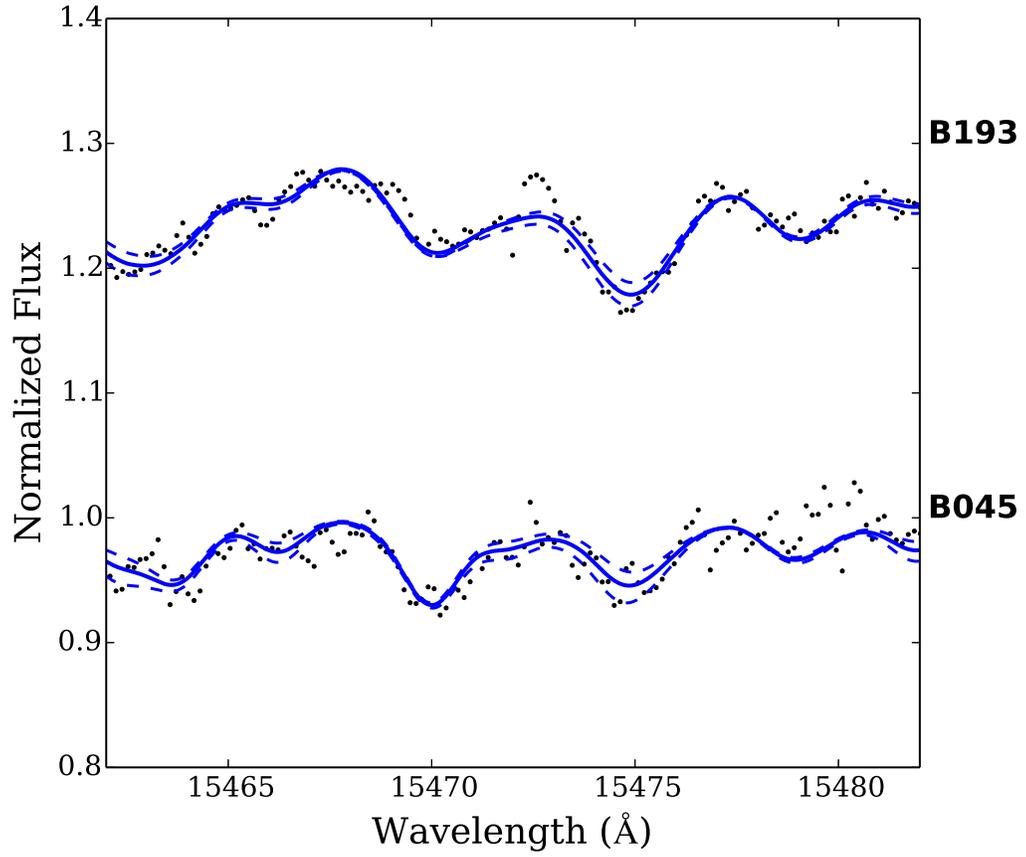}
\caption{Sample syntheses of CN features in two GCs, B193 and B045;
  these features are used to determine [N/Fe].   The solid lines show
  the best-fit syntheses, while the dashed lines show $\pm1\sigma$
  uncertainties in [N/Fe] of 0.20 dex in both cases.\label{fig:CNsynth}} 
\end{center}
\end{figure}

Carbon, nitrogen, and oxygen are crucial elements for GC studies.  In
the $H$-band, C, N, and O abundances can be determined from syntheses
of CO, CN, and OH molecular lines (see \citealt{Smith2013}) with an
assumed $^{12}$C/$^{13}$C ratio.  In  these IL spectra the S/N is not
sufficient and/or the blending is too severe to measure a precise
$^{12}$C/$^{13}$C ratio (though in some cases it is evident that
$^{12}$C/$^{13}\rm{C} < 20$). For all clusters, $^{12}$C/$^{13}\rm{C}
= 6$ is adopted to represent the approximate value for stars at the
tip of the RGB (e.g., \citealt{Gratton2000}).  Since most of the
molecular features are not strongly dependent on the precise
$^{12}$C/$^{13}$C ratio and the $H$-band IL spectrum is dominated by
tip of the RGB stars, this assumption should not have a strong effect
on these IL abundances (see Figure \ref{fig:C12C13}).  The final
[C/Fe], [N/Fe], and [O/Fe] ratios are listed in Table \ref{table:CNO},
while sample syntheses of N-sensitive CN features are shown in Figure
\ref{fig:CNsynth}.

\begin{figure}[h!]
\begin{center}
\centering
\subfigure{\includegraphics[scale=0.55,trim=1.25in 0in 0.25in 0.0in]{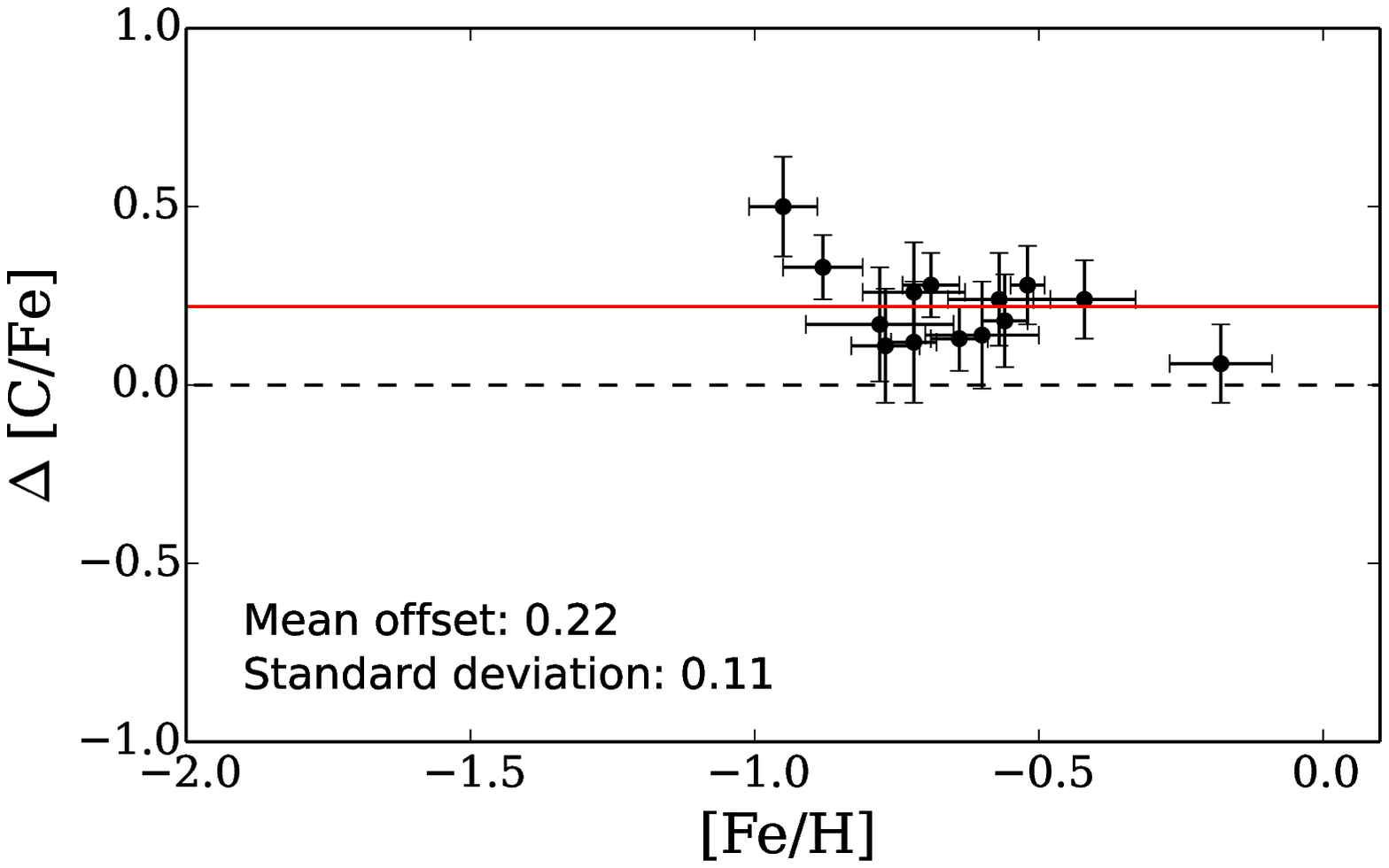}\label{fig:CComp}}
\subfigure{\includegraphics[scale=0.55,trim=0.5in 0in 1.25in 0.0in]{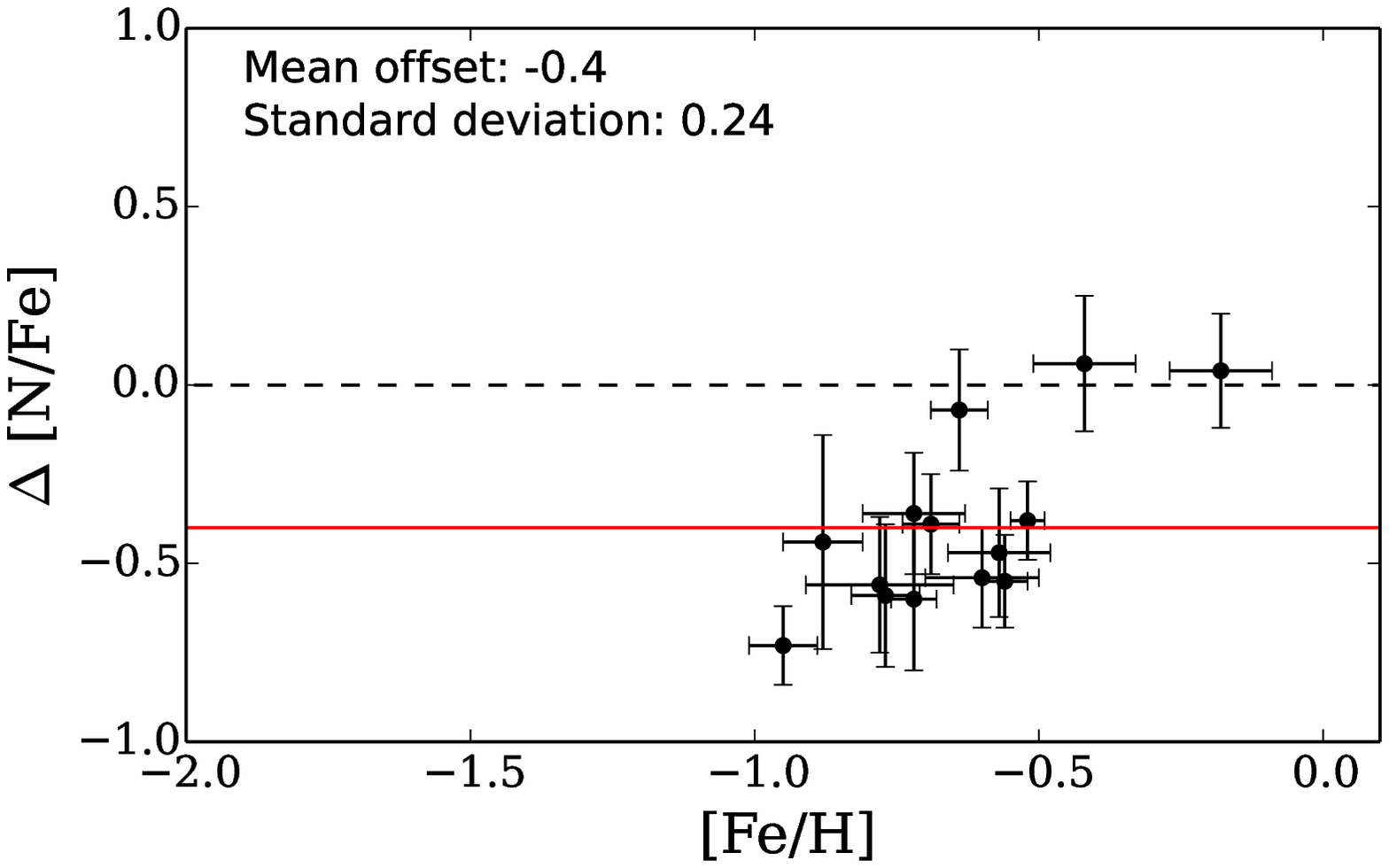}\label{fig:NComp}}
\caption{Comparisons of the $H$-band [C/Fe] (left) and [N/Fe] (right)
  ratios with the optical values from
  \citet{Schiavon2012,Schiavon2013}, as a function of $H$-band [Fe/H].
   The differences are given as optical$-$ $H$-band. The error bars
   represent $1\sigma$ random errors, the dashed line shows perfect
   agreement, and the solid red line shows the average offset (which
   is quoted in each panel).\label{fig:LightComp}}
\end{center}
\end{figure}

Figure \ref{fig:LightComp} shows a comparison between the $H$-band and
optical [C/Fe] and [N/Fe] ratios, as a function of [Fe/H].  Reliable
C, N, and O abundances can be difficult to obtain in the optical,
particularly in IL spectra.  C and N can be determined from CN and
C$_{2}$ features in the blue (e.g.,
\citealt{Worthey1994,TripiccoBell1995,Schiavon2007}), but the exact
abundances rely on an assumed O abundance \citep{GravesSchiavon2008}.
O can be determined via the forbidden lines at 6300 and 6363 \AA, but
this requires high resolution, high S/N spectra of low velocity
dispersion GCs because the lines are weak. The comparison optical
values are therefore from the Lick index analyses of
\citet{Schiavon2012,Schiavon2013}, and are only available for
metal-rich GCs ($[\rm{Fe/H}] \gtrsim -1$).  These [C/Fe] and [N/Fe]
ratios are determined from the C$_{2}$4668 and CN$_{1}$ and CN$_{2}$
($\sim 4140-4180$~\AA) Lick indices, assuming a fixed $[\rm{O/Fe}] =
0.3$ for all GCs.  Figure \ref{fig:LightComp} demonstrates that, in
general, the $H$-band [C/Fe] ratios are lower than the optical values
by 0.2 dex, on average, while the [N/Fe] ratios are  significantly
higher (by 0.5 dex on average, when the three GCs in agreement are
removed).  The [N/Fe] ratios agree for the massive, metal-rich GCs
B193, B163, and B225, though the [C/Fe] ratios are still slightly
offset for the latter two.  Note that the $H$-band abundances agree
with the expected abundances for tip of the RGB stars, which have
dredged up products from CNO cycling \citep{Gratton2000}.

\begin{figure}[h!]
\begin{center}
\centering
\includegraphics[scale=0.7]{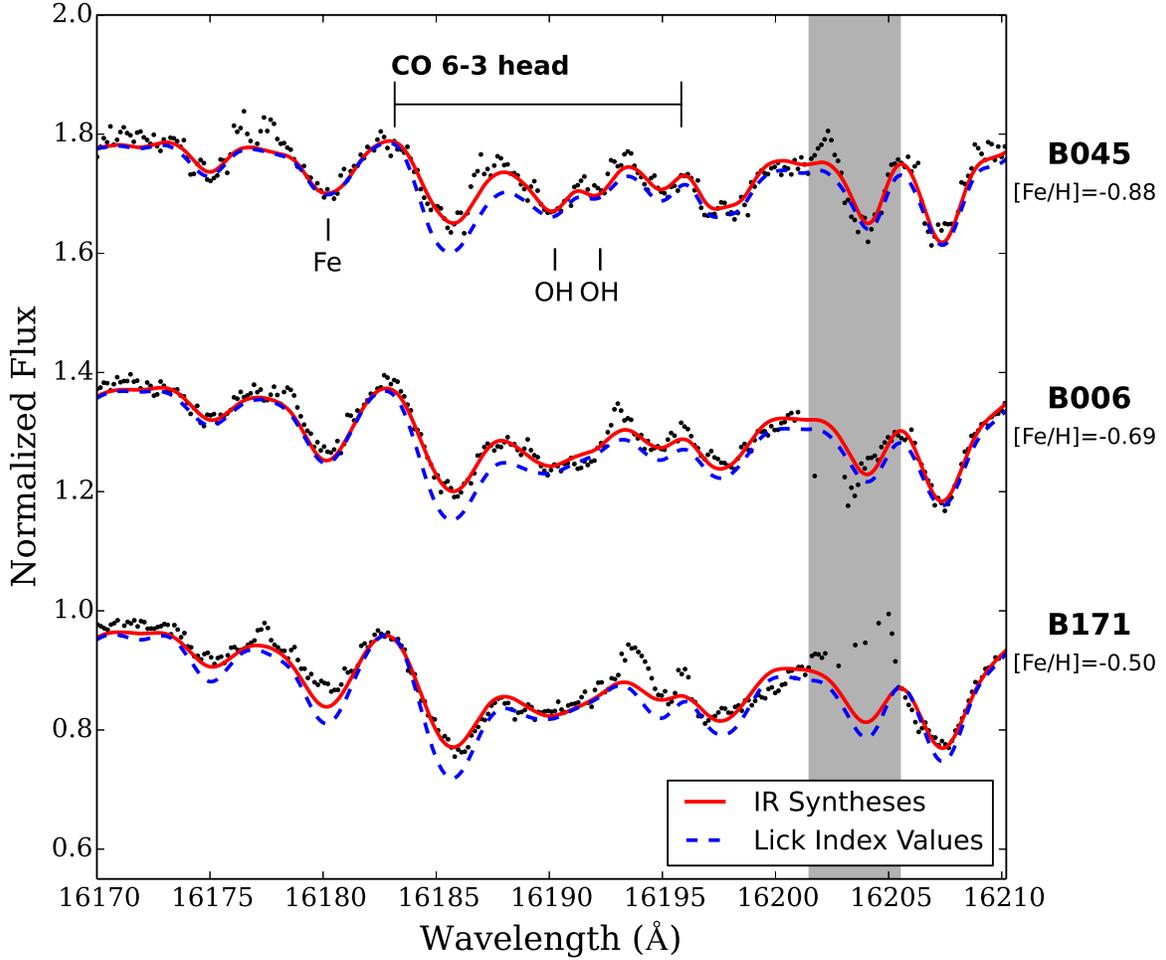}
\caption{Syntheses of the 16183 \AA \hspace{0.025in} CO bandhead,
  which is primarily sensitive to [C/Fe], in B171, B006, and B045.
  The grey area shows a region that was masked out in the data
  reduction pipeline (see Section \ref{subsec:DataReduction}). The
  black points show the data and the solid red lines show the
  best-fitting syntheses from this analysis.  The dashed blue lines
  show syntheses with the optical Lick index [Fe/H], [C/Fe], [N/Fe],
  and (assumed) [O/Fe] ratios.\label{fig:CO}}
\end{center}
\end{figure}

For the GCs with strong disagreement (i.e., all targets except the
most metal-rich GCs), the optical Lick index CNO values do not fit the
IR features. To demonstrate this, syntheses of a CO bandhead in B171,
B006, and B045 are shown in Figure \ref{fig:CO}, with the optimal
$H$-band CNO abundances and the Lick index C, N, and (assumed) O
abundances.  The 16183 \AA \hspace{0.025in} CO bandhead is primarily
sensitive to C; the syntheses in Figure \ref{fig:CO} suggest that the
Lick C abundances are too high to fit the $H$-band features,
particularly for the more metal-poor GCs in the sample.

This disagreement between the optical and the $H$-band is significant,
and will be addressed in Section \ref{subsubsec:CNspread}.

%

\clearpage
\subsection{Light Elements: Sodium, Magnesium, and Aluminum}\label{subsec:LightVary}
The ``light'' elements Na, Mg, and Al are also important for GC
studies because their abundances are known to vary between stars
within GCs (e.g., \citealt{Carretta2009}).  The $H$-band offers
complementary lines to the optical, with varying line strengths and
EPs.  While robust optical Na abundances can be derived from the
5682/5688 and 6154/6160 \AA \hspace{0.025in} doublets (and possibly
the NaD lines for metal-poor GCs) there is only one detectable weak
doublet in the $H$-band, which comes from a higher EP transition and
is not detectable in the most metal-poor GCs.  Mg has several medium
strength and strong lines in the optical; the $H$-band offers 2-6
more.  As mentioned in Section \ref{subsec:AbundDeterminations},
stronger lines can be utilized in the IR, making the strong $H$-band
Mg lines useful for GCs over a wide range of metallicities.  Optical
Al is primarily determined from the weak
6696/6698~\AA  \hspace{0.025in} lines, which are not detectable in the
lowest metallicity GCs (though there are additional lines in the blue
and red).  The $H$-band offers 2 strong and 1 moderately-strong Al
lines which can be detected even in low-metallicity GCs (note that
these strong \ion{Al}{1} lines have hyperfine structure components
that must be included to properly reproduce the strengths of the
lines).  The strongest \ion{Al}{1} line becomes prohibitively strong
at moderate metallicity ($[\rm{Fe/H}]~\sim~-1.2$), while for the
highest metallicity clusters all three Al lines are too strong.  For
clusters with only strong lines the REW limit was pushed up to $-4.5$;
as discussed in Section \ref{subsubsec:Strong}, this may introduce
systematic uncertainties of $\sim 0.1$ dex.

\begin{figure}[h!]
\begin{center}
\centering
\subfigure[]{\includegraphics[scale=0.55,trim=1.25in 0in 0.05in 0.0in]{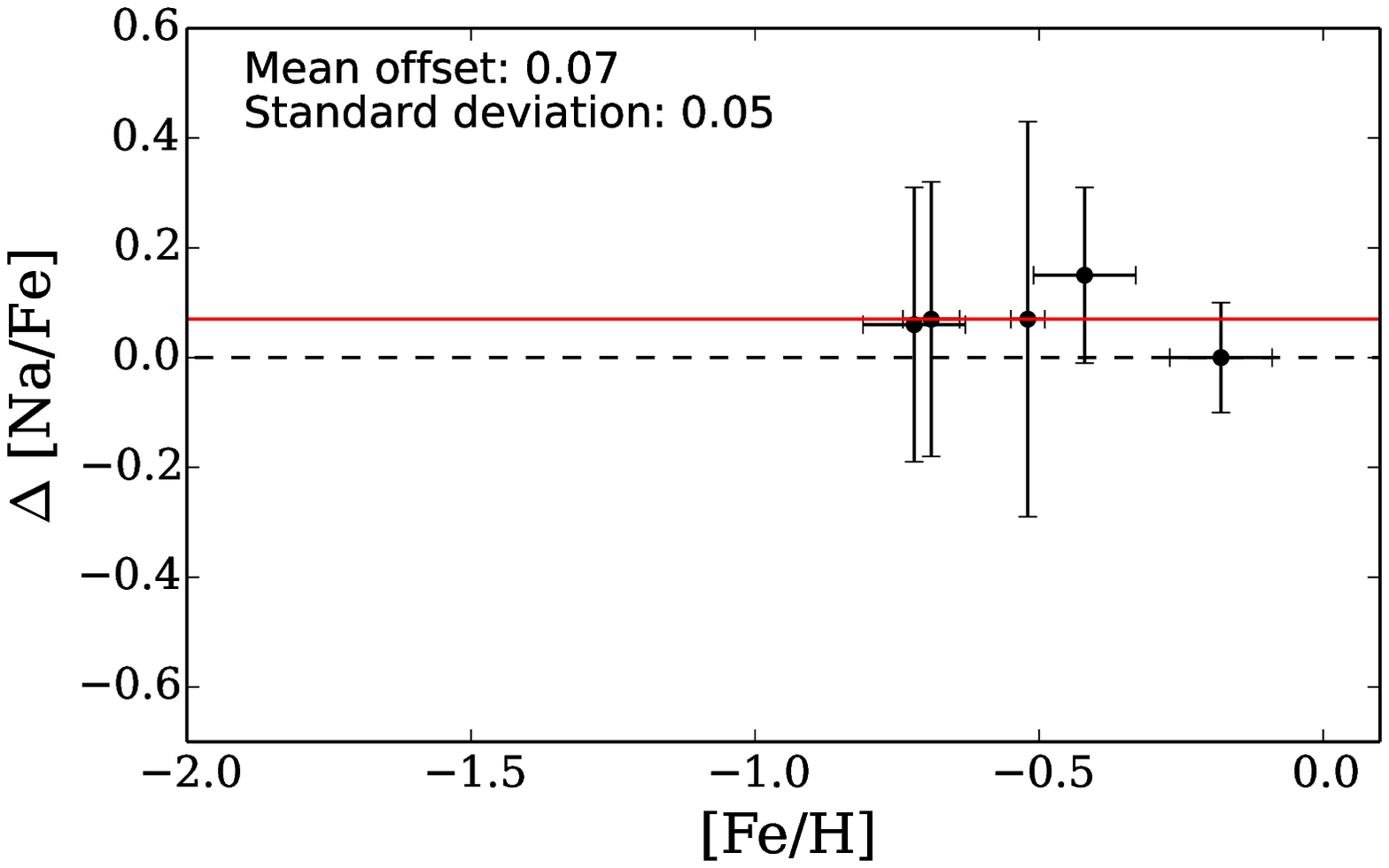}\label{fig:NaComp}}
\subfigure[]{\includegraphics[scale=0.55,trim=0.5in 0in 1.25in 0.0in]{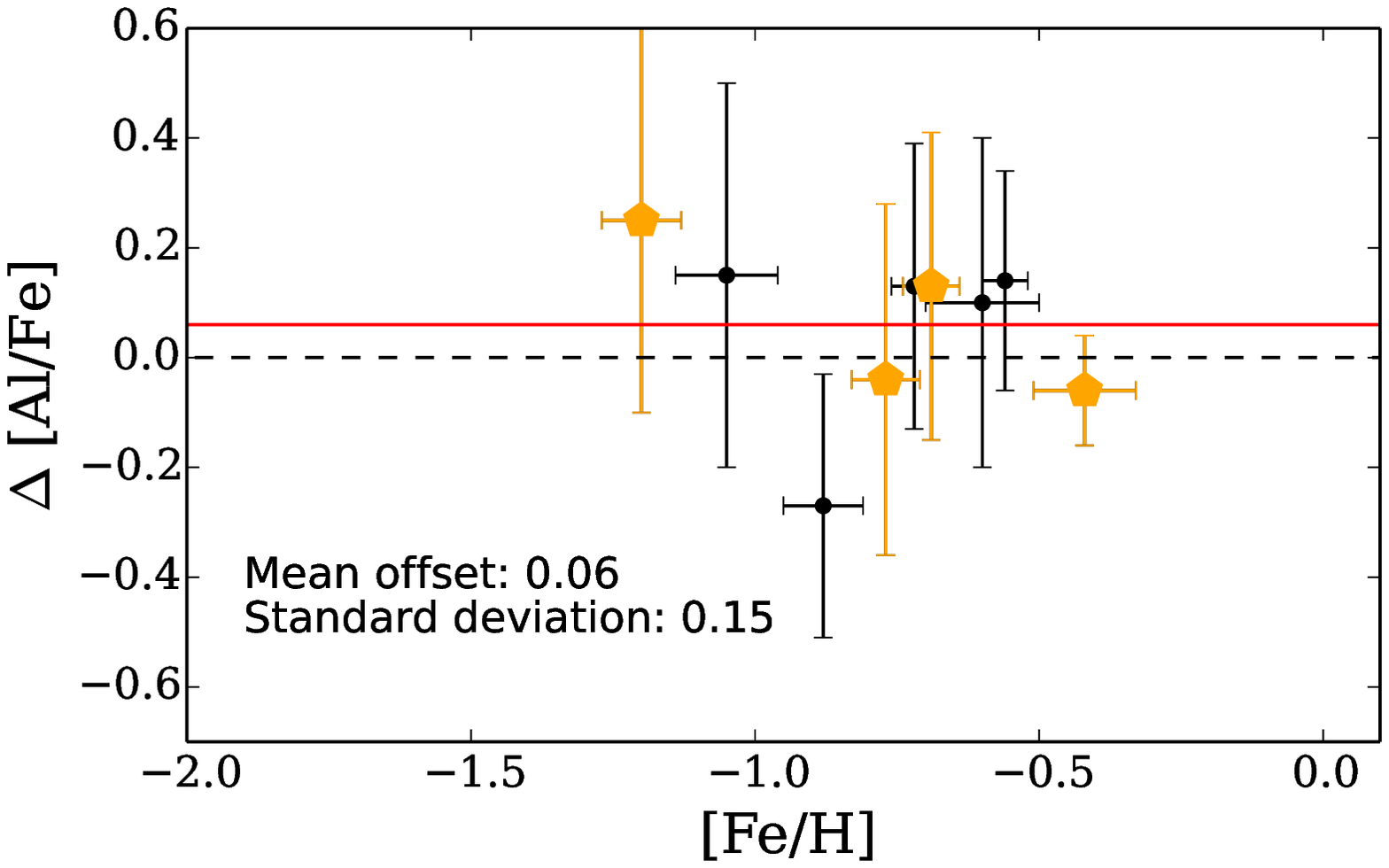}\label{fig:AlComp}}
\subfigure[]{\includegraphics[scale=0.55,trim=1.25in 0in 0.05in 0.0in]{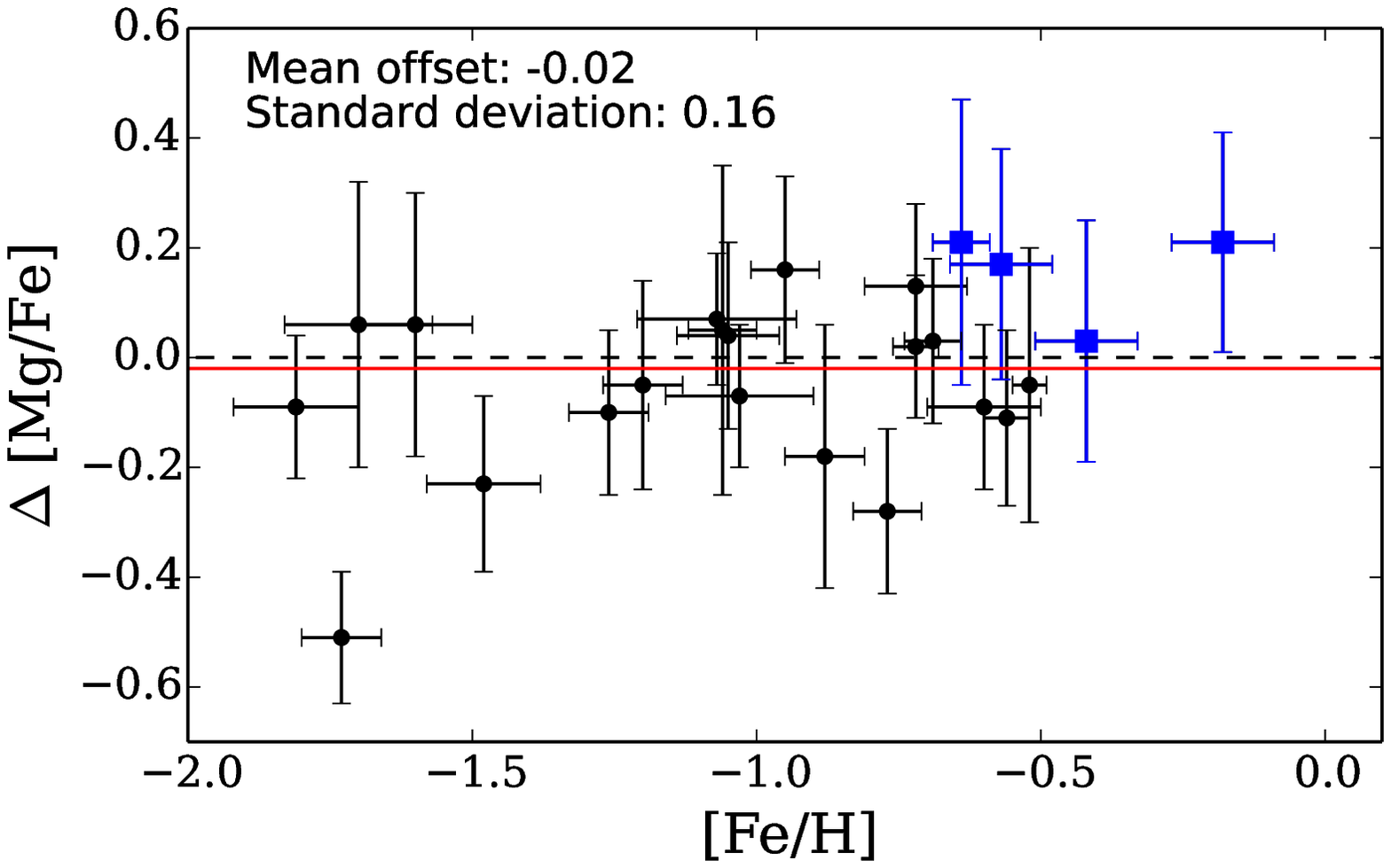}\label{fig:MgComp}}
\subfigure[]{\includegraphics[scale=0.55,trim=0.5in 0in 1.25in 0.0in]{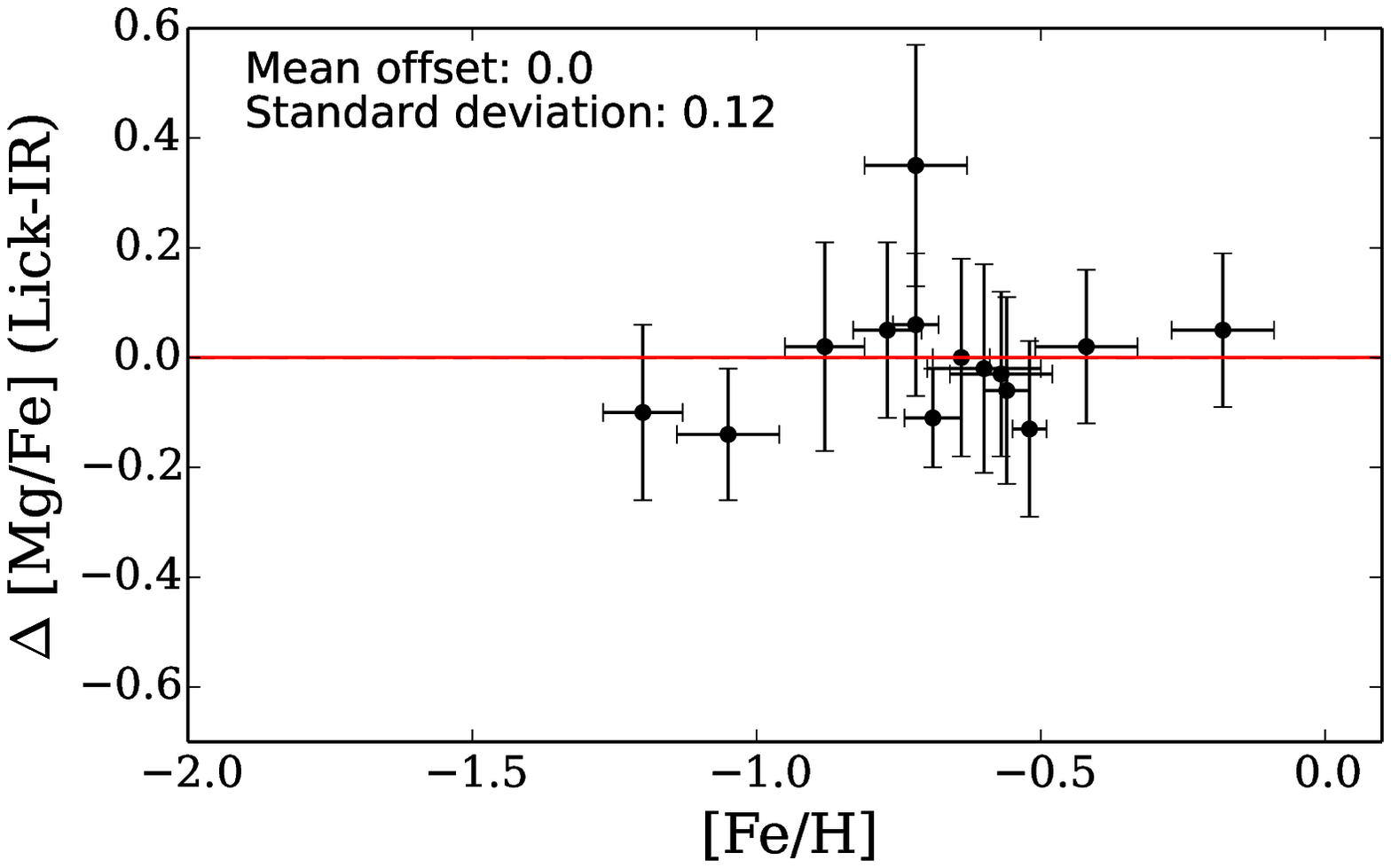}\label{fig:MgLRComp}}
\caption{Comparisons of the $H$-band [Na/Fe] (top left), [Al/Fe] (top
  right), and [Mg/Fe] (bottom left) with the optical values
  from \citet{Colucci2014} and Appendix \ref{appendix:OpticalAbunds},
  as a function of $H$-band [Fe/H].  The bottom right shows a
  comparison of $H$-band [Mg/Fe] with the Lick index [Mg/Fe] from
  \citet{Schiavon2013}.  Differences are optical $-$ $H$-band.   The
  error bars represent $1\sigma$ random errors, the dashed lines show
  perfect agreement, the solid red lines show the average offsets, and
  the quoted values list the average offsets.  Clusters whose optical
  abundances are derived from very strong lines are shown with blue
  squares, while clusters whose infrared abundances are derived from
  strong lines are shown with orange pentagons. \label{fig:LightVaryComp}}
\end{center}
\end{figure}

The $H$-band Na, Mg, and Al abundances are shown in Table
\ref{table:LightVary}, while Figure \ref{fig:LightVaryComp} shows the
optical vs. $H$-band comparisons.  There are few GCs with $H$-band and
optical [Na/Fe] abundances (because the $H$-band lines are weak and
difficult to detect).  For the four GCs with both optical and $H$-band
[Na/Fe] ratios, the agreement is decent.  The $H$-band also provides a
[Na/Fe] abundance for B193, which was not available in the optical.
Conversely, the weakness of the optical 6696/6698~\AA \hspace{0.025in}
lines means that there are only a few GCs with optical [Al/Fe] ratios.
For those ten GCs in common, the agreement between the optical and
$H$-band is generally good, though the errors are large and there is a
large scatter.  The $H$-band [Al/Fe] abundances are generally lower
than the optical, with one exception: B045's [Al/Fe] does not agree
with the optical value within its $1\sigma$ errors; its $H$-band value
is $\sim 0.2$ dex higher than its optical value.

Figure \ref{fig:MgComp} shows that though the Mg abundances have a
large scatter in the optical $-$ $H$-band difference, this scatter is
within the errors.  One cluster, B193, has a larger optical [Mg/Fe] by
$\sim 0.2$ dex---however, this cluster's optical Mg abundance was
derived with a single strong line (at 5528 \AA); it is thus possible
that the optical [Mg/Fe] is systematically offset by $>0.1$ dex (see
the discussion in Section \ref{subsubsec:Strong} and
\citealt{McWilliam1995b}).  Systematically higher [Mg/Fe] ratios are
also seen for the other metal-rich GCs whose Mg abundances were
derived with the same strong line.  Figure \ref{fig:MgLRComp} compares
these $H$-band Mg abundances to the lower resolution Lick index values
from \citet{Schiavon2013}, for metal-rich clusters only---the Mg
abundances agree within $1\sigma$ errors. One cluster, B403, has an
optical, Lick index [Mg/Fe] that is much higher than its $H$-band
value, by $\sim 0.4$ dex---however, this cluster's [Fe/H] is also in
disagreement by $>0.2$ dex, so this offset in [Mg/Fe] is not likely to
be significant.

Three clusters (B088, B240, and B235) have lower optical,
high-resolution Mg abundances (by $0.2-0.4$ dex) than in the
$H$-band. For all three cases, both optical and $H$-band abundances
are derived from 3-4 lines, each of which give consistent results.
B235's $H$-band [Mg/Fe] agrees with the lower-resolution Lick index
results from \citet{Schiavon2013}; the other two were not included in
that paper because they are too metal-poor.  B088 is offset by
$4\sigma$ according to the errors quoted in
\citet{Colucci2014}---however, these quoted optical errors are quite
small (0.02 dex due to age uncertainties), suggesting that their
stated error is too small.  Even if a larger error of 0.1 dex is
adopted for the optical, B088 is still offset by $>2\sigma$.  This
discrepancy will be investigated in Section
\ref{subsubsec:MgAlspread}.

\begin{table}
\centering
\begin{center}
\caption{Mean $H$-band abundances and random errors: Na, Mg, Al, Si, Ca, and Ti.\label{table:LightVary}\label{table:Alpha}}
  \newcolumntype{d}[1]{D{,}{\pm}{#1}}
  \begin{tabular}{@{}ld{3}cd{3}cd{3}cd{3}cd{3}cd{3}c@{}}
  \hline
 & \multicolumn{1}{c}{[Na/Fe]} & $N$ & \multicolumn{1}{c}{[Mg/Fe]} & $N$ & \multicolumn{1}{c}{[Al/Fe]} & $N$ & \multicolumn{1}{c}{[Si/Fe]} & $N$ & \multicolumn{1}{c}{[Ca/Fe]} & $N$ & \multicolumn{1}{c}{[Ti/Fe]} & $N$ \\
\hline
B232  & \multicolumn{1}{c}{$-$} & & -0.01,0.10 & 1 & \multicolumn{1}{c}{$-$}  & & 0.47,0.10 & 1 &  \multicolumn{1}{c}{$-$} & & \multicolumn{1}{c}{$-$} &  \\
B088  & \multicolumn{1}{c}{$-$} & &  0.03,0.10 & 3 & 0.51,0.20 & 1 &  0.38,0.09 & 3 &  0.31,0.20 & 1 & \multicolumn{1}{c}{$-$} & \\
B311  & \multicolumn{1}{c}{$-$} & & -0.04,0.14 & 2 & \multicolumn{1}{c}{$-$} & & 0.21,0.10 & 3 &  0.38,0.40 & 1 & \multicolumn{1}{c}{$-$} & \\
B012  & \multicolumn{1}{c}{$-$} & & -0.14,0.18 & 2 & 0.38,0.15 & 1& 0.43,0.11 & 2 &  \multicolumn{1}{c}{$-$} & & \multicolumn{1}{c}{$-$} &\\  
B240  & \multicolumn{1}{c}{$-$} & &  0.20,0.07 & 4 & 0.41,0.15 & 2& 0.22,0.06 & 4 &  \multicolumn{1}{c}{$-$} & & \multicolumn{1}{c}{$-$} & \\
B405  & \multicolumn{1}{c}{$-$} & &  0.07,0.12 & 3 & 0.35,0.20 & 1&  0.35,0.08 & 3 &  \multicolumn{1}{c}{$-$} & & \multicolumn{1}{c}{$-$} &  \\
B472  & \multicolumn{1}{c}{$-$} & &  0.14,0.10 & 4 & 0.49,0.20 & 1$^{a}$& 0.14,0.16 & 5 &  0.32,0.14 & 2 & \multicolumn{1}{c}{$-$} & \\
B386  & \multicolumn{1}{c}{$-$} & &  0.06,0.10 & 1 & 0.46,0.08 & 1& 0.46,0.08 & 3 &  0.26,0.20 & 1 & 0.36,0.10 & 1 \\
B312  & \multicolumn{1}{c}{$-$} & & -0.05,0.20 & 1 & 0.35,0.14 & 2& 0.52,0.07 & 3 &  \multicolumn{1}{c}{$-$} & & \multicolumn{1}{c}{$-$} &\\
B063  & \multicolumn{1}{c}{$-$} & &  0.34,0.08 & 3 & 0.29,0.20 & 1& 0.36,0.08 & 3 &  0.49,0.10 & 1 & 0.24,0.08 & 3  \\
B381  & \multicolumn{1}{c}{$-$} & &  0.22,0.10 & 1 & 0.47,0.15 & 1& 0.41,0.05 & 4 &  0.22,0.10 & 3 & \multicolumn{1}{c}{$-$} &  \\
B182  & \multicolumn{1}{c}{$-$} & &  0.19,0.05 & 2 & 0.19,0.11 & 2& 0.46,0.08 & 3 &  0.44,0.07 & 2 & 0.34,0.20 & 1 \\
B045  & \multicolumn{1}{c}{$-$} & &  0.22,0.15 & 2 & 0.32,0.15 & 1& 0.43,0.06 & 2 &  0.20,0.13 & 2 & 0.27,0.14 & 2 \\
B048  & \multicolumn{1}{c}{$-$} & & \multicolumn{1}{c}{$-$} & & \multicolumn{1}{c}{$-$} & & 0.48,0.10 & 2 &  0.32,0.07 & 2 & \multicolumn{1}{c}{$-$} &  \\
B235  & \multicolumn{1}{c}{$-$} & &  0.27,0.13 & 2 & 0.39,0.20 & 1& 0.40,0.05 & 4 &  0.37,0.07 & 2 & \multicolumn{1}{c}{$-$} & \\
B383  & \multicolumn{1}{c}{$-$} & &  0.22,0.09 & 2 & 0.22,0.14 & 2& 0.35,0.06 & 4 &  0.34,0.07 & 2 & \multicolumn{1}{c}{$-$} &  \\
B403  & 0.32,0.20             & 1 &  0.05,0.13 & 2 & \multicolumn{1}{c}{$-$} & & 0.42,0.10 & 1 &  0.22,0.06 & 2 & \multicolumn{1}{c}{$-$} &  \\
B006  & 0.39,0.14             & 2 &  0.43,0.05 & 2 & 0.43,0.15 & 1$^{a}$& 0.37,0.14 & 5 &  0.31,0.07 & 2 & 0.43,0.07 & 2 \\
B225  & \multicolumn{1}{c}{$-$} & &  0.24,0.14 & 2 & 0.64,0.14 & 2& 0.32,0.06 & 3 &  0.34,0.06 & 3 & 0.39,0.15 & 1 \\
B034  & \multicolumn{1}{c}{$-$} & &  0.30,0.10 & 1 & 0.40,0.20 & 1& 0.35,0.10 & 3 &  0.30,0.07 & 2 & 0.35,0.11 & 2 \\
B110  & \multicolumn{1}{c}{$-$} & &  0.23,0.11 & 2 & 0.32,0.14 & 1& 0.31,0.08 & 2 &  0.28,0.07 & 2 & \multicolumn{1}{c}{$-$} &  \\
B384  & \multicolumn{1}{c}{$-$} & &  0.26,0.13 & 2 & 0.16,0.09 & 1& 0.32,0.09 & 4 &  0.23,0.08 & 4 & 0.31,0.11 & 2 \\
B171  & 0.57,0.20             & 1 &  0.37,0.10 & 1 & \multicolumn{1}{c}{$-$} & & 0.27,0.08 & 5 &  0.30,0.07 & 2 & 0.35,0.10 & 2 \\
B163  & 0.57,0.10             & 1 &  0.22,0.10 & 1 & 0.56,0.05 & 2$^{a}$& 0.19,0.09 & 3 &  0.27,0.05 & 2 & 0.22,0.10 & 1 \\
B193  & 0.64,0.05             & 2 &  0.19,0.10 & 1 & 0.39,0.20 & 1$^{a}$ & 0.27,0.10 & 3 &  0.34,0.15 & 2 & 0.36,0.08 & 3  \\
 & & & & & & & & & & & & \\
\hline
\end{tabular}\\
\end{center}
\medskip
\raggedright $^{a}$ This $H$-band abundance was derived from at least
one strong line with $-4.7 < \rm{REW} -4.5$, which may lead to
systematic uncertainties of $\sim 0.1$ dex \citep{McWilliam1995b}.
\end{table}

\clearpage
\subsection{Alpha Elements: Silicon, Calcium, and Titanium}\label{subsec:Alpha}
The $\alpha$-elements Si, Ca, and Ti mainly form in massive stars and
are expelled into the interstellar medium by core-collapse
supernovae (which also produce iron).  (Mg is also an
$\alpha$-element; however, because it can vary between stars within a
single GC, it is discussed in Section \ref{subsec:LightVary} with O,
Na, and Al.)  Type Ia supernovae from lower mass stars also produce
copious amounts of iron but very little $\alpha$-elements.  The ratio
of [$\alpha$/Fe] therefore probes contributions from Type II vs. Ia
supernovae, and is valuable for galactic chemical evolution studies.
Because chemical evolution in dwarf galaxies proceeds differently than
in massive galaxies, the chemical abundances of dwarf galaxy stars and
clusters diverges with increasing metallicity (see, e.g.,
\citealt{Tolstoy2009}).  The [$\alpha$/Fe] ratios can therefore be
used for chemical tagging of GCs (e.g., \citealt{Hogg2016}), and are
essential for IL spectroscopy of extragalactic GCs (e.g.,
\citealt{Colucci2012,Colucci2014,Sakari2015}).

Again, the $H$-band provides complementary lines to the optical.  The
optical has numerous silicon lines, but they are often weak and
difficult to detect in metal-poor GCs.  There are 3 strong
\ion{Si}{1} lines in the $H$-band, and 6 more moderately strong
lines---note that the strongest \ion{Si}{1} lines at 15960 and 16095
\AA \hspace{0.025in} become too strong ($\rm{REW}\gtrsim -4.7$) at
high metallicity ($[\rm{Fe/H}] \gtrsim -0.7$).  The
optical has many detectable \ion{Ca}{1} lines of varying
strength---furthermore, these lines are also largely insensitive to
uncertainties in the underlying stellar populations (e.g., the
distribution of stars in temperature-$\log g$ space;
\citealt{Sakari2014}).  In the $H$-band there are three \ion{Ca}{1}
lines at high EP, though 2 are extremely blended, leading to large
random errors. \ion{Ti}{1} and \ion{Ti}{2} lines are readily available
in the optical, but both suffer from potential systematic
uncertainties.  The \ion{Ti}{1} lines may be affected by NLTE effects
\citep{Bergemann2011} and are strongly affected by uncertainties in
the underlying stellar populations \citep{Sakari2014}.  The
\ion{Ti}{2} lines are also affected by uncertainties in the stellar
populations, and many of the lines are located further in the blue
(where the S/N is often lower and where contributions from hotter
stars may have a larger effect).  The $H$-band offers a handful of
moderate strength \ion{Ti}{1} lines---though they are often blended
with other features, these lines may be less affected by uncertainties
in the underlying stellar populations.

\begin{figure}[h!]
\begin{center}
\centering
\subfigure[]{\includegraphics[scale=0.55,trim=1.25in 0.25in 0.05in 0.5in]{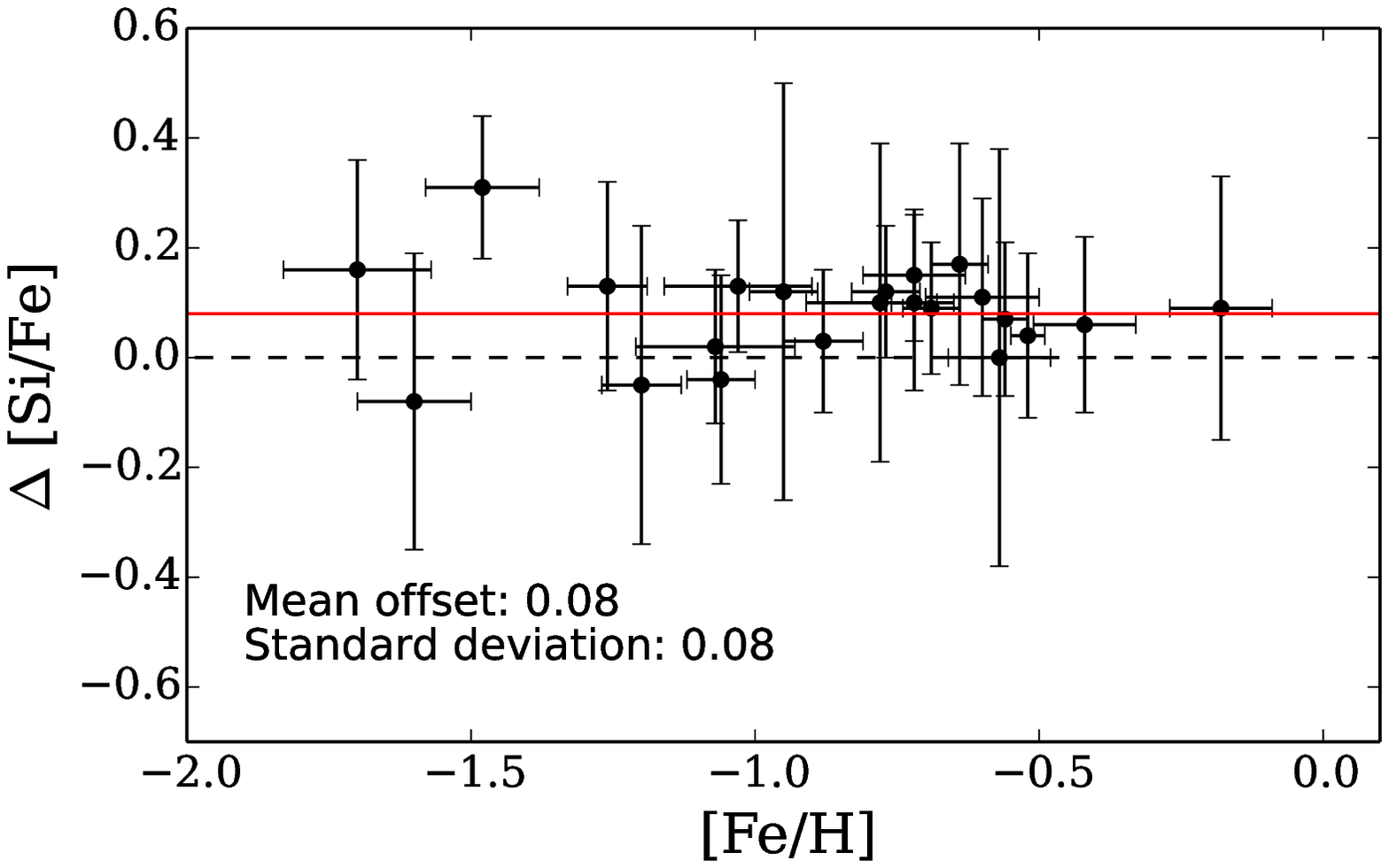}\label{fig:SiComp}}
\subfigure[]{\includegraphics[scale=0.55,trim=0.5in 0.25in 1.25in 0.5in]{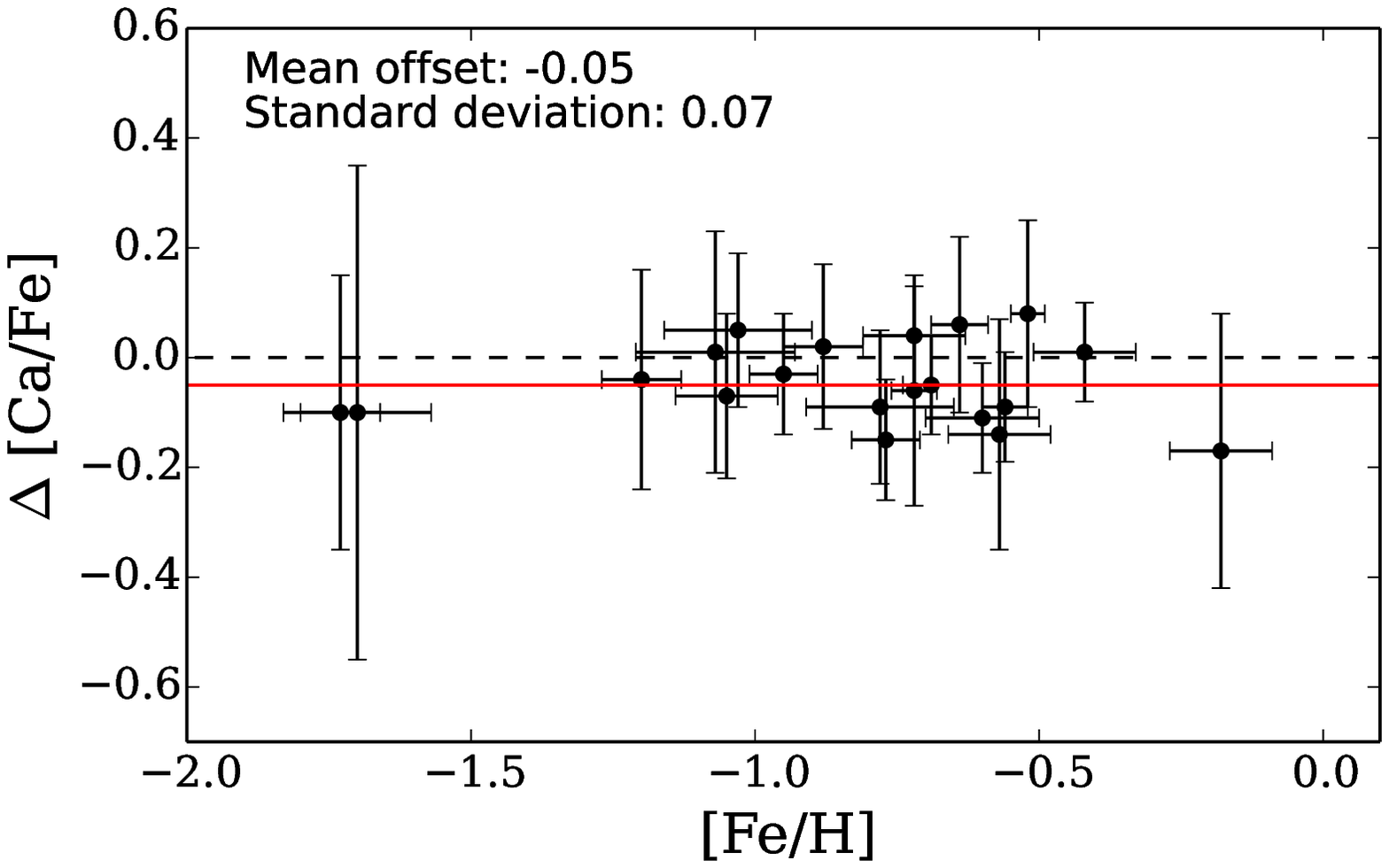}\label{fig:CaComp}}
\subfigure[]{\includegraphics[scale=0.55,trim=1.25in 0.25in 0.05in 0.15in]{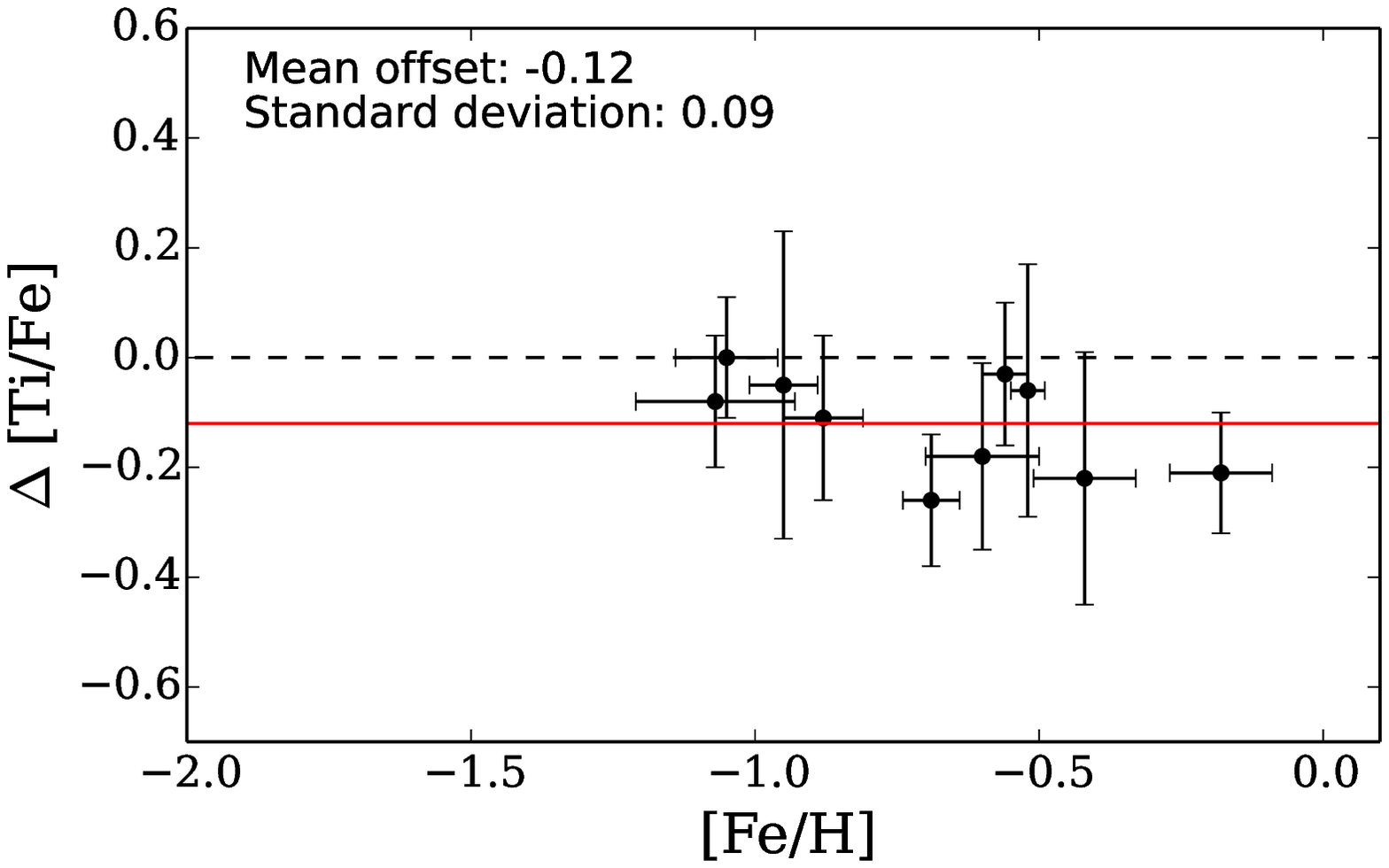}\label{fig:TiComp}}
\subfigure[]{\includegraphics[scale=0.55,trim=0.5in 0.25in 1.25in  0.15in]{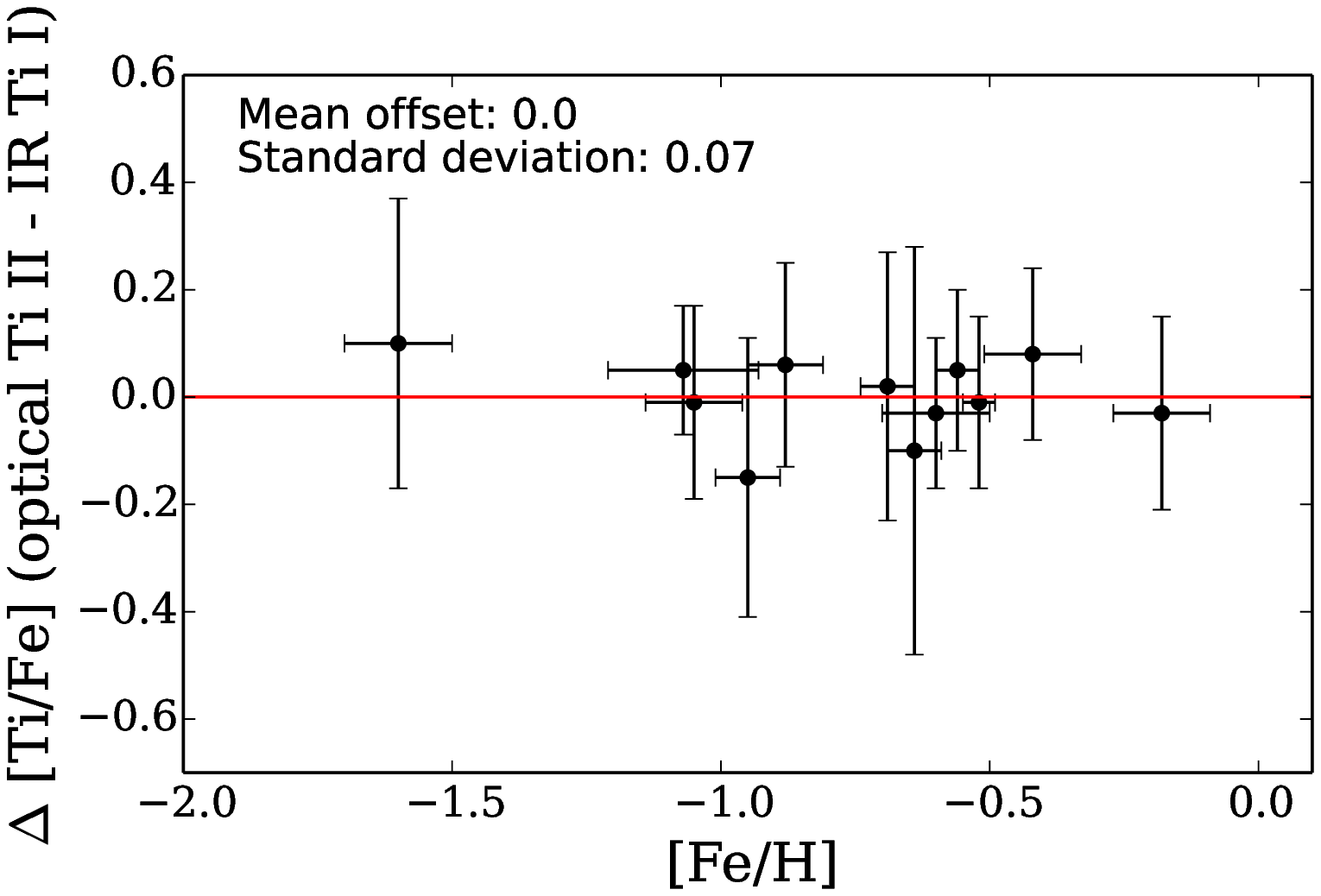}\label{fig:TiIIComp}}
\vspace{-0.1in}
\caption{Comparisons of the $H$-band [Si/Fe] (top left), [Ca/Fe] (top
  right), and [Ti/Fe] (bottom) ratios with the optical values from
  \citet{Colucci2014} and Appendix \ref{appendix:OpticalAbunds}, as a
  function of $H$-band [Fe/H].  The bottom left plot shows
  $\Delta$[\ion{Ti}{1}/Fe] while the bottom right plot shows the
  difference between optical [\ion{Ti}{2}/Fe] and $H$-band
  [\ion{Ti}{1}/Fe].  Differences are optical $-$ $H$-band.  The error
  bars represent $1\sigma$ random errors, the dashed lines show
  perfect agreement, the solid red lines show the average offsets, and
  the quoted values list the average offsets.\label{fig:AlphaComp}}
\end{center}
\end{figure}

The abundances of the $\alpha$-elements are listed in Table
\ref{table:Alpha} and are compared with the optical abundances in
Figure \ref{fig:AlphaComp}.  Figure \ref{fig:SiComp} compares
[Si/Fe].  The uncertainties in the optical Si abundances can be quite
large because the lines are often weak. There are three GCs whose
[Si/Fe] ratios do not agree between the optical and the $H$-band
within $1\sigma$ errors: B403, B381, and B240, all of which have
optical [Si/Fe] ratios that are $>0.5$ dex. The stronger $H$-band Si
lines may produce higher-precision Si abundances compared to the
weaker optical Si lines.

Calcium is an element which can be difficult to detect in the
$H$-band, particularly in the most metal-poor GCs.  The
four available lines are from high EP transitions which are clumped
together in a region at $\sim 16150$ \AA \hspace{0.025in} that is
superpersistence in some visits (see Table \ref{table:Targets}).  The
optical abundances, however, are well-constrained from measurements of
$2-13$ lines.  When $H$-band Ca lines can be detected, Figure
\ref{fig:CaComp} shows that the agreement between the optical and the
$H$-band is generally good, with the exception of B034 and B235, whose
$H$-band [Ca/Fe] ratios are $>1\sigma$ higher than the optical
values.

The $H$-band \ion{Ti}{1} lines are very weak and blended in these
spectra.  Table \ref{table:Alpha} shows that Ti is not detectable in
the most metal-poor clusters and is difficult to detect in GCs with
high velocity dispersions.  For the GCs with measurements in both the
$H$-band and the optical, the $H$-band [\ion{Ti}{1}/\ion{Fe}{1}]
ratios are generally higher than the optical values, as seen in Figure
\ref{fig:TiComp}.  B006, B034, and B193 are not in
agreement---however, all GCs show agreement between $H$-band
[\ion{Ti}{1}/\ion{Fe}{1}] and optical [\ion{Ti}{2}/\ion{Fe}{2}]
(Figure \ref{fig:TiIIComp}).  This suggests that the optical
\ion{Ti}{1} lines are systematically affected in some way that the IR
lines are not---again, this could indicate that the $H$-band lines are
less affected by NLTE effects (e.g., \citealt{GarciaHernandez2015}).

\vspace{0.1in}

\clearpage
\subsection{Potassium}\label{subsec:K}
\begin{figure}[h!]
\begin{center}
\centering
\includegraphics[scale=0.55]{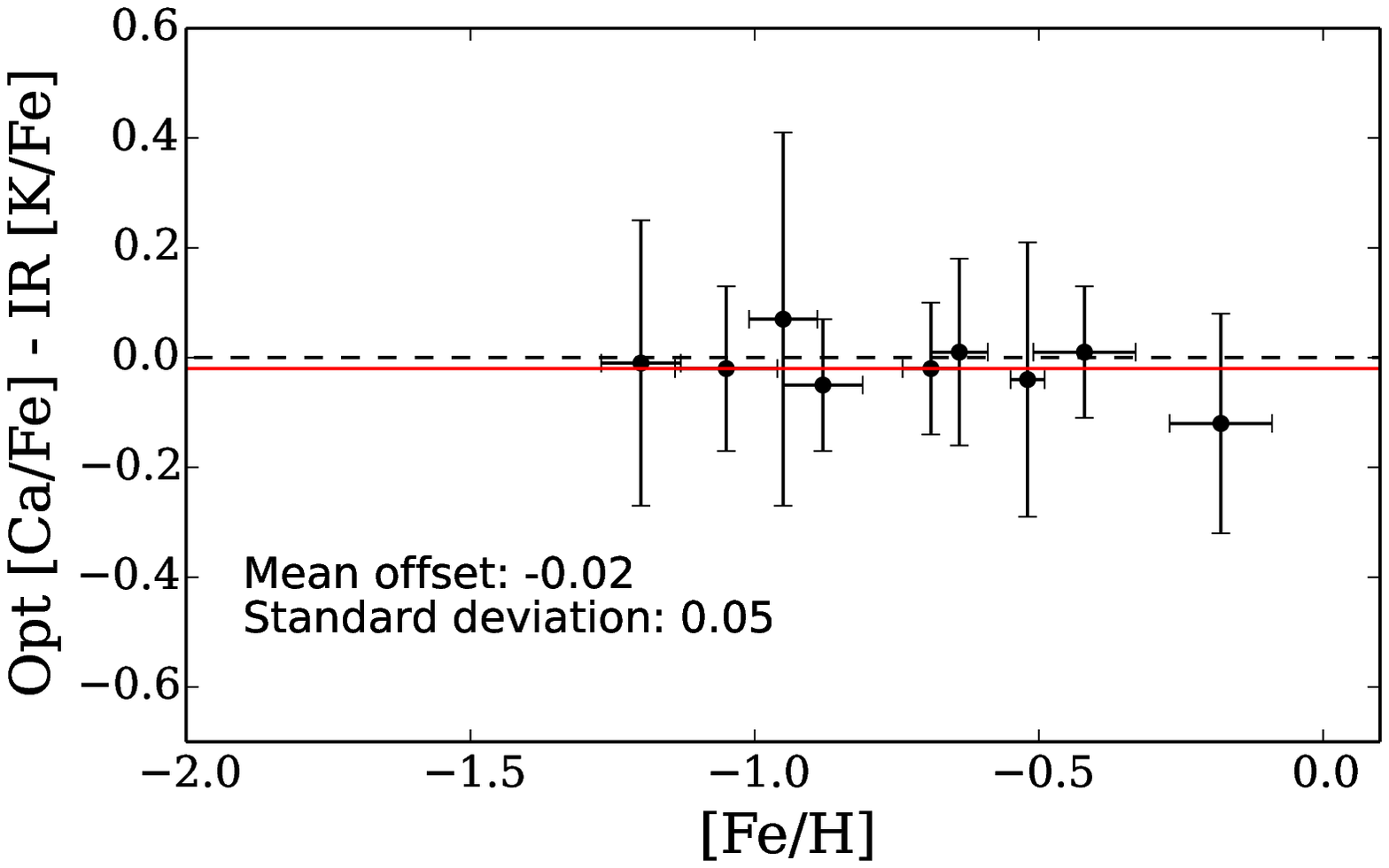}
\caption{Comparisons of the $H$-band [K/Fe] ratios to the optical
  [Ca/Fe] ratios, as a function of $H$-band [Fe/H].  The error bars
  represent $1\sigma$ random errors, the dashed line shows perfect
  agreement, the solid red line shows the average offset, and the
  quoted value is the average offset.\label{fig:KComp}}
\end{center}
\end{figure}

The $H$-band offers two moderate EP (2.67 eV) \ion{K}{1} lines at
15163 and 15168 \AA.  These lines lie on the blue edge of APOGEE's
spectral range, and they are not easily detectable in most of the
targets with low S/N.  K {\it is} detectable in only a handful of the
GCs, as shown in Table \ref{table:K}.  There are no optical, IL K
abundances to compare with, but [K/Fe] can be compared to the
$\alpha$-elements.  Potassium is not an $\alpha$-element, yet
abundance analyses of stars in the MW indicate that the chemical
evolution of [K/Fe] is similar to the evolution in [$\alpha$/Fe] (see,
e.g., \citealt{Zhang2006}).  The [K/Fe] ratios are compared to the
optical [Ca/Fe] ratios in Figure \ref{fig:KComp}; the agreement is
generally quite good.  The optical [Ca/Fe] ratios are chosen for this
comparison in lieu of Mg, Si, or Ti because they have the lowest
random and systematic errors and therefore likely to best represent
the cluster [$\alpha$/Fe] \citep{Sakari2014}.  K is therefore another
viable element for abundance analyses of extragalactic targets.  The
spectral lines are not easy to detect in spectra of this quality, but
the detections and errors would be vastly improved with higher S/N.

\begin{table}
\centering
\begin{center}
\caption{Potassium abundances.\label{table:K}}
  \newcolumntype{d}[1]{D{,}{\pm}{#1}}
  \begin{tabular}{@{}ld{3}c}
  \hline
 & \multicolumn{1}{c}{[K/Fe]} & N \\
\hline
B472  & 0.29,0.20 & 1 \\
B063  & 0.44,0.10 & 1 \\
B045  & 0.27,0.10 & 1 \\
B006  & 0.28,0.10 & 1 \\
B225  & 0.39,0.07 & 2\\ 
B171  & 0.42,0.15 & 1 \\
B163  & 0.27,0.08 & 2 \\
B193  & 0.29,0.10 & 2 \\
 & & \\
\hline
\end{tabular}\\
\end{center}
\medskip
\end{table}

\clearpage
\section{Discussion}\label{sec:Discussion}
The abundance results from Section \ref{sec:Abunds} can be summarized
as follows.

\begin{description}
\item[Iron: ] All clusters show good agreement between optical and
  $H$-band [Fe/H] ratios, most within 0.1 dex.  However, this is when
  the optical age and metallicity from the optical are adopted (though
  note that the isochrone [Fe/H] is refined if the $H$-band abundance
  differs enough from the optical).  It may be difficult to identify a
  viable isochrone from the $H$-band alone, as discussed in Section
  \ref{subsec:FeDiscussion}.\\

\item[CNO: ] The $H$-band [C/Fe] ratios are systematically lower than
  the optical Lick index values, while (with three metal-rich
  exceptions) the $H$-band [N/Fe] values are higher.  This discrepancy
  is explored further in Section \ref{subsec:MultiPopDiscussion}.\\

\item[Other elements: ] With the exception of a few outliers in each
  case, the $H$-band Na, Mg, Al, Si, and Ca abundances agree well with
  the optical values.  The $H$-band Ti abundance is in excellent
  agreement with the optical \ion{Ti}{2} abundance, suggesting that
  the $H$-band \ion{Ti}{1} lines are not as sensitive to NLTE effects
  as optical lines.  The outliers could be due to issues with specific
  lines (as a result of, e.g., S/N or atomic data) or could reflect
  problems with the models of the underlying populations (though see
  Sections \ref{subsec:FeDiscussion} and \ref{subsec:AGBratio}).  The
  higher precision Si and Al abundances and new O and K abundances
  enable the multiple populations in these M31 GCs to be probed in new
  ways, as discussed in Section \ref{subsec:MultiPopDiscussion}.  For
  elements that are not suspected to vary between stars in GCs, the
  general agreement between optical and $H$-band for all elements
  demonstrates the validity of the $H$-band for abundance analyses of
  unresolved targets (see Section \ref{subsec:ChemEvol}).\\ 
\end{description}
 
\vspace{-0.5in}

\subsection{Determining Isochrone Parameters in the Infrared}\label{subsec:FeDiscussion}
High-resolution optical analyses (e.g., 
\citealt{McWB,Colucci2009,Colucci2011a,Colucci2014,Sakari2013,Sakari2015})
utilize \ion{Fe}{1} lines to constrain the appropriate age and
metallicity of the stellar populations, by minimizing trends in iron
abundance with line wavelength, REW, and EP.  Improperly modeled
atmospheres will cause lines with different properties to have
systematically offset abundances---e.g., populations with too few hot
stars will require larger Fe abundances to match the strengths of the
highest EP lines.  This technique relies on a large sample of
\ion{Fe}{1} lines with a range of wavelengths, REWs, and EPs.  Even
in the optical, where a large selection of \ion{Fe}{1} lines is
readily available, the line-to-line scatter prohibits better precision
in cluster ages than $\sim1-5$ Gyr \citep{Colucci2014}.  Section
\ref{subsec:Fe} demonstrates that the $H$-band offers only a few
detectable, moderately blended \ion{Fe}{1} lines ($\sim2-13$,
depending on metallicity, S/N, and cluster velocity dispersion).  The
severe blending with nearby lines, especially molecular features, at
this resolution renders equivalent width analyses extremely
difficult. Furthermore, these \ion{Fe}{1} lines are primarily from
high EP transitions, with most having ${\rm EP} > 5$ eV; only two have
${\rm EP} < 5$~eV, and those lines are not detectable in all GCs. The
$H$-band \ion{Fe}{1} lines have a variety of strengths, but only the
strongest lines are detectable in the most metal-poor, high velocity
dispersion GCs.  This limits the usefulness of the $H$-band for
determining GC age without the aid of optical data.  Note that while
there are other, better techniques for determining GC age (e.g.,
isochrone fitting of partially resolved GCs, \citealt{Mackey2013}; or
fitting age-sensitive features in the optical,
\citealt{Caldwell2009,Caldwell2011}) this section is only concerned
with whether the $H$-band spectra alone are sufficient for identifying
an appropriate isochrone (and therefore for performing a detailed
chemical abundance analysis).

\begin{figure}[h!]
\begin{center}
\centering
\subfigure[Age: 12 Gyr]{\includegraphics[scale=0.4,trim=0in 0.25in 0.0in 0.25in,clip]{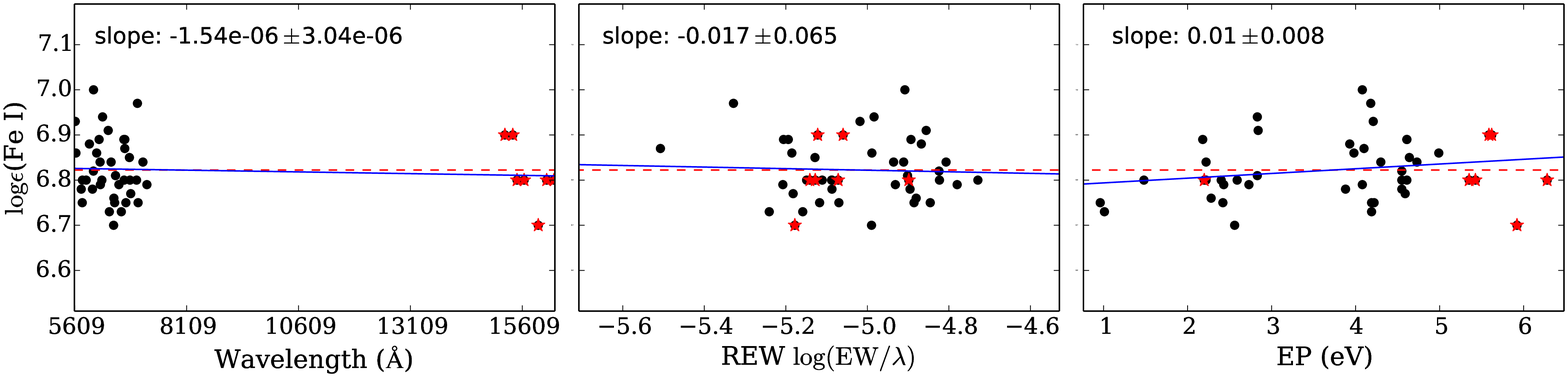}\label{fig:B006Trendsa}}
\subfigure[Age: 5 Gyr]{\includegraphics[scale=0.4,trim=0in 0.25in 0.0in 0.25in,clip]{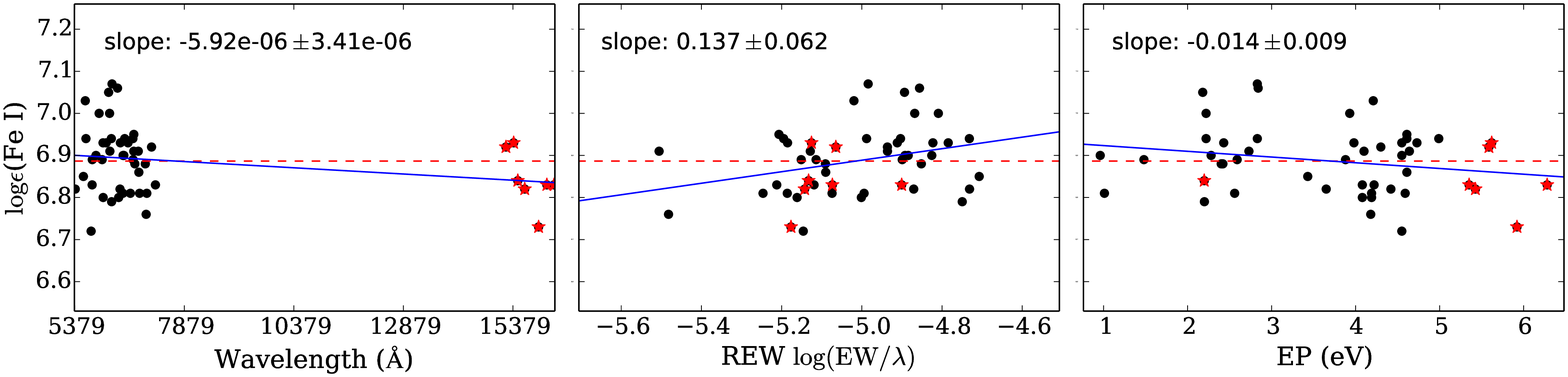}\label{fig:B006Trendsb}}
\caption{The trends in \ion{Fe}{1} abundance with wavelength, REW, and
  EP, for the cluster B006. 
  {\it Top:} Abundances with the assumed optimal age of 12 Gyr.
  {\it Bottom: } Abundances with a younger age of 5 Gyr.  The optical
  lines (from Appendix \ref{appendix:OpticalAbunds})  are shown with
  black circles, while the IR lines are shown as red stars.  The
  dashed red line shows the average \ion{Fe}{1} abundance, while the
  solid blue line shows the linear least squares fit to the points.
  The slopes are listed in each panel.   For the 12 Gyr case, the
  $H$-band lines maintain the flat slopes from the optical, while
  adding range in wavelength and EP. The slopes from the optical
  points in the 5 Gyr case indicate that the age is inappropriate for
  B006; however, the $H$-band points do not have the parameter range
  necessary to constrain the cluster age.\vspace{-0.25in}}
\end{center}
\end{figure}

Figure \ref{fig:B006Trendsa} shows that for B006
($[\rm{Fe/H}]~=~-0.69$) the $H$-band \ion{Fe}{1} lines agree well with
the optical lines, and add more range to the wavelength and EP plots.
On their own, however, the $H$-band lines do not provide a sufficient
range in REW or EP to flatten the slopes and constrain GC age.  This
is demonstrated in Figure \ref{fig:B006Trendsb}, where a younger age
of 5 Gyr is adopted.  The optical lines indicate that 5 Gyr is not
likely to be an appropriate age for B006.  The $H$-band lines by
themselves, on the other hand, cannot be used to rule out an age of 5
Gyr.  The situation becomes even worse for clusters with low S/N
and/or low [Fe/H], in which fewer \ion{Fe}{1} lines are detectable.
Minimizing trends in the $H$-band \ion{Fe}{1} abundances is therefore
not a viable way to constrain the cluster age.

\begin{figure}[h!]
\begin{center}
\centering
\subfigure[]{\includegraphics[scale=0.5]{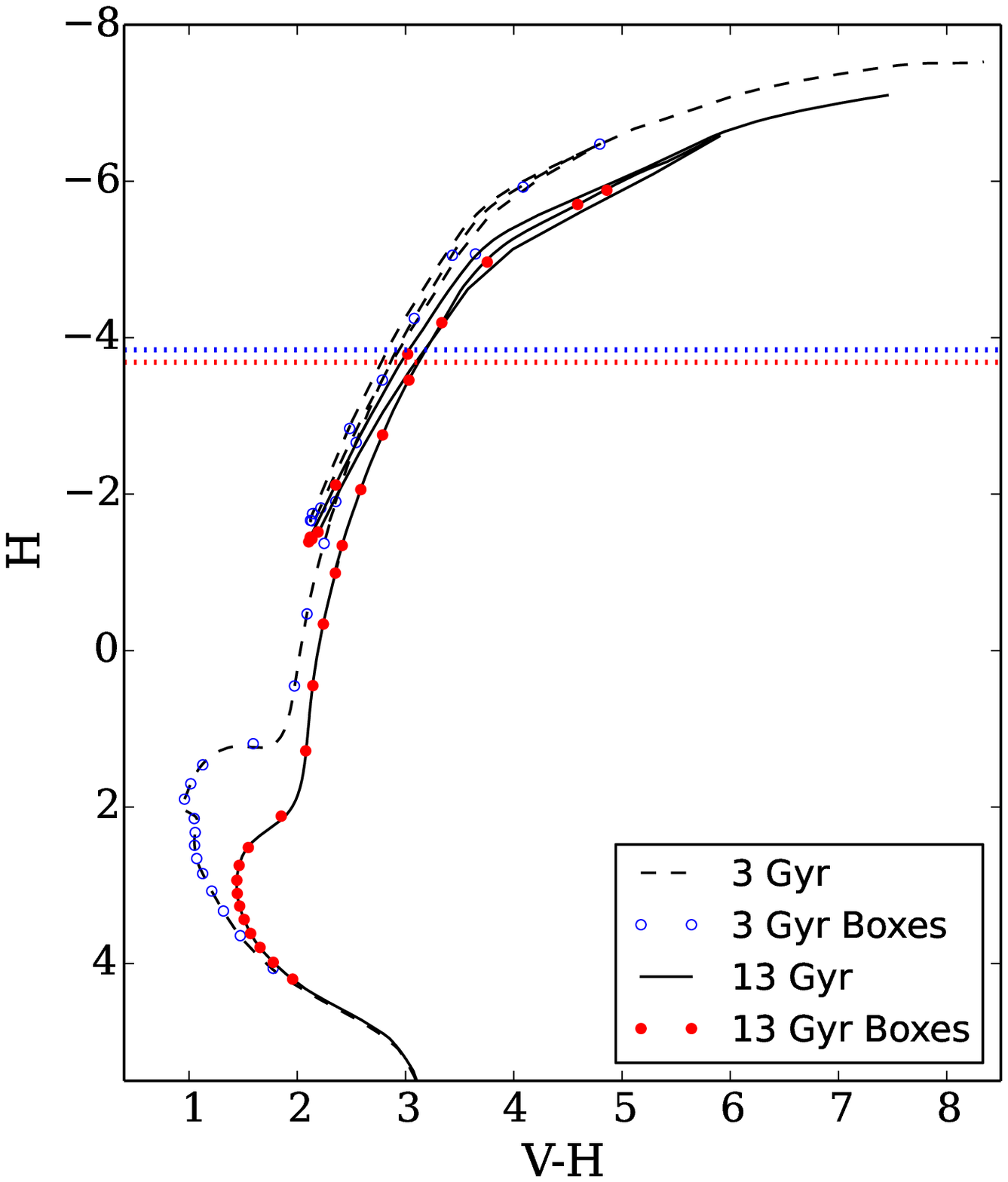}\label{fig:HbandIsos}}
\subfigure[]{\includegraphics[scale=0.5]{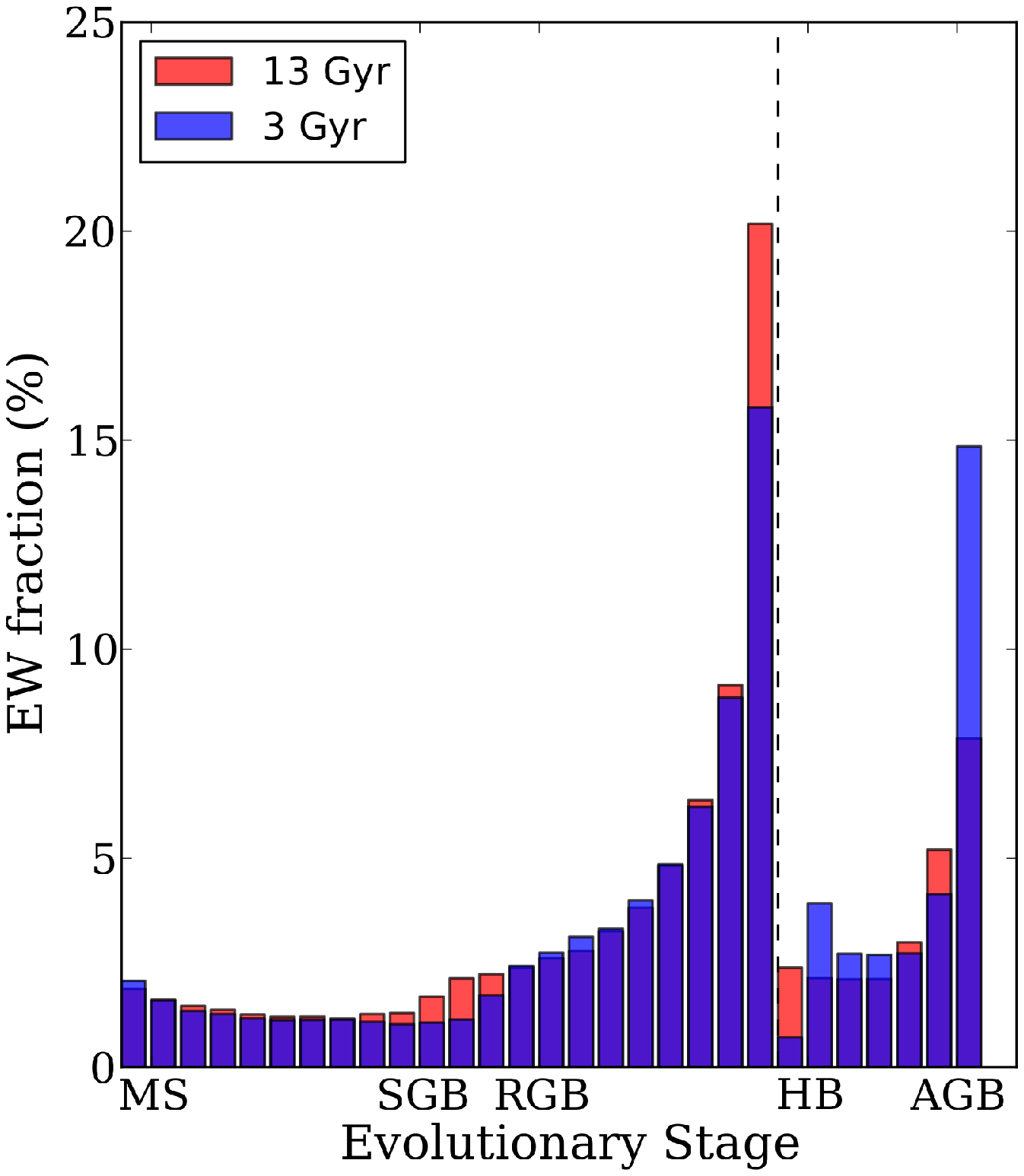}\label{fig:EWfrac}}
\caption{Age effects in the $H$-band.  {\it Left: } BaSTI isochrones
  at $[\rm{Fe/H}] = -0.6$ for ages of 13 Gyr (solid line) and 3 Gyr
  (dashed line).  The H-R Diagram boxes are shown on top of the
  isochrones (see Section \ref{subsec:ModelAtms}). The half-light
  $H$-band levels are shown with dotted lines, and are well above the
  HB stars.  {\it Right: } The contributions to the strength of an
  \ion{Fe}{1} line (at 15207 \AA) from the two isochrones with
  different ages, per H-R Diagram box.  Both examples illustrate that
  the $H$-band IL is dominated by the brightest RGB and AGB
  stars.\label{fig:Age}}
\end{center}
\end{figure}

However, even with the younger age of 5 Gyr, the $H$-band abundances
still converge on roughly the same metallicity ($[\rm{Fe/H}]~=~-0.66$),
suggesting that the $H$-band is less sensitive to cluster age than the
optical (at least for GCs older than $\sim3$ Gyr).  This can be
understood simply by considering the relative contributions from
various sub-populations in the $H$-band.  Figure \ref{fig:HbandIsos}
shows sample isochrones for a GC with $[\rm{Fe/H}]~=~-0.6$ at ages of
13 Gyr and 3 Gyr.  Figure~\ref{fig:EWfrac} then shows the fractional
contribution from each box to the strength of an \ion{Fe}{1} line (at
15207~\AA).  These two figures illustrate that the $H$-band continuum
level {\it and} \ion{Fe}{1} line strength are dominated by the tip of
the RGB and AGB stars.  The turnoff stars have very little effect on the
\ion{Fe}{1} line, suggesting that cluster age will be less important for
determining [Fe/H] in the $H$-band (though note, however, that the RGB
is slightly offset in the 3 Gyr case, and that the younger isochrone's
AGB has a greater effect on the \ion{Fe}{1} line).

\begin{figure}[h!]
\begin{center}
\centering
\includegraphics[scale=0.8]{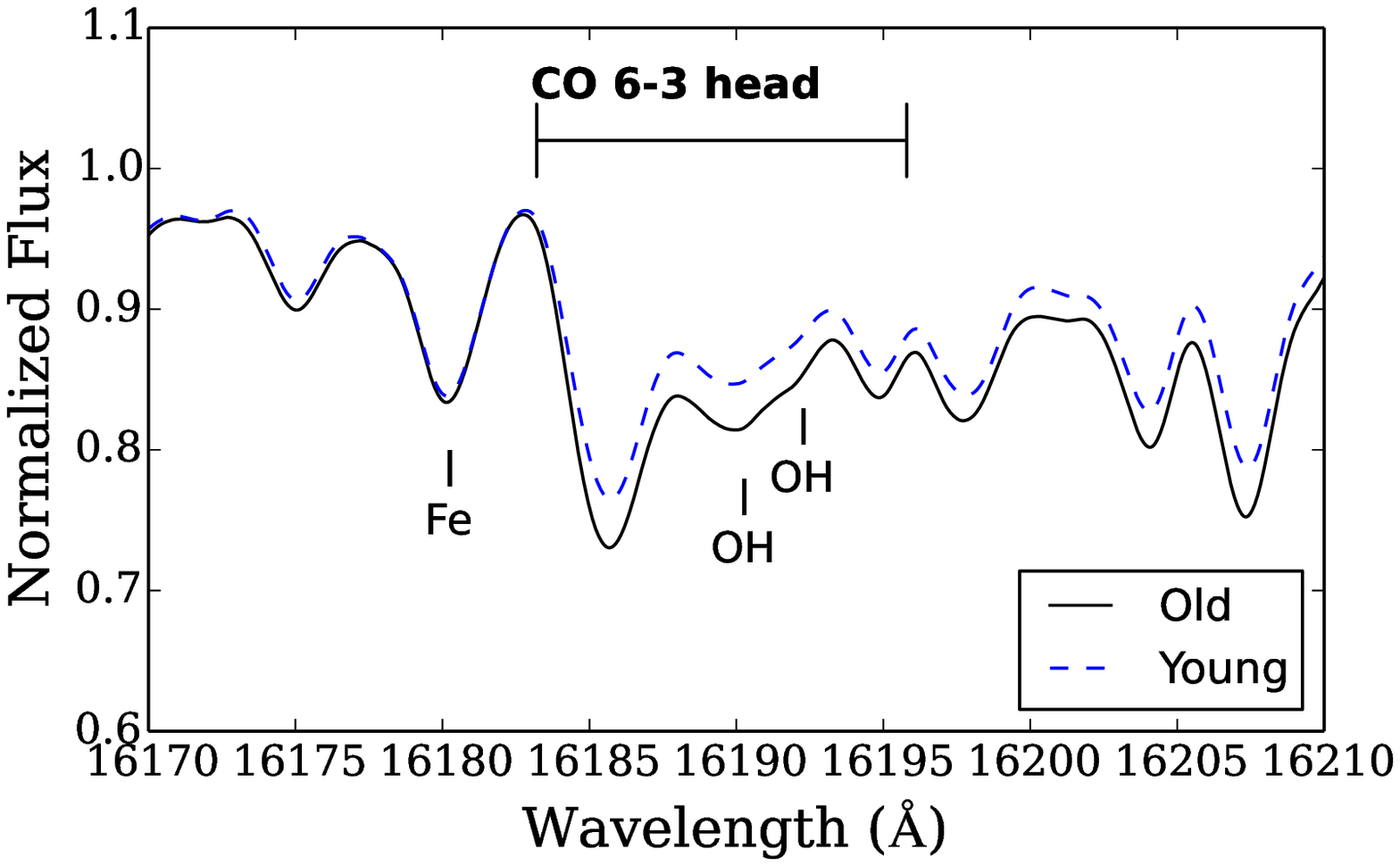}\label{fig:COyoungandold}
\caption{Syntheses of a CO bandhead in a GC with $[\rm{Fe/H}] = -0.5$,
  utilizing two different isochrones, an old 13 Gyr one (the solid
  black line) and a younger 3 Gyr one (the dashed blue line).  The
  younger isochrone requires a larger C abundance to reproduce the
  strength of this feature, but a corresponding increase in [Fe/H]
  means that the relative [C/Fe] ratio changes by only 0.05
  dex.\label{fig:COAge}}
\end{center}
\end{figure}

To test the effects of GC age on the abundances of the other elements,
$H$-band abundances were determined for synthetic metal-rich
($[\rm{Fe/H}]~=~-0.5$) and metal-poor ($[\rm{Fe/H}]~=~-1.7$) clusters,
each with ages of 13 and 3 Gyr.  The total abundance differences for
the two clusters are provided in Table \ref{table:Age}.  The
differences in relative [X/Fe] ratios are all $\lesssim 0.1$ dex.
Figure \ref{fig:COAge} then shows syntheses of a CO bandhead in B171
(starting at 16183 \AA), where the [C/Fe] from the old population is
adopted. Lowering the age has slightly weakened the CO
features---bringing the line strengths into agreement requires
increasing [C/H] by 0.15 dex for the younger isochrone.  A similar
increase in [Fe/H] (by 0.1 dex) means that $\Delta[\rm{C/Fe}] = 0.05$
for the 3 Gyr case.  The effect of having a younger age by 10 Gyr is
therefore not significant for the abundances determined in this
analysis, at least for clusters older than $\sim 3$
Gyr.\footnote{Though it is not shown here, the effects of adopting a
  blue horizontal branch are similarly negligible.}

To summarize, $H$-band IL spectra do not have enough \ion{Fe}{1} lines
to constrain GC age in the same way as the optical.  However, for GCs
older than $\sim 3$ Gyr, the precise age has a much smaller impact on
$H$-band abundances than on optical abundances.  As long as an
appropriate isochrone metallicity is selected, an age that is off by
as much as 10 Gyr will only lead to negligible abundance offsets in
old GCs.

\begin{table}
\centering
\begin{center}
\caption{Abundance offsets when the stellar ages are lowered from 13
  Gyr to 3 Gyr.\label{table:Age}}
 \newcolumntype{d}[1]{D{.}{.}{#1} }
 \begin{tabular}{@{}ld{3}d{3}d{3}d{3}d{3}d{3}d{3}d{3}d{3}d{3}@{}}
  \hline
 & & \multicolumn{9}{l}{$\Delta[\rm{X/Fe}]$} \\
 & \multicolumn{1}{c}{$\Delta[\rm{Fe/H}]$} & \multicolumn{1}{c}{C} & \multicolumn{1}{c}{N} & \multicolumn{1}{c}{O} & \multicolumn{1}{c}{Na} & \multicolumn{1}{c}{Mg} & \multicolumn{1}{c}{Al} & \multicolumn{1}{c}{Si} & \multicolumn{1}{c}{Ca} & \multicolumn{1}{c}{Ti}\\
\hline
MR & 0.10 & 0.05 & 0.0  & 0.10 & 0.0 & 0.0 & 0.05 & -0.05 & 0.05 & 0.05\\
MP & 0.0  & 0.0  & 0.05 & 0.0  & \multicolumn{1}{c}{$-$} & 0.05 & <0.05 & 0.0 & <0.05 & \multicolumn{1}{c}{$-$} \\  
 && & & & \\
\hline
\end{tabular}\\
\end{center}
\medskip
\raggedright  .\\
\end{table}

\vspace{-0.3in}

\subsection{Signatures of Multiple Populations}\label{subsec:MultiPopDiscussion}
All bona fide, classical MW GCs are known to host significant
star-to-star abundance variations.  All GCs have stars which exhibit a
Na/O anticorrelation, while some massive, metal-poor GCs also show
Mg/Al anticorrelations (e.g., \citealt{Carretta2009}).  Some GCs have
significant CN variations (such as 47~Tuc; \citealt{Briley2004}).  A
few of the most massive GCs host spreads in heavy neutron-capture
elements such as Ba and Eu (e.g., M15; \citealt{Sneden1997}), while an
even smaller number host Fe variations (e.g., \citealt{Carretta2010}).
As \citet{Colucci2014} have pointed out, interpreting the integrated
abundances of GCs in terms of abundance variations within the clusters
is non-trivial.  The correlations are observed between individual
stars in a given cluster, while IL abundances represent a
flux-weighted average from a single cluster.  However, deviations from
the ``primordial'' abundances are inferred to exist in distant,
partially resolved M31 GCs based on, e.g., enhanced integrated
[Na/Fe], deficient integrated [Mg/Fe], and/or abundance correlations
with cluster mass \citep{Colucci2014,Sakari2015,Schiavon2013}.

These $H$-band abundances offer the chance to study new elements, like
O, and to probe lines in a wavelength regime that is dominated by
different stars than the optical.  This section explores what can be
learned about multiple populations in distant GCs from a combination
of optical and $H$-band IL spectroscopy, focusing on CN variations
(Section \ref{subsubsec:CNspread}), the Na/O anticorrelation (Section
\ref{subsubsec:NaOspread}), the Mg/Al anticorrelation (Section
\ref{subsubsec:MgAlspread}), and Fe spreads (Section
\ref{subsubsec:Fespread}).

\subsubsection{CN}\label{subsubsec:CNspread}
This section examines the possibility that the disagreements between
the $H$-band and optical, Lick index CN abundances could be due to the
presence of strong star-to-star CN variations within the GC.  In
particular, the different wavelength regimes may be sensitive to
different stellar sub-populations because the $H$-band is only
sensitive to the tip of the RGB stars while the optical is more
sensitive to hot stars.

\paragraph{Variations up the RGB}
In addition to the variations that occur between the multiple GC
populations, surface C and N abundances (and $^{12}$C/$^{13}$C ratios)
also vary up as a star evolves up the RGB (see, e.g.,
\citealt{Gratton2000}).  The IR is predominantly sensitive to the
evolved tip of the RGB stars, while the bluer, optical Lick indices
may be more sensitive to the less-evolved RGB stars which have
``normal'' (i.e., pre-dredge up) C and N abundances.

To test this, synthetic clusters were created with [C/Fe], [N/Fe], and
$^{12}\rm{C}/^{13}\rm{C}$ variations along the upper RGB, one
metal-rich ($[\rm{Fe/H}]~=~-0.5$), the other metal-poor
($[\rm{Fe/H}]~=~-1.7$).  The variations from \citet{Gratton2000} are
adopted: initial values of $[\rm{C/Fe}] = 0$, $[\rm{N/Fe}] = 0$,
and $^{12}\rm{C}/^{13}\rm{C} =  40$ are adopted for all lower RGB stars;
above the RGB bump the $^{12}\rm{C}/^{13}\rm{C}$ is a step function
down to 5, while [C/Fe] and [N/Fe] linearly change along the RGB to
final values of -0.5 and 1.0, respectively.  All of the evolved HB and
AGB stars are assumed to have the same values as the tip of the RGB
stars.  Note that these values are for more {\it metal-poor} field
stars---metal-rich stars are expected to show smaller variations.
However, this exercise provides a good test of the most extreme
changes that could occur. These composite synthetic spectra are then
treated as the observed spectra, and best-fitting integrated
abundances are determined from the composite spectrum.  The derived
[C/Fe], [N/Fe], and $^{12}\rm{C}/^{13}\rm{C}$ ratios are shown in
Table \ref{table:RGBvariations}, and match the abundances that were
input for the tip of the RGB, HB, and AGB stars.  This demonstrates
that the $H$-band integrated [C/Fe], [N/Fe], and
$^{12}\rm{C}/^{13}\rm{C}$ ratios are largely insensitive to the
variations up the RGB and reflect the abundances of the most evolved
RGB stars.

\begin{table}
\centering
\begin{center}
\caption{Integrated abundances when [C/Fe], [N/Fe] and
  $^{12}$C/$^{13}$C variations up the RGB are adopted.\label{table:RGBvariations}}
 \newcolumntype{d}[1]{D{.}{.}{#1} }
 \begin{tabular}{@{}lcd{6}d{7}@{}}
  \hline
 && \multicolumn{1}{c}{$[\rm{Fe/H}] = -0.5$} & \multicolumn{1}{c}{$[\rm{Fe/H}] = -1.7$} \\
\hline
$\Delta[\rm{C/Fe}]$  && -0.50  & -0.50^{a}  \\
$\Delta[\rm{N/Fe}]$  &&  1.0   & 0.95  \\
$^{12}\rm{C}/^{13}\rm{C}$ && \multicolumn{1}{c}{$5-6$} & \multicolumn{1}{c}{$-$} \\
 && & \\
\hline
\end{tabular}\\
\end{center}
\medskip
\raggedright  $^{a}$ Note that some of the CO features are
undetectable in the spectra of the most metal-poor clusters.\\
\end{table}

\begin{figure}[h!]
\begin{center}
\centering
\includegraphics[scale=0.55,trim=0.5in 0in 1.25in 0.0in]{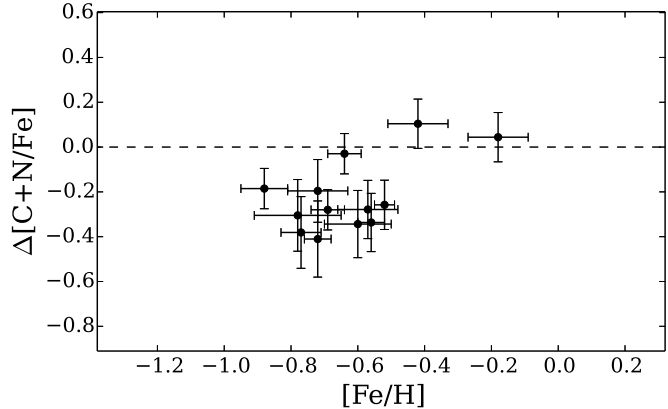}
\caption{$\Delta$[(C+N)/Fe] (optical $-$ $H$-band) as a function of
  $H$-band [Fe/H].  The dashed line shows perfect agreement.}\label{fig:CN} 
\end{center}
\end{figure}

If the optical IL features are more sensitive to the less-evolved
stars, they should have higher [C/Fe] and lower [N/Fe] ratios, {\it
  but the total amount of $[(\rm{C}+\rm{N})/\rm{Fe}]$ should remain
  the same}, since the abundance changes occur as C is converted into
N.  Note that this comparison assumes that the optical C and N are
sensitive to the same stellar populations, which may not be true given
the wavelength separation between the C$_{2}$ and CN indices.  Figure
\ref{fig:CN} compares the $[(\rm{C}+\rm{N})/\rm{Fe}]$ ratios, which
are not equal between the optical and IR.  This suggests that the
offsets cannot {\it only} be due to the optical Lick ratios probing
stars further down the main sequence.  Other systematic possibilities
are that the Lick C and N abundances are not tracing the same
populations or that there are other unknown systematic offsets between
the optical and $H$-band.

\paragraph{Varying sensitivities to multiple populations between the
  $H$-band and optical}
Another possibility is that the optical and $H$-band have different
sensitivities to the multiple populations within the GCs.  C, N, and O
variations {\it are} observed between stars in Galactic GCs. For
instance, 47~Tuc exhibits a strong bimodality in CN-weak and CN-strong
stars that persists through the main sequence
(e.g., \citealt{Briley2004}), and all MW GCs show variations in
O (typically seen as a Na/O anticorrelation; e.g.,
\citealt{Carretta2009}).   As with the variations up the RGB, the
effects of multiple CN populations can be examined by creating a
synthetic population with multiple populations and deriving a
single IL abundance.  The clusters are assumed to have C, N, O, and
Na variations---none of these elements are expected to significantly
change the temperature of the RGB \citep{Vandenberg2012}, so the same
isochrones are used for each population.  Note that variations up the
RGB were not included within each sub-population.

\begin{figure}[h!]
\begin{center}
\centering
\subfigure[]{\includegraphics[scale=0.55,trim=1.25in 0in 0.05in 0.0in]{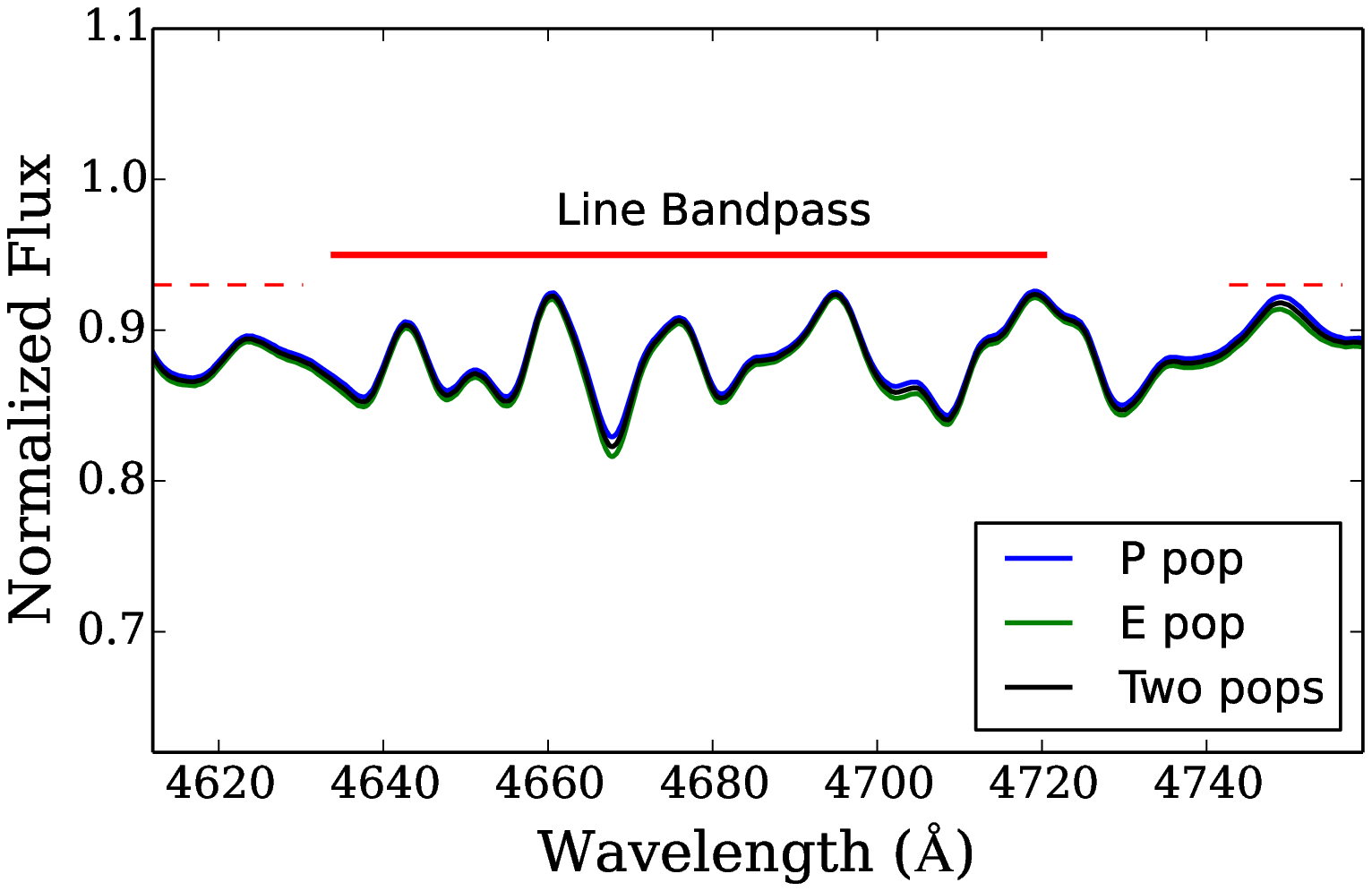}\label{fig:C4668TwoPops}}
\subfigure{\includegraphics[scale=0.55,trim=0.5in 0in 1.25in 0.0in]{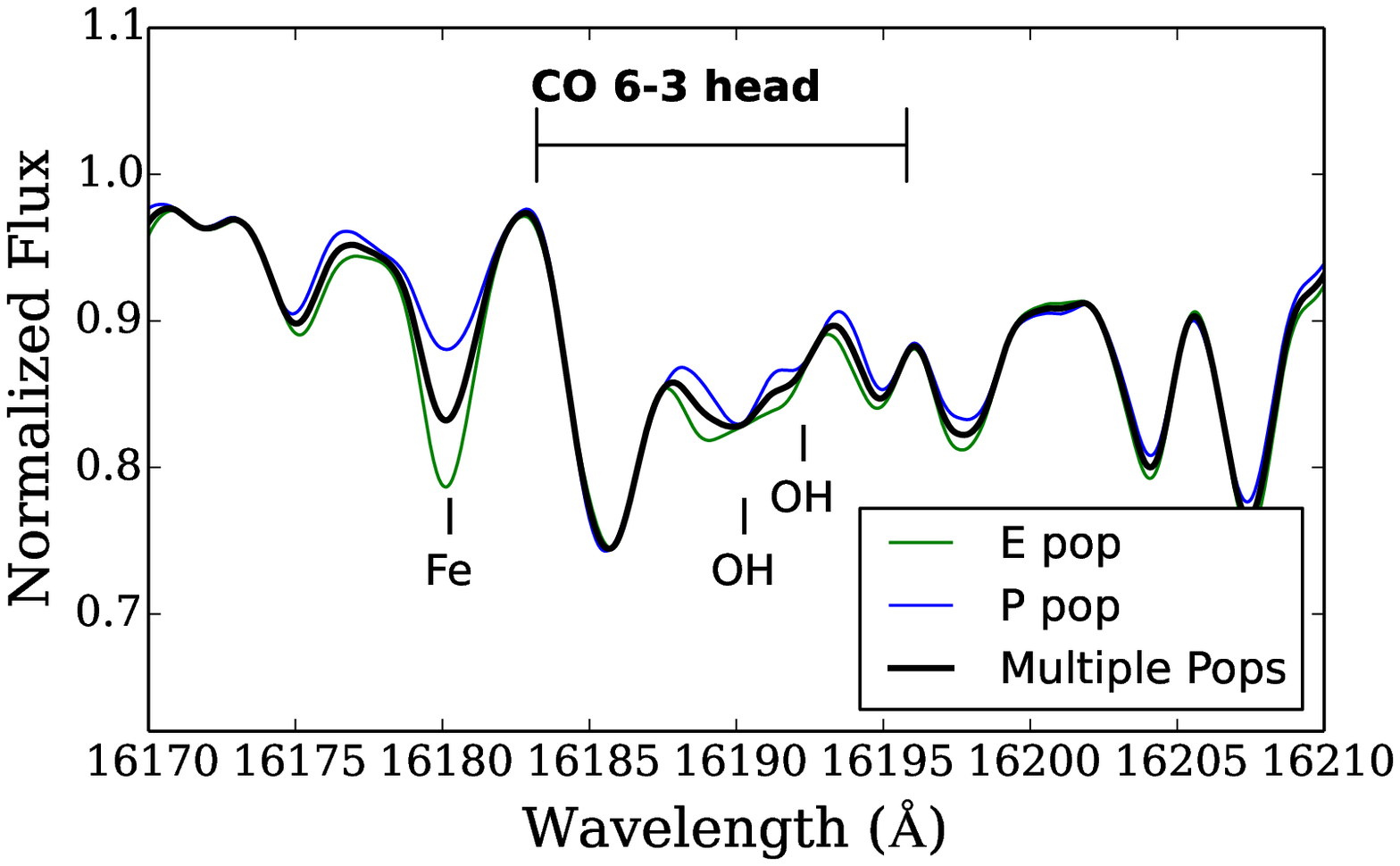}\label{fig:CIR}}
\caption{Syntheses of C-sensitive features in a metal-rich
  cluster ($[\rm{Fe/H}] = -0.5$) with multiple populations.   {\it
    Left: } The 4668 \AA \hspace{0.025in} feature at a Lick index
  resolution of 5 \AA.  {\it Right: } The 16183 \AA \hspace{0.025in}
  bandhead with B171's line broadening.  The blue lines show syntheses
  with abundances typical of a GC's ``primordial'' population, while
  the green lines show syntheses with the abundances typical of the
  ``extreme'' population (enhanced N, lower C).  The black line shows
  syntheses with a population composed of 60\% extreme, 40\%
  primordial.\label{fig:CMultiPops}}
\end{center}
\end{figure}

Figure \ref{fig:CIR} shows the syntheses of GCs with multiple
populations.  In both the optical and the $H$-band, the IL values fall
in between the range from the ``primordial'' and ``extreme''
populations.  The best-fitting values from the optical and $H$-band
are identical, suggesting that the optical vs. $H$-band offset is
likely not due to varying sensitivity to GC multiple populations.

\begin{figure}[h!]
\begin{center}
\centering
\includegraphics[scale=0.55]{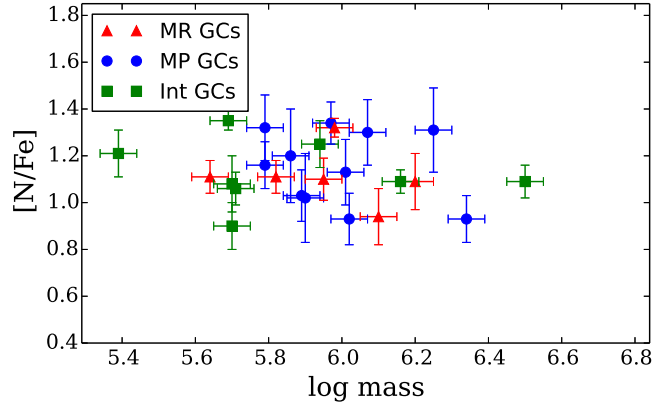}
\caption{[N/Fe] versus cluster mass.  The points are grouped according
to GC metallicity.  The metal-poor (MP) GCs have $[\rm{Fe/H}] < -1.2$, the
metal-rich (MR) ones have $[\rm{Fe/H}] > -0.8$, and the
intermediate-metallicity (Int) ones fall in between.  There is no
convincing trend with mass in any of the subgroups, though the mass
range is not very large.\vspace{-0.25in}}\label{fig:MassDiffN}
\end{center}
\end{figure}

\paragraph{Trends with cluster mass}
In their Lick index analysis of MW GCs, \citet{Schiavon2013} detected
a metallicity-dependent trend of increasing [N/Fe] with increasing
cluster mass.  They interpreted this as a signature of multiple
populations, where more massive GCs are able to form a larger fraction
of ``second generation'' stars.  Trends with mass are also be
investigated with the $H$-band abundances. Figure \ref{fig:MassDiffN} shows
[N/Fe] versus cluster mass, utilizing the cluster masses from
\citet{Schiavon2013}.  The clusters are grouped by metallicity, as in
Schiavon et al.  No convincing trend is seen in any of the metallicity
groups.  This is unsurprising given that the $H$-band is dominated by
tip of the RGB stars: all the tip of the RGB stars have enhanced
[N/Fe] because of normal stellar evolutionary processes, and the
$H$-band [N/Fe] ratios therefore should not depend on cluster
mass. Note, however, that the mass range of these GCs is not as large
as the Schiavon et al. sample, and these results may change if more
GCs are included.

\subsubsection{Na/O}\label{subsubsec:NaOspread}
\begin{figure}[h!]
\begin{center}
\centering
\subfigure[]{\includegraphics[scale=0.55,trim=1.25in 0.25in 0.05in 0.50in]{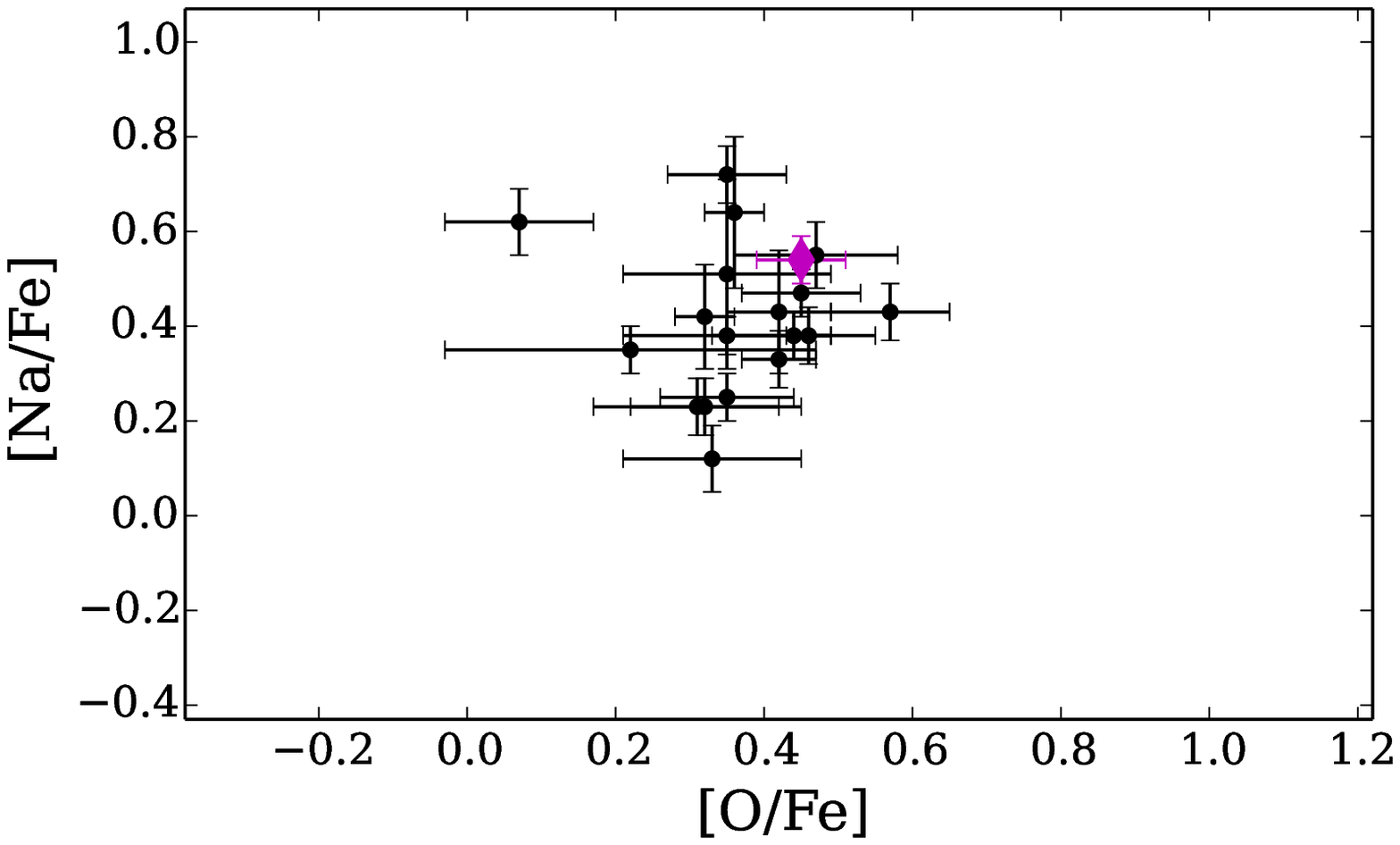}\label{fig:CompNaO}}
\subfigure[]{\includegraphics[scale=0.55,trim=0.5in 0.25in 1.25in 0.50in]{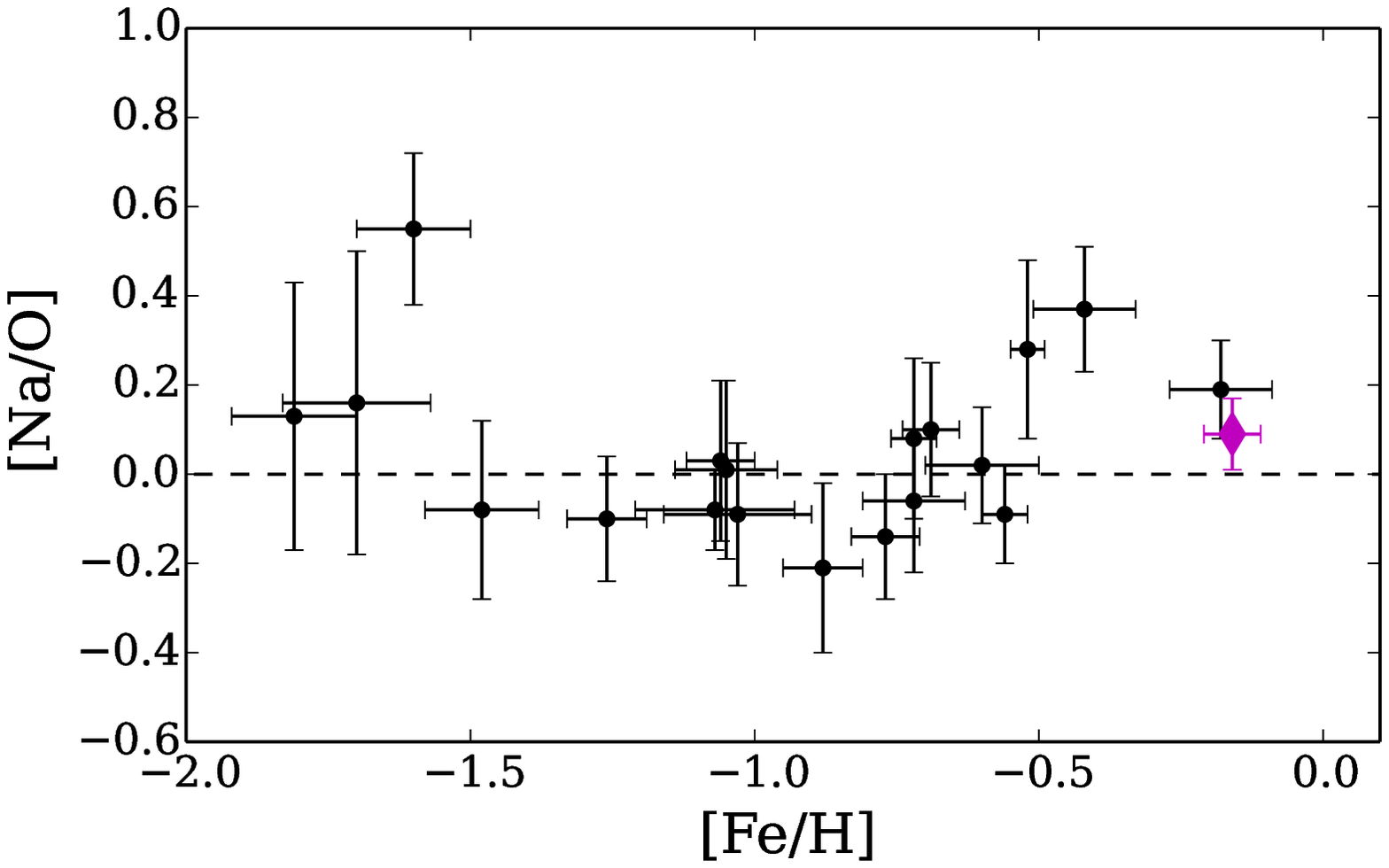}\label{fig:NaOFe}}
\subfigure[]{\includegraphics[scale=0.55,trim=1.25in 0.25in 0.05in 0.15in]{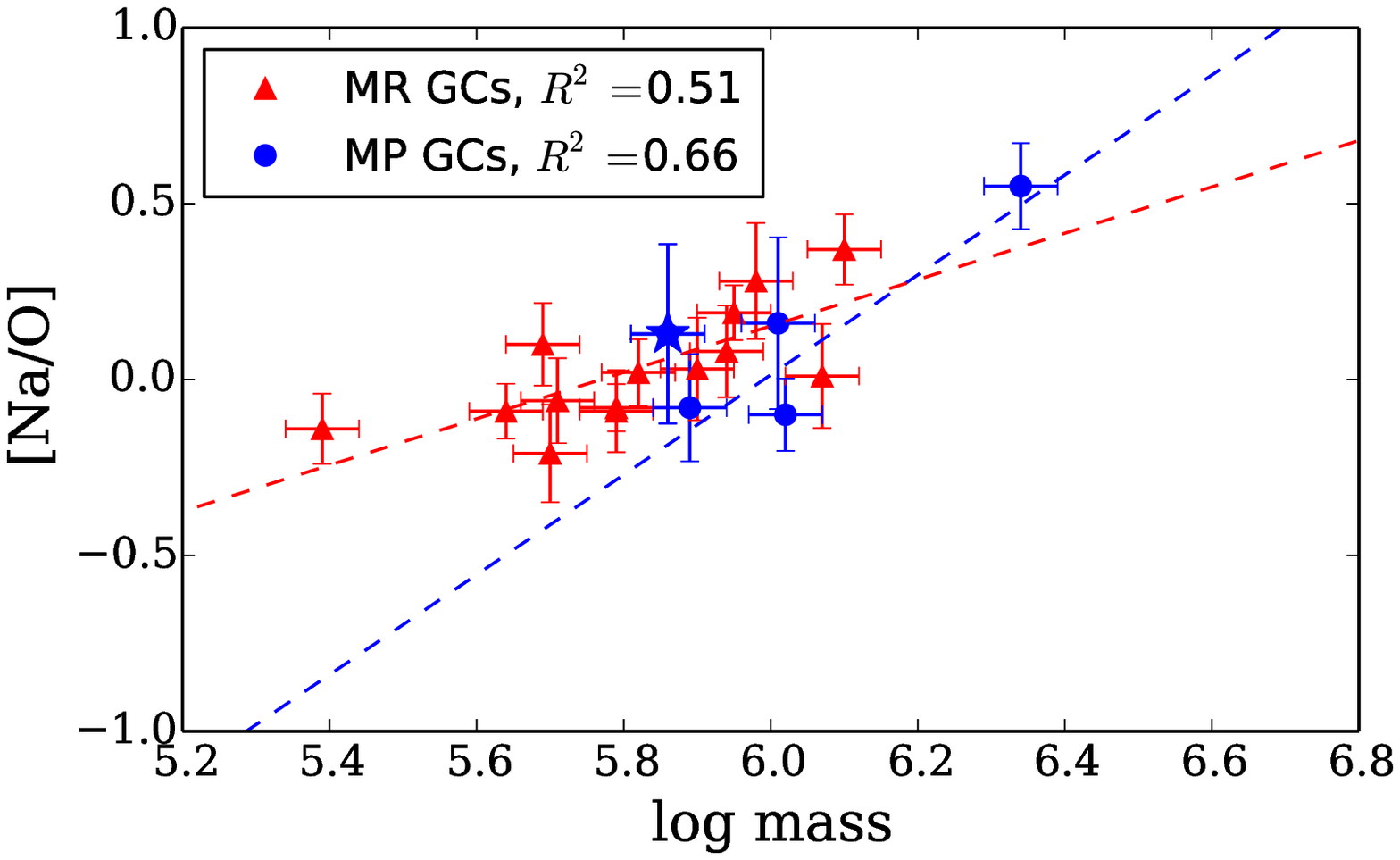}\label{fig:NaOMass}}
\subfigure[]{\includegraphics[scale=0.55,trim=0.5in 0.25in 1.25in 0.15in]{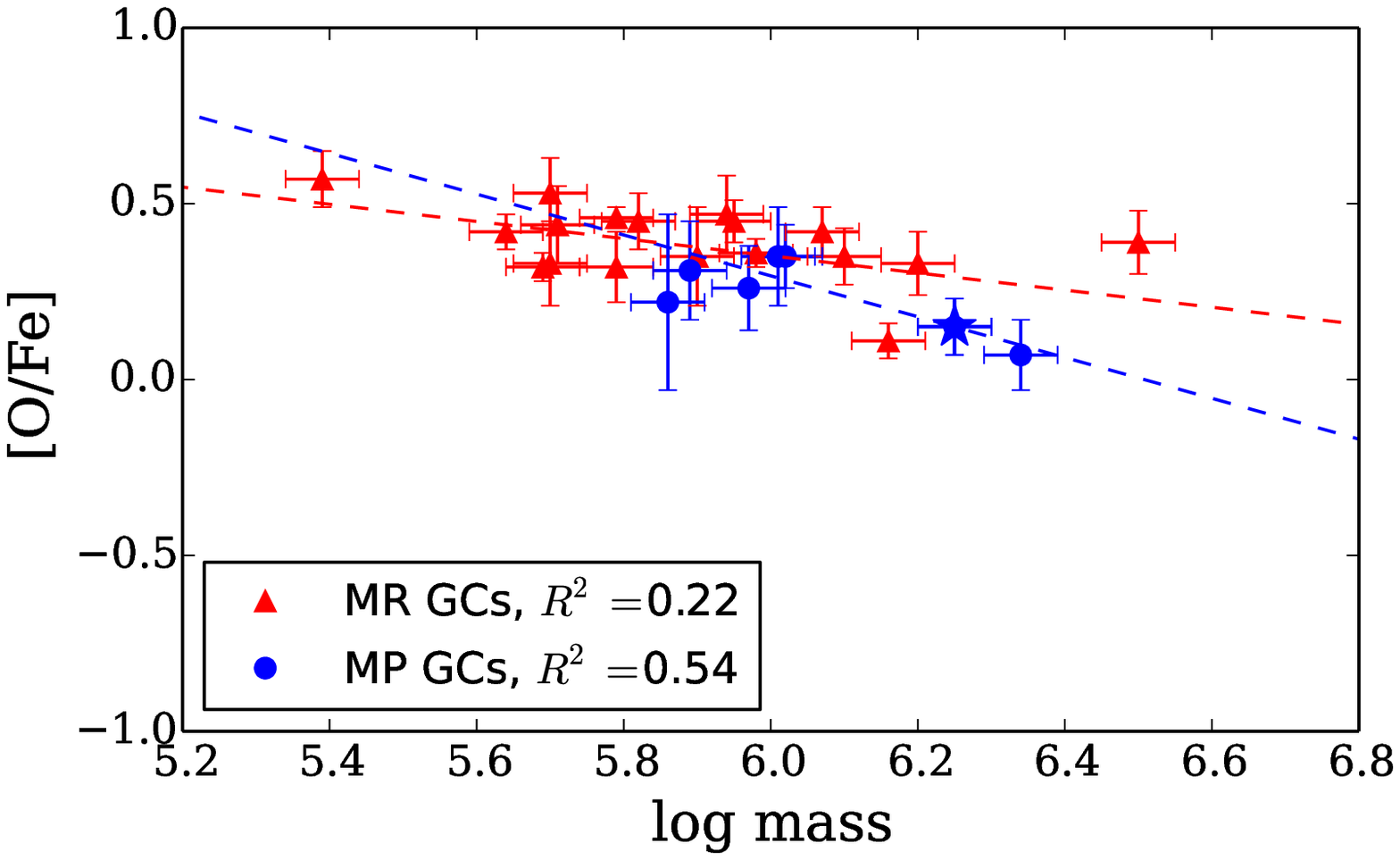}\label{fig:OMass}}
\caption{{\it Top Left: } Optical [Na/Fe] vs. $H$-band [O/Fe].  The
  magenta diamond shows B193, whose $H$-band [Na/Fe] is utilized in
  lieu of an optical abundance.  
  {\it Top Right:} [Na/O] (with optical Na and $H$-band O) vs. cluster
  [Fe/H].  The dashed line shows equal [Na/Fe] and [O/Fe].
  {\it Bottom Left: } [Na/O] vs. cluster mass (from
  \citealt{Schiavon2013}.  The clusters are grouped by metallicity:
  metal-poor (MP) GCs have $[\rm{Fe/H}] < -1.2$.  The dashed lines
  show fits to each metallicity group, with coefficients of
  determination shown. 
{\it Bottom Right: } [O/Fe] vs. cluster mass.\vspace{-0.25in}\label{fig:NaO}}
\end{center}
\end{figure}

The Na/O anticorrelation is well-established within MW GCs
\citep{Carretta2009}.  Extragalactic GCs have been inferred to host
Na/O anticorrelations since many have high integrated [Na/Fe]
\citep{Colucci2014,Sakari2015}.  Figure \ref{fig:CompNaO} shows
optical [Na/Fe] vs. $H$-band [O/Fe] (except for B193, whose $H$-band
[Na/Fe] is used in lieu of an optical value).  There is no clear
anticorrelation; however, in IL an anticorrelation would not be
expected.  All the clusters lie within the range of individual MW GC
stars (see \citealt{Carretta2009}) in the region represented by the
``intermediate'' populations (see the descriptions in
\citealt{Carretta2009HB}).  Only one GC, B012, falls in the region of
the ``extreme'' population.

\vspace{-0.25in}
\paragraph{Trends with metallicity and mass}
Figures \ref{fig:NaOFe} and \ref{fig:NaOMass} show the $H$-band IL
[Na/O] ratios versus [Fe/H] and cluster mass (from
\citealt{Schiavon2013}).  Note that not all of the GCs have [Na/Fe]
abundances, particularly two of the most massive GCs, B088 and B225.
There have been some hints that the extent of the Na/O anticorrelation
may change with metallicity (e.g., Figure 5 in
\citealt{Carretta2009}); however \citet{Carretta2010a} argue that
[Fe/H] most strongly correlates with the fraction of
``extreme'' stars (i.e., those that are the most enhanced in Na and
deficient in O). The extent of the Na/O anticorrelation is also
expected to vary with cluster mass, since the relative numbers of
stars in various populations may depend on mass (e.g.,
\citealt{Carretta2009HB,Carretta2010a,Schiavon2013}).  Figure
\ref{fig:NaOFe} shows no obvious trend with metallicity, though Figure
\ref{fig:NaOMass} hints at a possible trend with GC mass.  Again,
the clusters are grouped by metallicity; the metal-rich GCs appear to
have the strongest correlation of [Na/O] with mass.  This abundance
trend follows the trend found by \citet{Colucci2014}, which showed an
increasing integrated [Na/Fe] with absolute magnitude and velocity
dispersion (as expected since their Na abundances were utilized for
this comparison).  Figure \ref{fig:OMass} then examines the trend in
integrated [O/Fe]. The trend is weak for the metal-rich clusters, but
may be stronger for the handful of metal-poor clusters.

\vspace{-0.25in}
\subsubsection{Mg/Al}\label{subsubsec:MgAlspread}
\begin{figure}[h!]
\begin{center}
\centering
\subfigure[]{\includegraphics[scale=0.55,trim=1.25in 0.25in 0.05in 0.3in]{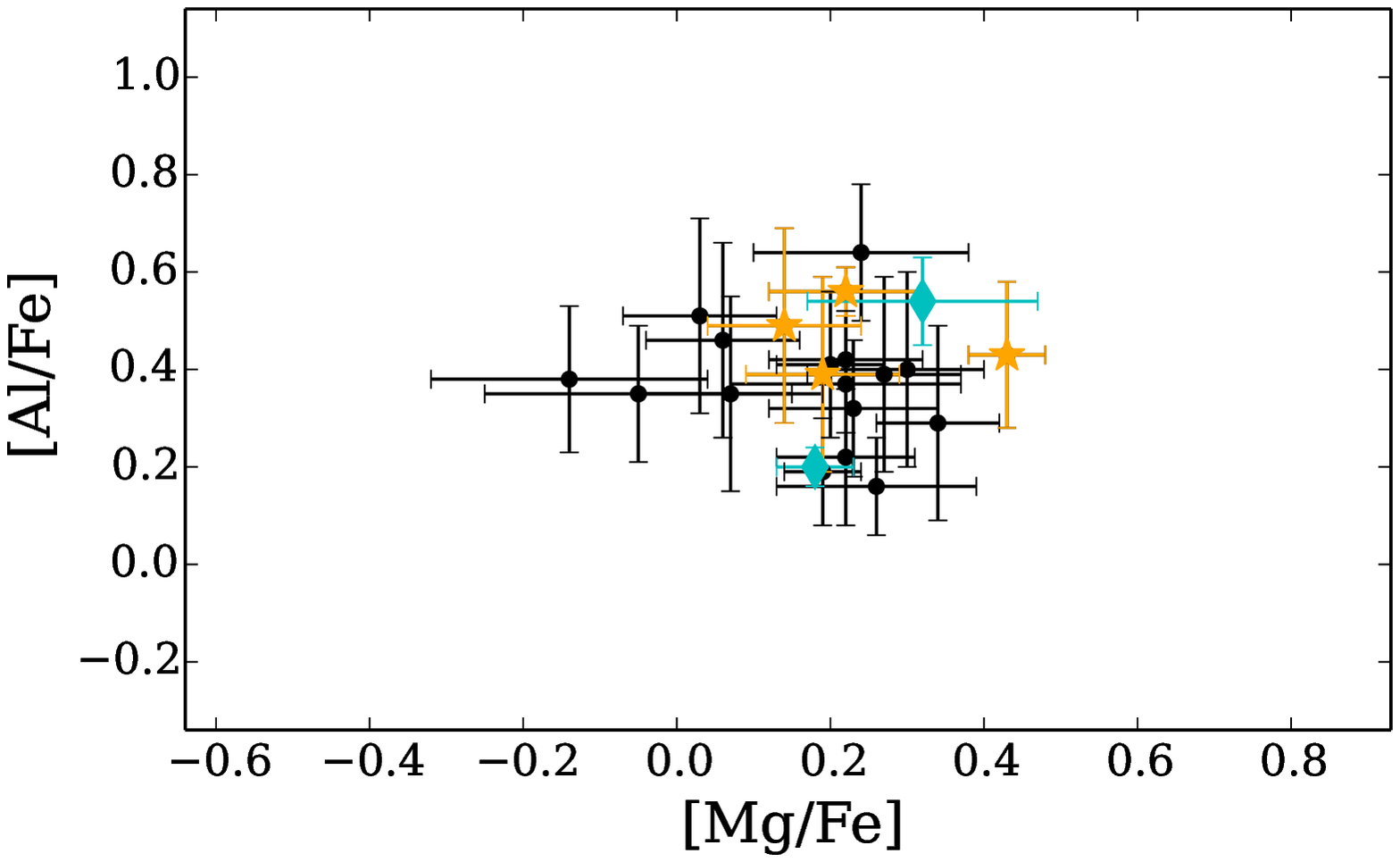}\label{fig:CompMgAl}}
\subfigure[]{\includegraphics[scale=0.55,trim=0.5in 0.25in 1.25in 0.3in]{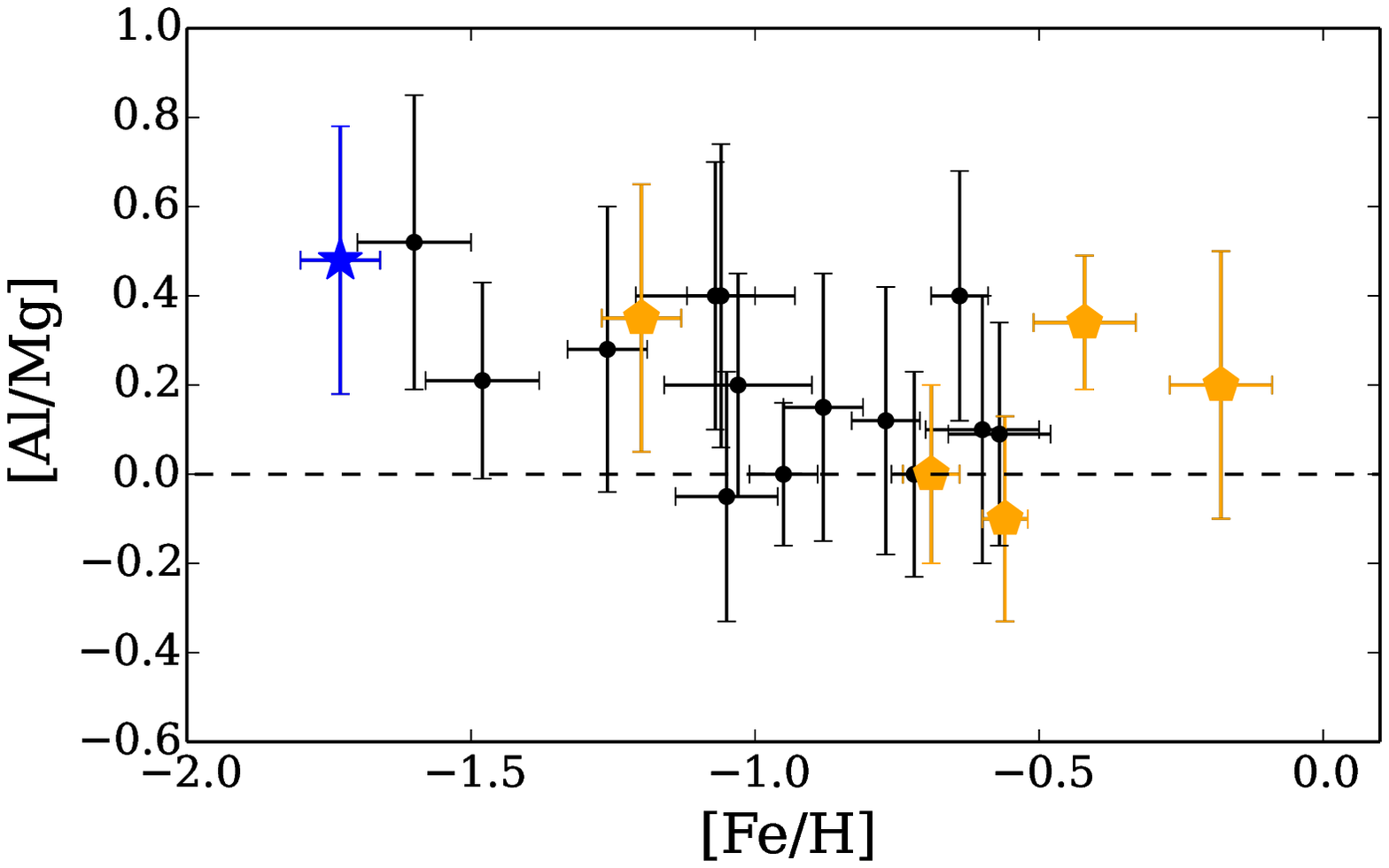}\label{fig:MgAlFe}}
\subfigure[]{\includegraphics[scale=0.55,trim=0.0in 0.25in 0in 0.15in]{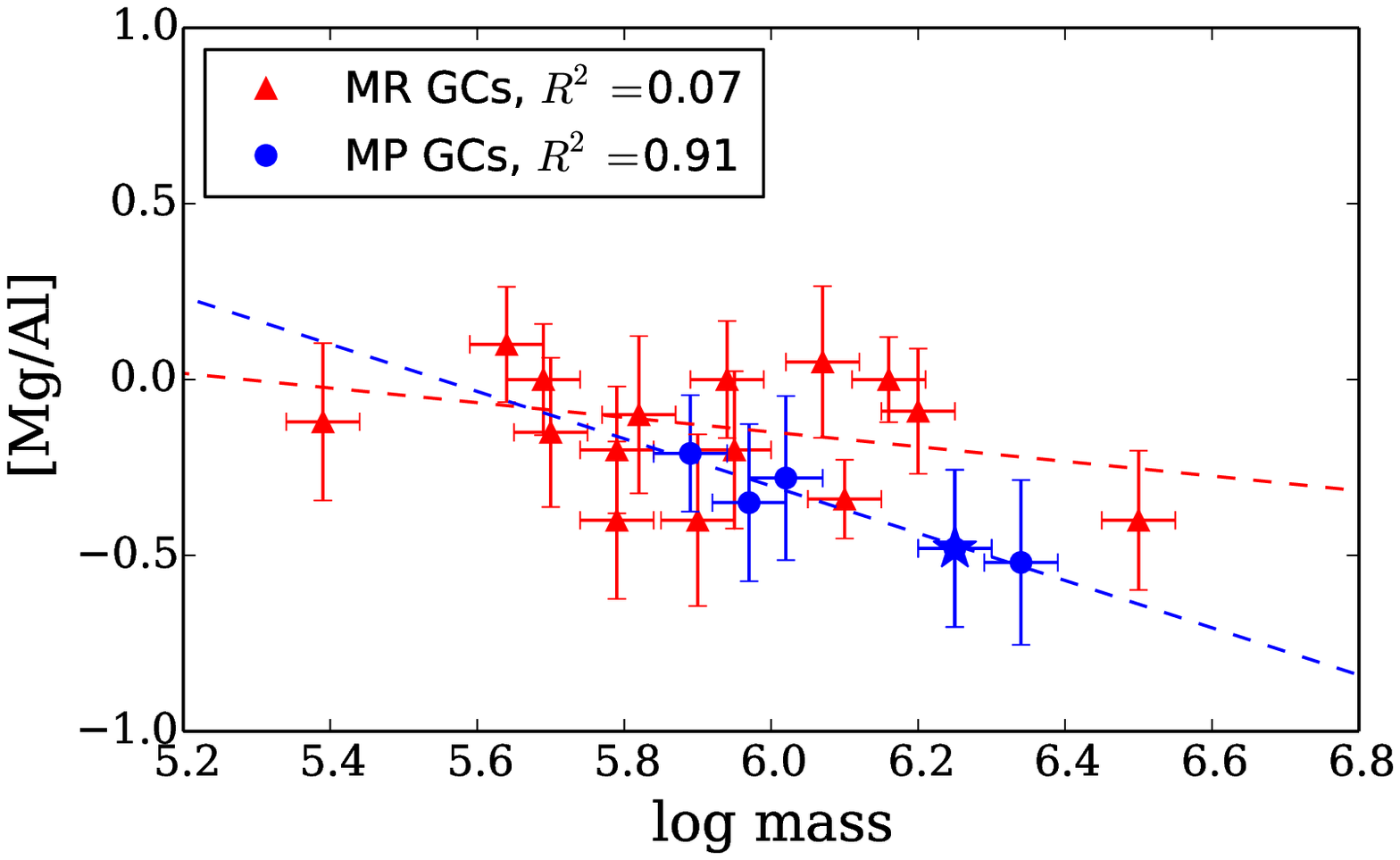}\label{fig:MgAlMass}}
\caption{{\it Top Left: } $H$-band [Al/Fe] vs. [Mg/Fe].  The cyan
  diamonds show B171 and B403, whose optical [Al/Fe] abundances are
  utilized in lieu of $H$-band values.  Again, the orange pentagons
  show clusters whose [Al/Fe] ratios are determined from strong lines.
  {\it Top Right:} [Al/Mg] vs. cluster $H$-band [Fe/H].  The dashed
  line shows perfect agreement.
  {\it Bottom: } [Al/Mg] vs. cluster mass.  Metallicity bins and fits
  are as in Figure \ref{fig:NaOMass}.  In panels $b$ and $c$, B088 is
  shown with a blue star.\label{fig:MgAl}}
\end{center}
\end{figure}

Mg/Al anticorrelations have been observed in the optical within the
most massive, metal-poor ($[\rm{Fe/H}] \lesssim -1.2$) Galactic GCs
(e.g., \citealt{Carretta2009}) and in the IR for several additional
clusters \citep{Meszaros2015}.  Figure \ref{fig:CompMgAl} shows
$H$-band [Al/Fe] vs. [Mg/Fe]; the orange pentagons show the GCs whose
[Al/Fe] was determined from strong lines ($-4.7<$REW$<-4.5$) while the
cyan diamonds show clusters whose optical [Al/Fe] ratios are used
(since they do not have $H$-band Al abundances).  There is no clear
anticorrelation in the IL ratios.  

\vspace{-0.5in}
\paragraph{Trends with metallicity and mass}
Figure \ref{fig:MgAlFe} then compares [Mg/Al] versus $H$-band [Fe/H].
No trends are seen in any of the sub-groups, though if
optical [Mg/Fe] ratios are used, B088 creates a slight trend at the
metal-poor end.  The [Mg/Al] ratio is then plotted against cluster
mass in Figure \ref{fig:MgAlMass}, again grouped into metallicity bins
(since the presence of the Mg/Al anticorrelation may be
metallicity-dependent; \citealt{Carretta2009}).  Note that
\citet{Colucci2014} found a positive trend in [Mg/Fe] and [Al/Fe] with
mass in metal-rich GCs, but a negative trend in massive metal-poor
GCs.  With $H$-band Al and Mg, no significant trends in [Mg/Al]
are seen in the metal-rich population; although there is a correlation
in the metal-poor population, this is due to only five GCs, one of
which is B088, whose optical and $H$-band Mg abundances are very
discrepant.  More metal-poor clusters are needed to assess whether or
not such a trend with mass exists.

\vspace{-0.35in}
\paragraph{The case of B088}
Recall that B088's [Mg/Fe] shows significant disagreement between the
optical and the $H$-band.  \citet{Colucci2014} find a very low optical
$[\rm{Mg/Fe}] = -0.48\pm 0.02$, while the $H$-band value from this
paper is higher by $4\sigma$.  B088 is a massive ($\log M = 6.25$;
\citealt{Schiavon2013}), metal poor ($[\rm{Fe/H}] = -1.71$) GC, and
may therefore host a significant Mg/Al anticorrelation---for instance,
the more metal-poor, similarly massive Milky Way GC M15 has known
Mg/Al variations and a low, optical IL [Mg/Fe] \citep{Sakari2013}.
One way to create an offset in abundance between the optical and the
$H$-band might be to introduce a temperature spread on the RGB and
examine spectral lines with varying sensitivities to the two
populations.  One way to create a temperature split on the RGB is
varying [Mg/Fe] \citep{Vandenberg2012}.

\begin{figure}[h!]
\begin{center}
\centering
\subfigure{\includegraphics[scale=0.55,trim=1.25in 0in 0.05in 0.0in]{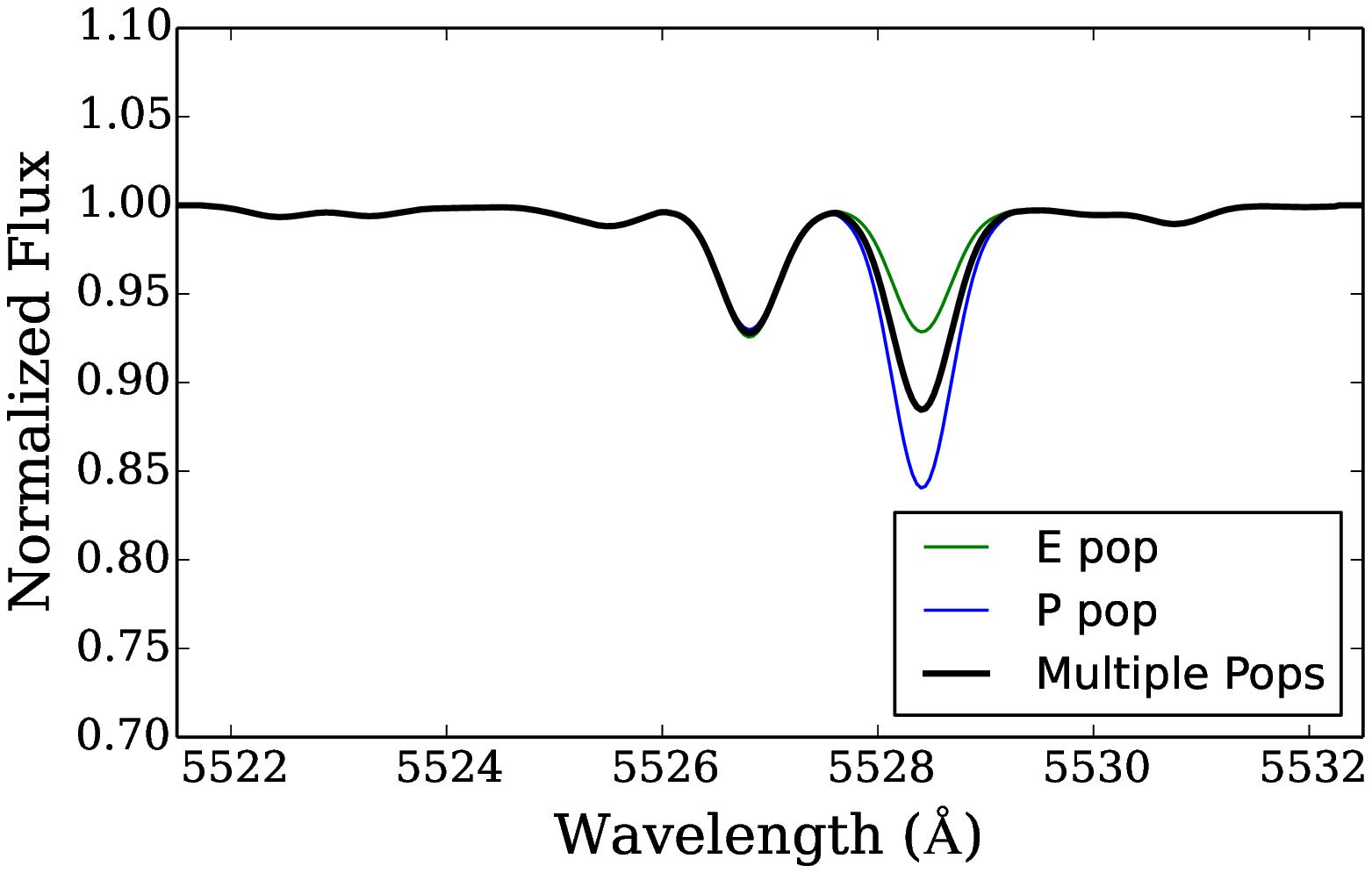}\label{fig:MgOp}}
\subfigure{\includegraphics[scale=0.55,trim=0.5in 0in 1.25in 0.0in]{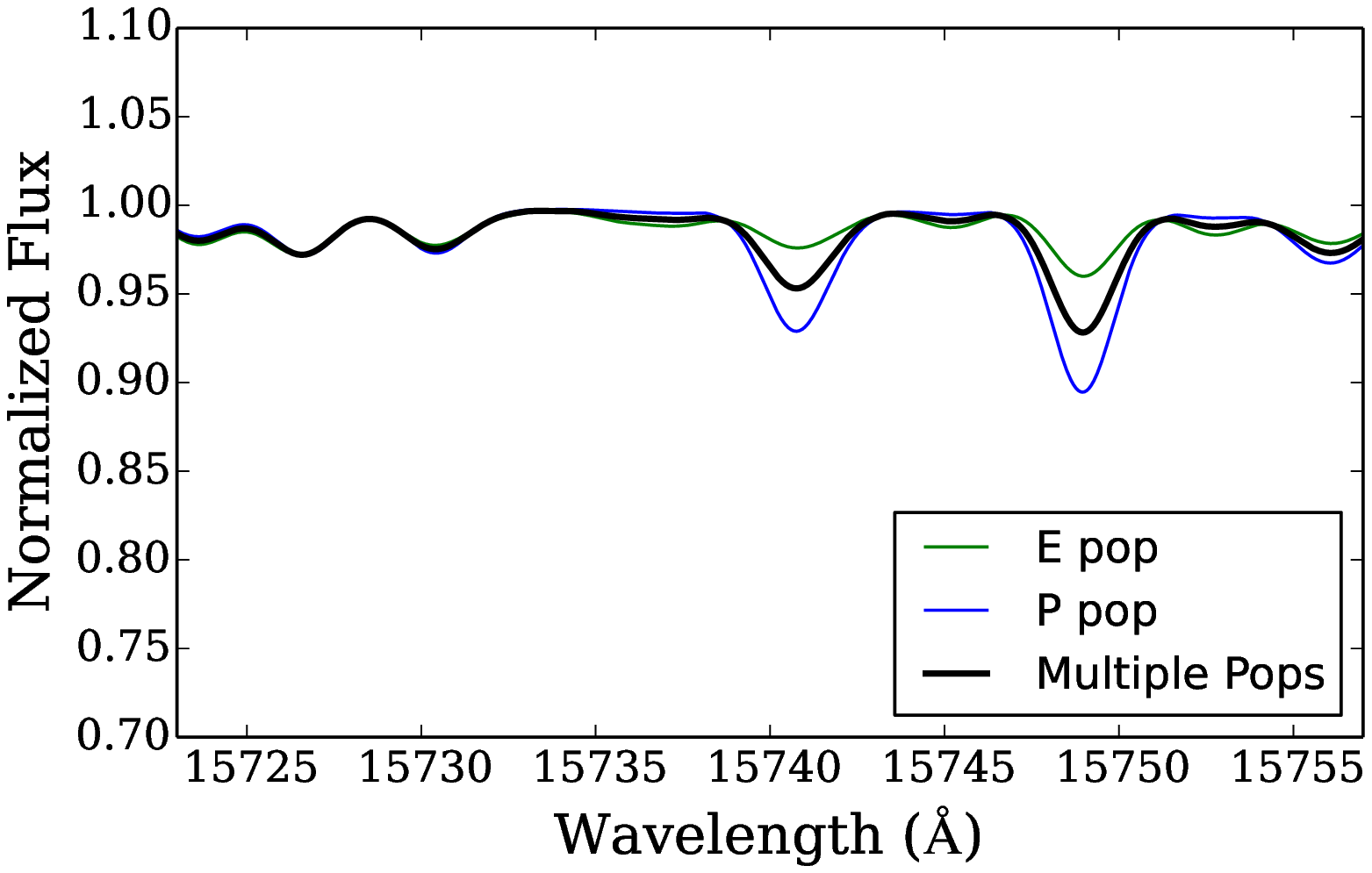}\label{fig:MgIR}}
\caption{Syntheses of \ion{Mg}{1} lines in a B088-like cluster, with
  $[\rm{Fe/H}] = -1.84$.  {\it Left: } The optical 5528
  \AA \hspace{0.025in} line.  {\it Right:} The infrared 15740 and
  15748 \AA \hspace{0.025in} spectral lines.  The blue lines show syntheses
  with abundances typical of a GC's ``primordial'' population, while
  the green lines show syntheses with the abundances typical of the
  ``extreme'' population.  The black line shows syntheses with a
  population composed of 50\% of each.\label{fig:MgMultiPops}}
\end{center}
\end{figure}

To test if Mg variations can explain the disagreement between the
optical and the IR, two synthetic populations were created.  The
``primordial'' (P) population has $[\rm{O/Fe}]=[\rm{Mg/Fe}] = +0.4$,
while the ``extreme'' (E) population has $[\rm{O/Fe}]=-0.2$ and
$[\rm{Mg/Fe}] = -0.6$.  The cluster is assumed to have 50\% of each
population---this assumption is unrealistic given the small numbers of
``extreme'' stars found in most GCs, but again, this should test the
maximal offsets that could occur.  Isochrones with $\alpha$-element
enhancement are utilized for the P population, while solar-scaled ones
are selected for the E population. Figure \ref{fig:MgMultiPops} shows
that the synthesis with hypothetical multiple populations falls in
between those for the E and P populations alone.  Most importantly,
when the synthetic, multiple-population spectra are analyzed as single
spectra, the $H$-band and the optical converge on the same value for
the two lines, $[\rm{Mg/Fe}]~=~-0.01$.  This indicates that the
disagreement between the optical and the $H$-band is likely {\it not}
due to different sensitivities to the multiple populations, {\it if
  the only variations are in O, Na, Mg, and Al.}

\vspace{-0.4in}
\subsubsection{Possible Fe Spreads?}\label{subsubsec:Fespread}
Another way to introduce temperature shifts on the RGB is by invoking
a metallicity spread within the cluster.  Indeed, B088's {\it Hubble
  Space Telescope} color-magnitude diagram \citep{Perina2009} hints at
the possibility of a metallicity spread on the RGB, while its high
ellipticity ($\epsilon = 0.28$; \citealt{Barmby2007}) suggests that it
is not a typical GC.  Again, synthetic populations were created to
test this effect, one with $[\rm{Fe/H}] = -1.31$ and the other with
$[\rm{Fe/H}] = -2.14$; the two sub-populations were assumed to have
equal numbers of stars.  Treating the synthetic composite spectrum as
an observed spectrum, the $H$-band lines that were detectable in B088
(all from high EP transitions) converge on an integrated
$[\rm{Fe/H}]~=~-1.55$.  Utilizing the optical EWs from
\citet{Colucci2014}, the optical values converge on an average
$[\rm{Fe/H}] = -1.6$.  Even with an [Fe/H] spread introduced, the
average IL [Fe/H] ratios are approximately the same between the
optical and the $H$-band, and fall between the two ``real'' values, in
agreement with the values derived for B088.  However, there is an
important caveat here: the introduction of a metallicity spread has
changed the slopes with REW and (to a lesser extent) EP in the
optical, and this may affect the derived GC age.

Without altering the GC age, however, introducing an Fe spread could
lead to an offset in \ion{Mg}{1} between the optical and the IR.
Comparisons between the 5528 \AA \hspace{0.025in} and IR 15740, 15748,
and 15764 \AA \hspace{0.025in} lines suggest that an offset of
$\Delta[\rm{Mg/Fe}]~\sim~0.2$ dex can be created just from invoking a
metallicity spread.  Thus, offsets between the optical and IR {\it may}
probe the existence of Fe spreads within GCs, {\it provided that
  temperature sensitive lines are utilized.}  More detailed test of
this phenomenon are needed to verify if this is an accurate test of Fe
spreads within GCs.

\vspace{-0.25in}
\subsection{The AGB/RGB ratio}\label{subsec:AGBratio}
Though the $H$-band light is dominated by tip of the RGB stars, AGB
stars are also a major contributor to the IL.  Among other things,
the models of the underlying stellar populations that are utilized in
this paper rely on a) the assumption of an IMF to populate the
isochrones and b) prescriptions for how the AGB is modeled and how
long stars will remain on the AGB.  Recently, APOGEE $H$-band
abundances of Galactic GC giant stars combined with ground-based
photometry suggest that the ratio of AGB to RGB stars may be higher in
the $H$-band than previously determined from the optical
(\citealt{GarciaHernandez2015}, Meszaros et al., {\it in prep.};
however, possible selection effects are very difficult to evaluate).
AGB stars are likely to have a strong effect on an $H$-band IL
spectrum, and may be a source of systematic errors in the integrated
abundances. To test the effect of a high AGB/RGB ratio in the
$H$-band, the number of AGB stars was manually increased in a GC with
$[\rm{Fe/H}] = -0.5$, and the effects on the derived abundances were
determined.  In general, when the number of AGB stars increases, the
IL spectral lines become stronger---this means that lower abundances
are needed to fit  the observed spectra.  However, for reasonable
AGB/RGB ratios, these offsets are $\lesssim 0.05$ dex. Significant
abundance differences ($\sim 0.1$ dex) are not evident until the
AGB/RGB ratio becomes fairly high ($\sim 50$\%)---even then, the
offsets in differential [X/Fe] ratios are lower than 0.1 dex.  It
therefore seems that the AGB/RGB ratio is not likely to significantly
affect the derived IL abundances.

Stochastic sampling of the AGB and RGB may have a larger effect on the
$H$-band abundances, even in these massive clusters; this will be
explored in a future paper.

\subsection{The Chemical Evolution of M31's GCs}\label{subsec:ChemEvol}
\begin{figure}[h!]
\begin{center}
\centering
\includegraphics[scale=0.55]{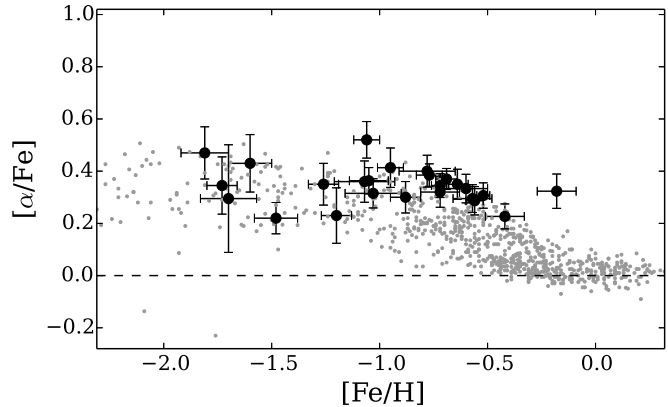}
\caption{[$\alpha$/Fe] vs. [Fe/H] in the M31 GCs (black circles) and
  MW field stars (grey points; from the sources in \citealt{Venn2004},
  with supplements from \citealt{Reddy2006}).\label{fig:aFe}}
\end{center}
\end{figure}

In most GCs, Si, Ca, Ti, and Fe are not expected to vary between stars
within the GC, and their integrated abundances should reflect the
abundances of a GC's birth environment.  $H$-band Si, Ca, Ti, and Fe
can therefore be used for chemical tagging and chemical evolution
studies, as in the optical (e.g.,
\citealt{Colucci2012,Colucci2013,Colucci2014,Sakari2015}).  Figure
\ref{fig:aFe} shows $H$-band integrated [$\alpha$/Fe] vs. [Fe/H] for
the M31  GCs, where $\alpha$ is an average of Si, Ca, and Ti (or one
or two of the elements if all three are not available).  Also shown
are MW field stars, from \citet{Venn2004} and \citet{Reddy2006}.  M31
field stars are not shown because the field stars are too faint for
high quality analyses; however, the M31 field stars are expected to
follow the MW field stars, especially in this metallicity range.  As
expected from optical analyses, the M31 GCs do generally track the MW
field stars.  As in the optical, B193's [$\alpha$/Fe] is slightly
higher than MW field stars at the same $[\rm{Fe/H}]\sim -0.2$ (see
\citealt{Colucci2014}); this indicates that B193 may have formed in an
environment that experienced rapid chemical enrichment.

However, high precision Ca and Ti abundances can be obtained from the
optical along with abundances of a wider variety of elements---is the
optical therefore the preferred wavelength regime for IL analyses, or
does the $H$-band offer any significant advantages over the optical?
Section \ref{subsec:FeDiscussion} demonstrated that $H$-band
abundances are less sensitive to GC age (and HB morphology) than the
optical.  In situations when GC age is poorly constrained, systematic
uncertainties may be lower in the $H$-band.  Furthermore, GCs that are
heavily obscured and reddened by dust will be much easier to observe
in the IR than in the optical.

It is also worth noting that there are additional spectral lines in
the $J$ and $K$ bands.  Expanded IR IL spectroscopy with more
wavelength coverage would increase the number of observed elements and
the precision in a single abundance.

\section{Conclusions}\label{sec:Conclusion}
This paper has presented an $H$-band integrated light spectroscopic
analysis of 25 bright GCs that are associated with M31. The target GCs
span a wide range in metallicity (from $[\rm{Fe/H}] = -1.8$ to
$-0.2$), a moderate range in total mass (from $\log \rm{mass} \sim
5.4$ to $\sim 6.5$), and a small range in age (from $\sim 6.5$ to $14$
Gyr).   All the GCs were previously targeted for high-resolution
optical spectroscopy, enabling a comparison between $H$-band and
optical abundances.  The primary results from this study are as
follows.

\begin{enumerate}
\item The $H$-band offers a wide variety of spectral lines that
  complement the optical.  In addition to the handful of additional
  \ion{Fe}{1} lines, the $H$-band offers intermediate and strong
  \ion{Mg}{1}, \ion{Al}{1}, and \ion{Si}{1} lines and weaker Na, Ca,
  Ti, and K lines.  Molecular CO, CN, and OH features allow
  determinations of C, N, and O abundances.  $^{12}$C/$^{13}$C ratios
  cannot be well constrained in these spectra, but could be measurable
  in higher quality spectra from metal-rich targets.

\item The \ion{Fe}{1} lines in the $H$-band provide Fe abundances that
  are in excellent agreement with the optical abundances.  Although
  there is a small offset ($\sim 0.05$) between the optical and
  $H$-band [\ion{Fe}{1}/H] ratios, this may reflect different NLTE
  corrections between the optical and IR (e.g.,
  \citealt{GarciaHernandez2015}).  Despite the small offset, the
  $H$-band lines agree with the optical trends in Fe abundance with
  line wavelength, reduced equivalent width, and excitation potential.
  However, the parameter ranges are smaller amongst the $H$-band
  lines, and it may not be possible to constrain the GC age without
  complementary optical data (photometric or spectroscopic).  However,
  the $H$-band abundances ratios are relatively insensitive to GC age,
  at least for GCs older than $\sim 3$ Gyr.

\item The $H$-band [C/Fe] and [N/Fe] abundances reflect typical tip of
  the RGB stellar abundances.  However, they are systematically offset
  from the C and N abundances derived from the optical 4668
  \AA \hspace{0.025in} Lick index.

\item With a few exceptions, the abundances of Na, Mg, Al, Si, and Ca
  are in excellent agreement between the optical and the $H$-band.
  The $H$-band \ion{Ti}{1} abundances agree well with the optical
  \ion{Ti}{2} abundances, suggesting that the $H$-band lines may be
  less sensitive to NLTE effects than the optical ones.

\item The $H$-band offers new [O/Fe] and [K/Fe] abundances that are
  not available in the optical.  The K abundances trace Ca, as
  predicted from stars in the MW.

\item With the detailed $H$-band abundances, multiple populations in
  extragalactic GCs can be explored in new detail. The integrated
  [Na/O] ratio is found to roughly correlate with cluster mass,
  suggesting that the relative numbers of ``second generation'' stars
  may increase with cluster mass.  No convincing similar trend was
  found for [Mg/Al].

\item As expected for GCs associated with a massive spiral galaxy, the
  $H$-band [$\alpha$/Fe] ratios track MW field stars, demonstrating
  that $H$-band IL spectroscopy can be utilized for chemical tagging
  analyses of unresolved targets.
\end{enumerate}

Thus, $H$-band integrated light spectroscopy will be a valuable tool for
studying more distant, unresolved stellar populations, particularly
those that are highly reddened.

\acknowledgements
The authors thank the anonymous referee for suggestions that improved
the manuscript.  The authors also thank Chris Sneden for developing
the {\it synpop} version of {\tt MOOG}, Andrew McWilliam for
discussions regarding integrated light analyses, and Masen Lamb for
discussions regarding IR syntheses.  C.M.S. acknowledges funding from
the Kenilworth Foundation.  
D.B. acknowledges support from grant RSF 14-50-00043.
CAP is thankful to the Spanish MINECO for funding through grant AYA2014-56359-P.
T.C.B. acknowledges partial support for
this work from grants PHY08-22648; Physics Frontier Center/Joint
Institute or Nuclear Astrophysics (JINA), and PHY 14-30152; Physics
Frontier Center/JINA Center for the Evolution of the Elements
(JINA-CEE), awarded by the US National Science Foundation.
D.A.G.H. was funded by the Ram\'{o}n y Cajal fellowship number
RYC-2013-14182 and he acknowledges support provided by the Spanish
Ministry of Economy and Competitiveness (MINECO) under grant and
AYA-2014-58082-P.
J.S. acknowledges partial support from NSF grant AST-1514763 and the
Packard Foundation.

\clearpage
\footnotesize{

}

\clearpage
\normalsize

\appendix

\section{New Optical Abundances for Five M31 Clusters}\label{appendix:OpticalAbunds}
In addition to the $H$-band data presented in this paper, optical
abundances are derived for five additional clusters.  These data
supplement the literature data from \citet{Colucci2009,Colucci2014}.

\subsection{Observations and Data Reduction}
The new optical spectra were obtained in 2009 and 2010 with the High
Resolution Spectrograph (HRS; \citealt{HRSref}) on the Hobby-Eberly Telescope
(HET; \citealt{HETref,HETQueueref}) at McDonald Observatory in Fort
Davis, TX; details are shown in Table \ref{table:OpticalObservations}.
The observations and data reduction were carried out in the same way
as the M31 targets in \citet{Sakari2015}.  Briefly, the $1\arcsec$
slit was used, yielding a spectral resolution of $R = 30,000$.  The
600 gr/mm cross disperser was positioned to allow spectral coverage
from $\sim 5320-6290$ and $\sim 6360-7340$ \AA \hspace{0.025in} in the
blue and the red, respectively.  The large $3\arcsec$ fibers were used
to cover the GCs past their half-light radii; fibers located
$10\arcsec$ from the central fiber provided simultaneous sky and
background observations, which were subtracted during the data
reduction.  Hot stellar standards were also observed for removal of
telluric features.

The data were reduced in IRAF, utilizing variance weighting during
spectral extraction.  Normalizations were performed utilizing
continuum fits to an extremely metal-poor star, with additional
low-order polynomial fits (see \citealt{Sakari2013}).  Individual
stellar spectra were cross-correlated with a high-resolution Arcturus
spectrum \citep{Hinkle2003} and shifted to the rest frame.  The
individual observations were combined with average sigma-clipping,
weighted by flux.  Velocity dispersions were then determined from a
final cross-correlation with Arcturus, as described in
\citet{Sakari2013}.

\begin{table}
\centering
\begin{center}
\caption{Observation information and derived isochrones for new optical data.\vspace{0.1in}\label{table:OpticalObservations}}
  \begin{tabular}{@{}llcccccc@{}}
  \hline
 & Observation & $t_{\rm{exp}}$ & S/N$^{a}$  & $v_{\rm{helio}}$ & $\sigma$ & Isochrone & Age\\
 & Dates       & (s) & (6000 \AA) & (km s$^{-1}$)  & (km
  s$^{-1}$) & [Fe/H]  & (Gyr)\\
\hline
B006  & 2009 Oct 11, 14, 18, 19, 20 & 10565 & 140 & $-236.3 \pm 0.5$ & $10.56 \pm 0.4$ & $-0.70$ & 12 \\
B063  & 2009 Nov 15, 25, & 11997 & 200 & $-304.2 \pm 0.5$ & $15.15 \pm 0.6$ & $-1.01$ & 14 \\
 & \phantom{2009} Dec 12, 13, 16 & & & & & & \\
B171  & 2009 Dec 17, 18, 20, 21 & 16200 & 250 & $-267.7 \pm 0.5$ & $16.81 \pm 0.5$ & $-0.69$ & 13 \\
      & 2010 Jan 10, 18 & & & & & & \\
B311  & 2009 Oct 24, 27, & 11016 & 160 & $-515.6 \pm 1.0$ & $12.70 \pm 0.4$ & $-1.62$ & 13 \\
 & \phantom{2009}  Nov 12, 17 & & & & & & \\
B472  & 2009 Dec 23, 2010,  & 10800 & 120 & $-120.9 \pm 0.5$ & $14.37 \pm 0.6$ & $-1.01$ & 11 \\
 & \phantom{2009} Jan 4, 5, 13 & & & & & & \\
 & & & & & & & \\
\hline
\end{tabular}\\
\end{center}
\medskip
\raggedright $^{a}$ S/N is per resolution element; there are 2.7
pixels per resolution element.\\
\end{table}

\subsection{Best-fitting Isochrones}\label{subsec:OpticalIsos}
Several of these clusters have been partially resolved with {\it HST},
allowing constraints to be placed on age and metallicity---however,
these CMDs cannot resolve the inner regions, nor do they reach the
main sequence turnoff.  As in \citet{McWB},
\citet{Colucci2009,Colucci2011a,Colucci2014}, and \citet{Sakari2015},
appropriate isochrone parameters can be determined spectroscopically
by minimizing line-to-line trends in \ion{Fe}{1} abundance with
wavelength, reduced equivalent width, and excitation potential.  As
described in Section \ref{subsec:ModelAtms}, BaSTI isochrones are
utilized, since they model the evolved HB and AGB stars.  The
parameters of the spectroscopically-derived isochrones are shown in
Table \ref{table:OpticalObservations}---the values agree well with the
literature values.

\subsection{Detailed Abundances}\label{subsec:OpticalAbunds}
The abundances of elements with clean, unblended lines were determined
through equivalent width (EW) analyses---this includes Fe, Si, Ca, Ti,
and Ni. EWs were measured with the automated program {\tt DAOSPEC}
(\citealt{DAOSPECref}; also see \citealt{Sakari2013}), which are
provided in Table \ref{table:EWs}.  Abundances were calculated with
the IL EW analysis task {\it abpop} (in the June 2014 version of the
code {\tt MOOG}) and are shown in Table \ref{table:OpticalAbunds}.  Errors are
calculated as in \citet{Shetrone2003} and \citet{Sakari2015}.

\begin{table}
\centering
\begin{center}
\caption{Line list for optical abundances.\label{table:EWs}}
  \newcolumntype{d}[1]{D{,}{\pm}{#1}}
  \begin{tabular}{@{}lcccccccc}
  \hline
 & & & & \multicolumn{5}{l}{EW (m\AA)} \\
Wavelength (\AA) & Element & EP (eV) & $\log gf$ & B006 & B063 & B171 & B311 & B472 \\
\hline
 & & & & & & & & \\
 5324.191 & 26.0 & 3.211 & -0.103 & $-$   & $-$   & $-$   & 110.9 & $-$\\
 5367.476 & 26.0 & 4.420 &  0.443 & $-$   & $-$   & 140.4 & $-$   & 93.3\\
 5369.974 & 26.0 & 4.371 &  0.536 & $-$   & $-$   & $-$   &  69.4 & $-$\\
 5383.380 & 26.0 & 4.312 &  0.645 & $-$   & $-$   & $-$   &  73.3 & 114.0\\
 5389.486 & 26.0 & 4.42  & -0.410 & 73.0  &  59.3 & 100.7 & $-$   & 44.8\\
 5393.176 & 26.0 & 3.240 & -0.715 & $-$   & 114.1 & $-$   & $-$   & 105.1\\
 5405.785 & 26.0 & 0.990 & -1.852 & $-$   & $-$   & $-$   & 125.0 & $-$\\
 5424.080 & 26.0 & 4.320 &  0.520 & $-$   & 109.8 & $-$   & $-$   & 102.8\\
 5429.706 & 26.0 & 0.958 & -1.881 & $-$   & $-$   & $-$   & 149.2 & $-$\\
 & & & & & & & & \\
\hline
\end{tabular}\\
\end{center}
\medskip
\raggedright  {\bf Notes: } Table \ref{table:EWs} is published in its
entirety in the electronic edition of \textit{The Astrophysical Journal}. A portion is shown here for guidance
regarding its form and content.\\
\end{table}

The abundances of the other elements were derived from spectrum
syntheses with {\tt MOOG}'s {\it synpop} routine (see Section
\ref{subsec:Synpop}).  Abundances of Na, Mg, and Eu were
determined with the complete line lists from \citet{Sakari2013}.
Additionally, Ba and Al lines were synthesized with new line lists
that include atomic lines, isotopic information, hyperfine structure,
and molecular lines, where appropriate.  The best fits were identified
by eye, accounting for uncertainties in line profiles and continuum
placement.

As in \citet{Sakari2013,Sakari2014,Sakari2015}, all optical [Fe/H] and
[X/Fe] ratios are calculated differentially, line-by-line, relative to
the solar abundances derived with the same techniques.  Three of the
targets overlap with \citet{Colucci2014}---the values are generally in
good agreement.

\begin{table}
\centering
\begin{center}
\caption{Optical abundances.\label{table:OpticalAbunds}}
  \newcolumntype{d}[1]{D{,}{\pm}{#1}}
  \begin{tabular}{@{}ld{3}cd{3}cd{3}cd{3}cd{3}c}
  \hline
Element & \multicolumn{2}{l}{B006} & \multicolumn{2}{l}{B063} & \multicolumn{2}{l}{B171} & \multicolumn{2}{l}{B311} & \multicolumn{2}{l}{B472} \\
 & \multicolumn{1}{c}{Abundance} & $N$ & \multicolumn{1}{c}{Abundance} & $N$ & \multicolumn{1}{c}{Abundance} & $N$ & \multicolumn{1}{c}{Abundance} & $N$ & \multicolumn{1}{c}{Abundance} & $N$ \\
\hline
$[$\ion{Fe}{1}/H$]$ & -0.73,0.02 & 39 & -1.10,0.02 & 35 & -0.53,0.02 & 37 & -1.79,0.04 & 21 & -1.17,0.03 & 48 \\
$[$\ion{Fe}{2}/H$]$ & -0.68,0.14 &  3 & -1.25,0.03 &  2 & -0.57,0.04 &  3 & -1.63,0.04 &  2 & -1.22,0.15 & 1 \\
$[$\ion{Na}{1}/\ion{Fe}{1}$]$ & 0.42,0.11 & 2 & 0.43,0.13 & 2 & 0.64,0.16 & 2 & 0.51,0.20 & 1 & \multicolumn{2}{c}{---}  \\
$[$\ion{Mg}{1}/\ion{Fe}{1}$]$ & 0.46,0.10 & 1 & 0.38,0.09 & 2 & 0.32,0.15 & 1 & 0.02,0.12 & 2 & 0.09,0.09 & 2 \\
$[$\ion{Al}{1}/\ion{Fe}{1}$]$ & 0.56,0.13 & 2 & 0.44,0.15 & 1 & 0.54,0.09 & 2 &\multicolumn{2}{c}{---} & 0.74,0.15 & 1 \\
$[$\ion{Si}{1}/\ion{Fe}{1}$]$ & 0.46,0.05 & 2 & \multicolumn{2}{c}{---} & 0.31,0.07 & 4 & 0.37,0.10 & 2 & 0.09,0.13 & 2 \\
$[$\ion{Ca}{1}/\ion{Fe}{1}$]$ & 0.26,0.02 & 4 & 0.42,0.05 & 4 & 0.38,0.10 & 3 & 0.28,0.05 & 7 & 0.28,0.06 & 6 \\
$[$\ion{Ti}{1}/\ion{Fe}{1}$]$ & 0.17,0.05 & 4 & 0.24,0.03 & 2 & 0.29,0.13 & 3 & 0.25,0.10 & 1 & \multicolumn{2}{c}{---}  \\
$[$\ion{Ti}{2}/\ion{Fe}{2}$]$ & 0.45,0.18 & 2 & 0.23,0.10 & 1 & 0.34,0.06 & 2 & 0.26,0.08 & 1 & 0.45,0.04 & 2 \\
$[$\ion{Ni}{1}/\ion{Fe}{1}$]$ & 0.02,0.05 & 6 &-0.11,0.03 & 4 &-0.09,0.07 & 5 &-0.23,0.08 & 2 &-0.19,0.18 & 3 \\
$[$\ion{Ba}{2}/\ion{Fe}{2}$]$ & 0.31,0.10 & 2 & 0.15,0.13 & 2 &-0.40,0.20 & 2 & \multicolumn{2}{c}{---} & -0.19,0.18 & 2 \\
$[$\ion{Eu}{2}/\ion{Fe}{2}$]$ & 0.64,0.15 & 1 & 0.40,0.25 & 1 & 0.33,0.35 & 1 & \multicolumn{2}{c}{---} &  0.67,0.15 & 1 \\
 & & & & & & & & & & \\
\hline
\end{tabular}\\
\end{center}
\medskip
\end{table}

\end{document}